\documentclass{article}

\usepackage{microtype}
\usepackage{graphicx}
\usepackage{subfigure}
\usepackage{booktabs} %

\usepackage[accepted]{icml2018}

\icmltitlerunning{Adversarial GMM}

\newcount\Colorflag
\usepackage{amsthm,amsmath,amssymb}
\usepackage{color}
\usepackage{nicefrac}
\usepackage{comment}
\usepackage{bbm}



\usepackage[]{color-edits}
\addauthor{vs}{blue}
\addauthor{gl}{red}

\newcommand{\E}{\mathbb{E}}

\newcommand{\R}{\mathbb{R}}

\renewcommand{\Pr}{\ensuremath{\mathrm{Pr}}}

\newtheorem{theorem}{Theorem}[]

\newtheorem{definition}[]{Definition}

\newcommand{\mcR}{{\mathcal R}}

\newcommand{\mcA}{{\mathcal A}}
\newcommand{\mcH}{{\mathcal H}}

\newcommand{\mcN}{{\mathcal N}}

\newcommand{\mcX}{{\mathcal X}}

\newcommand{\mcG}{{\mathcal G}}
\newcommand{\mcF}{{\mathcal F}}

\newcommand{\ba}{\begin{array}}
\newcommand{\ea}{\end{array}}
\newcommand{\bs}{\begin{align}\begin{split}\nonumber}
\newcommand{\bsnumber}{\begin{align}\begin{split}}
\newcommand{\es}{\end{split}\end{align}}

\newcommand{\ldot}[2]{\langle #1, #2 \rangle}

\newcommand{\MSE}{\ensuremath{\text{MSE}}}

\usepackage{hyperref}

\usepackage{lipsum}

\newcommand\blfootnote[1]{%
  \begingroup
  \renewcommand\thefootnote{}\footnote{#1}%
  \addtocounter{footnote}{-1}%
  \endgroup
}

\begin{document}

\twocolumn[
\icmltitle{Adversarial Generalized Method of Moments}

\icmlsetsymbol{equal}{*}

\begin{icmlauthorlist}
\icmlauthor{Greg Lewis}{equal,msrnb}
\icmlauthor{Vasilis Syrgkanis}{equal,msr}
\end{icmlauthorlist}

\icmlaffiliation{msrnb}{Microsoft Research and NBER}
\icmlaffiliation{msr}{Microsoft Research}

\icmlcorrespondingauthor{Greg Lewis}{glewis@microsoft.com}
\icmlcorrespondingauthor{Vasilis Syrgkanis}{vasy@microsoft.com}

\icmlkeywords{Deep Learning, Econometrics, Causal Inference, Generalized Method of Moments}

\vskip 0.3in
]

\printAffiliationsAndNotice{\icmlEqualContribution} %

\begin{abstract}
We provide an approach for learning deep neural net representations of models described via conditional moment restrictions. Conditional moment restrictions are widely used, as they are the language by which social scientists describe the assumptions they make to enable causal inference. We formulate the problem of estimating the underling model as a zero-sum game between a modeler and an adversary and apply adversarial training. Our approach is similar in nature to Generative Adversarial Networks (GAN), though here the modeler is learning a representation of a function that satisfies a continuum of moment conditions and the adversary is identifying violating moments. We outline ways of constructing effective adversaries in practice, including kernels centered by k-means clustering, and random forests. We examine the practical performance of our approach in the setting of non-parametric instrumental variable regression.
\end{abstract}

\section{Introduction}
\blfootnote{Prototype code of the presented methods can be found here: \href{https://github.com/vsyrgkanis/adversarial_gmm}{https://github.com/vsyrgkanis/adversarial\_gmm}}

Understanding how policy choices affect social systems requires an understanding of the causal relationships between those policies and the outcomes of interest.
To measure these causal relationships, social scientists look to either field experiments, or quasi-experimental variation in observational data.
Most of the observational studies rely on assumptions that can be formalized in moment conditions.
This is the basis of the estimation approach  known as generalized method of moments (GMM), for which Lars Hansen won a Nobel Prize.

While GMM is an incredibly flexible estimation approach, in practice its usage is confined to some special cases.
One reason for this is that the underlying independence (randomization) assumptions often imply an infinite number of moment conditions.
Imposing all of them is infeasible with finite data, but it is hard to know which ones to select.
For some special cases, asymptotic theory provides some guidance, but it is not clear that this guidance translates well when the data is finite and/or the models are non-parametric.  
Given the increasing availability of data and new machine learning approaches, researchers and data scientists may want to apply adaptive non-parametric learners such as neural nets and trees to these GMM estimation problems, but this requires a way of selecting moment conditions \emph{that are adapted to the hypothesis class of the learner.}

This paper offers a solution.
We take the problem of selecting moment conditions and make it part of the overall estimation problem.
Specifically, we consider environments in which underlying parameters are partially identified by a set of conditional moment restrictions.
We show that the identified set is equivalently characterized as the solution to a zero-sum game between two players: a modeler and an adversary (Theorem 1).  
The modeler aims to minimize a GMM criterion function over a finite set of unconditional moments characterized by test functions, while the adversary tries to pick those test functions so as to maximize the criterion.
We show that under some relatively weak conditions, the solution to this zero-sum game can be found by an adaptive best response process, where in each round the modeler and adversary simultaneously update their parameters and the test functions as a best response to the last round of play.   
In this way the adversary continually provides feedback to the modeler, similar to the approach taken in Generative Adversarial Networks (GANs).
This feedback is adapted to the current weaknesses of the modeler.  

Under some conditions, training can be done using stochastic gradient descent, the workhorse tool in training deep neural nets (DNNs).
So one possible ``deep'' architecture is one in which the modeler learns a neural net while the adversary simultaneously learns the bandwidth of a set of Gaussian kernels over the space of instruments, and how to weight them.
We prove that such an architecture has good generalization performance, in the sense that it identifies neural nets that with high probability do not substantially violate the moment conditions out of sample (Theorem 2).
We also suggest architectures in which the adversary learns random forests or neural nets, which may be more suitable in high dimensions.

The paper concludes with simulation experiments for the case of non-parametric instrumental variables (IV).  
Our estimation approach performs better than a number of leading alternatives including a two-stage polynomial basis expansion and the deep IV approach of \citet{deepiv}.

\paragraph{Related literature.}
There is a large literature on GMM  \cite{Hansen1982}, including asymptotic theory on the efficient (infeasible) instruments in the non-parametric case \cite{Chamberlain1987} as well as some practical guidance on how to choose them by use of an auxiliary model \cite{GallantTauchen96}. 
There has been little work combining machine learning with generalized method of moments.  
A recent exception is the generalized random forest approach of \citet{Athey2016}.
There, a random forest is used to detect heterogeneity in treatment effects across a covariate set. Those treatment effects themselves are then solved for on each leaf using a ``local'' GMM estimation equation.  
Our paper complements their work: we show how to solve such GMM estimation problems when the both the set of available moment conditions and the hypothesis space may be large. 

Moreover, our paper has strong connection to the Generative Adversarial Network (GAN) literature \cite{Goodfellow14,Goodfellow17,arjovsky2017wasserstein}, as we use adversarial training to learn a model of the world. However, unlike GANs our modeler is learning a counterfactual model rather than a distribution of data points, and the adversary is identifying conditional moment restrictions that are violated, rather than trying to distinguish between true and fake samples as in DCGANs \cite{Goodfellow14}. Our problem is similar in spirit to Wasserstein GANs \cite{arjovsky2017wasserstein}, where the adversary is trying to identify moments/functions of the distribution of the generator, that differ from the corresponding moments of the true distribution of data. In our setting, the adversary is trying to identify conditional moment restrictions that are violated by the model of the modeler. For that reason, both the architecture of the modeler and that of the critic are inherently different from the ones used in GANs. Our work also has connections with recent work on CausalGANs for learning graphical models \cite{Kocaoglu2017}. Even though causal GANs also attempt to learn causal relationships about the world, the applications and the architecture is inherently different from our work. In particular, \citet{Kocaoglu2017} look at primarily vision tasks and learning distributions of images that satisfy a probabilistic graphical model structure. Our goal is to enable deep learning for economic applications, where the causal estimation problems are typically formulated as a set of conditional moment restrictions, rather than a probabilistic graphical model constraint.

\section{The Conditional Moment Problem}

We consider the problem of estimating a flexible econometric model $h$ that satisfies a set of conditional moment restrictions:
\begin{equation}\label{eqn:cond-moments}
     \E[\rho(z; h) | x] = 0
\end{equation}
where $x\in \mcX\subseteq \R^{d}$, $z\in \R^{p}$, $h\in \mcH$ for $\mcH$ a compact convex hypothesis space  and $\rho: \R^{p} \times \mcH \rightarrow \R^{q}$. 
The truth is some model $h_0$ that satisfies all the moment restrictions.  
We will denote with $\mcH_{I}$ the set of all such models (the \emph{identified set}). 
If $\mcH$ is small relative to the set of conditional moment restrictions implied by the model, then $h_0$ may be uniquely pinned down by~(\ref{eqn:cond-moments}).
This is the case of \emph{point identification}.

But since the influential work of \citet{Manski1989}, there has been increasing interest in the case where the moment restrictions only suffice for partial identification.
In this case all that can be learned, even with infinite data, is the set $\mcH_{I}$.  
Our goal is to find a model/hypothesis $h\in \mcH_I$, with probability going to one as the size of the dataset goes to infinity.
Since $h$ is observationally equivalent to $h_0$ --- at least with respect to the moments of interest --- this seems to be all one can reasonably hope for.
It also sidesteps the tricky question of whether the flexible DNN representations we consider are point identified.

\subsection{Equilibrium Formulation}
We start our analysis with the observation that we can replace our conditional moment restrictions with a infinite set of unconditional moments of the form:
\begin{equation}
    \mcH_{I} \equiv \{h \in \mcH: \forall f \in \mcF: \E[\rho(z; h) f(x) ] =0\}
\end{equation}
where $\mcF$ is the set of functions $f : \mcX \rightarrow [0,\infty)$. It is easy to see that this is equivalent using as test functions the Dirac functions at each $x \in \mcX$.

Any hypothesis $h \in \mcH_{I}$ must then also be a solution of the following minimax loss minimization problem:
\begin{equation}\label{eqn:minimax}
    \min_{h\in \mcH} \max_{f \in \mcF} \left(\E[\rho(z; h) f(x)] \right)^2
\end{equation}
The equivalence follows from the fact that any model that is a zero for all test functions must also be so for the worst possible test function; and that the minimum of the squared loss occurs at zero.  Conversely, any solution to the minimax problem must attain a zero for every test function, and so is in the identified set.  

This characterization of the identified set is impractical.  We would like to use the data to learn a model $h$ that has good worst-case generalization performance over test functions $f$, but the max operator makes this objective function hard to optimize.  One of the main insights of the paper is that we can instead turn this into a zero-sum game, where the minimizing player, the \emph{modeler}, wants to choose a model $h\in \mcH$ so as to minimize the violation of the moment conditions and the maximizing player, the \emph{adversary}, wants to choose a distribution $\sigma$ over test functions $f \in \mcF$, so as to maximize the expected violation of moments. We can find a Nash equilibrium of this game by an iterative best response algorithm, which in our case will be alternating stochastic gradient descent.

Von Neumann's classic minimax theorem provides conditions under which the Nash equilibrium is in fact a solution to the minimax problem defined by Equation \eqref{eqn:minimax}. The main requirement is that the modeler's problem is convex and the adversary's is concave. To simplify exposition we will assume the former throughout the paper.
On the other hand, the adversary's problem is unlikely to be concave in the original choice set $\mcF$, and so we allow the adversary to play a mixed strategy $\sigma$. 
Thus we have reduced the problem of finding an $h$ in the identified set to finding an equilibrium pair $(h^*, \sigma^*)$ of the zero-sum game.

Transforming the problem to a zero-sum game makes training more feasible. Still, the hypothesis class for the adversary is too large to generalize well or estimate from finite data.  For example, how should one evaluate a Dirac delta test function at a point $x$ where no data has been observed? We solve this in the usual way: by restricting the hypothesis class to something more reasonable, with hyper-parameters that adapt to the size of the data. To achieve this we introduce the notion of a set of $\gamma$-test functions. 
In the algorithms that follow below, we consider an adversary that learns test functions based on kernels or decision trees or neural nets. 
\begin{definition}[$\gamma$-test functions] A set of functions $\mcF$ is a set of $\gamma$-test functions if for any conditional moment $\E[\rho(z; h)|x]$ there exists a function $\bar{f}$ in the convex closure $\bar{\mcF}$ of $\mcF$, such that if the violation of the unconditional moment $\E[\rho(z;h) \bar{f}(x)]$ is smaller than $\epsilon$, then the violation of the conditional moment is smaller than $\gamma(\epsilon)$, i.e.  $\forall x\in \mcX, \exists \bar{f} \in \bar{\mcF}$: 
\begin{equation}
   \left| \E[\rho(z; h)\bar{f}(x)]\right| \leq \epsilon \implies  \left|\E[\rho(z; h) | x]\right| \leq \gamma(\epsilon)
\end{equation}
\end{definition}

For instance, as we show in the supplementary material, if the conditional moments are $\lambda$-Lipschitz in the conditioning variable $x$, then a simple class of test functions are all uniform Kernel test functions of the form $f(x; x_0) = 1\{\|x-x_0\|_{\infty}\leq h\}$ for $x_0$ being a point on a discretization of the space $\mcX$ to multiples of some small number $h$. This constitutes a set of $\gamma$-test functions with $\gamma(\epsilon) = h\lambda + \frac{\epsilon}{\mu h^{d}}$, for $\mu > 0$ a lower bound on the density of $x$. If we have a target $\epsilon$ in mind, we can set the optimal bandwidth $h=\left(\epsilon/\lambda\mu\right)^{1/(d+1)}$, to get $\gamma(\epsilon)=2\lambda^{d/(d+1)}\left(\epsilon/\mu\right)^{1/(d+1)}$. 

We are now ready to state our first theorem, which argues that if hypotheses that approximately satisfy all the moment conditions are close to the identified set, all $\epsilon$-equilibria of the zero-sum game defined by a class of $\gamma$-test functions have at most $\sqrt{\epsilon}$ violations of the unconditional moments (since the value of the game is the square of the maximum moment violation), at most $\gamma(\sqrt{\epsilon})$ violations of the conditional moments and so are close to the identified set: 
\begin{theorem}[Approximate Set Identification]\label{thm:apx-id}
Suppose that for any $h\in \mcH$:
\begin{equation}\label{eqn:distance-violation}
    \left| \E[\rho(z; h) | x]\right| < \epsilon \; \forall x \; \Rightarrow d(h, \mcH_I) \leq \kappa(\epsilon)
\end{equation}
where $d(h, \mcH_I)$ denotes the distance of $h$ to $\mcH$ with respect to some metric on space $\mcH$. Then the Hausdorff distance between the set of hypotheses in the support of any $\epsilon$-equilibrium of the zero-sum game defined by a set $\mcF$ of $\gamma$-test functions and the identified set defined by Equation \eqref{eqn:cond-moments} is at most $\kappa(\gamma(\sqrt{\epsilon}))$.
\end{theorem}

\paragraph{Remark.}
Even though we phrase our approach for conditional moment problems, we can also use the same approach for any over-identified unconditional moment problem, so as to perform optimal moment weighting.
Suppose that instead we are given a set of $m$ moments restrictions of the form:
\begin{equation}
    \forall i=1,\ldots, m: \E[\rho_i(z; h)] = 0
\end{equation}
In this case there is no need to construct an $\epsilon$-cover of our test functions.
Instead, the adversary will best respond by randomizing over the unconditional moment restrictions.

\subsection{Example: Non-Parametric IV Regression}\label{sec:iv-example}
Before moving to the estimation part of our results, we offer a concrete example of a conditional moment problem and how the approximate identification theorem applies. 
We consider the case of IV regression where the data $z$ are a tuple $(x, w, y)$, where $x$ is an instrument, $w$ is a treatment and $y$ is an outcome. 
The structural model of the data that we envision takes the form:
\begin{align}
    y =~& h(w) + e\\
    w =~& f(x) + g(e) + v
\end{align}
where $\E[e] = 0$ and $e \perp x$.
Notice that $w$ is correlated with $e$ via the treatment equation, so that the treatment variable is endogenous.
We have the constraint that $\E[e | x]=0$. 
This enables causal inference about the relationship between $w$ and $y$ in the presence of the latent confounder $e$. Our goal is to estimate $h(x)$ based on this set of conditional moment restrictions: $\E[ y - h(w) | x] = 0$.
Hence, in this setting $\rho(z; h) = y - h(w)$ for $z = (y,w)$.

We investigate whether there is a metric that satisfies our conditions in Theorem \ref{thm:apx-id}. Let $h^*\in \mcH_I$ be a model in the identified set. Let $h$ be a model for which all conditional moments are violated by at most $\epsilon$. Then this implies the following bound on the difference between $h$ and $h^*$:
\begin{align*}
    \epsilon \geq~& \left|\E[y - h(w) | x]\right| \\
    = ~& \left|\E[y - h(w) | x]\right| + \left|\E[y - h^*(w) | x]\right|  \\
    \geq~& \left|\E[h^*(w) - h(w) | x]\right|
\end{align*}
where the second line follows since $\left|\E[y - h^*(w) | x]\right| = 0$ and the third line follows by the triangle inequality.
Thus we can define the metric in the $\mcH$ space as: $\|h - h^*\| = \sup_{x} \left| \E[h^*(w) - h(w) | x] \right|$.
If $x$ and $w$ are perfectly correlated and in one-to-one correspondence, then observe that the latter metric corresponds to simply the infinity norm: $\|h - h^*\|_{\infty} = \sup_{w} \left|h(w) - h^*(w)\right|$. The looser the instrument is correlated with the treatment, then the looser our metric will be. If $x$ is independent of $w$, then our guarantee on being close to the identified set weakens to matching the function $h^*(w)$ in expectation over $w$. For instance, predicting the unconditional mean of $y$'s would be a good such function $h(w)$. Our approach has the nice property that the finite-sample guarantees adapts to the strength of the instrument. The stronger the instrument the better the metric with respect to which we are close to the identified set and the closer we are in finding a function $h$ that is observationally equivalent to some model in the identified set.

\section{Estimation}

We now consider the estimation problem of the hypothesis $h$ from a finite set of data points $\{(z_1, x_1), \ldots, (z_n, x_n)\}$ drawn i.i.d. from the data generating distribution. We will take the standard route to estimation and replace the population problem with the finite sample approximation of it, i.e. for any function let $\E_n$, denote the expectation with respect to the empirical distribution of data, i.e. $\E_n[\rho(z; h)f(x)] = \frac{1}{n}\sum_{i=1}^n \rho(z_i; h) f(x_i)$. Moreover, for simplicity and so as to enable some of the algorithmic approaches that we will invoke in this section, we will restrict ourselves to finite classes of test functions %
We will then consider the zero-sum game defined by the empirical loss function: for $\theta \in \Theta$ and $\sigma \in \Delta(\mcF)$:
\begin{equation}
    L_n(h, \sigma) = \sum_{f \in \mcF_{\epsilon}} \sigma_f  \left(\E_n\left[\rho(z; h) f(x)\right]\right)^2
\end{equation}
To solve this game we will consider flexible parameterizations of the function $h$, denoted by $h_\theta$ for $\theta \in \Theta$, $\Theta$ finite-dimensional.
We assume that the  parameterization of the modeler is flexible enough such that there always exists $\theta^*\in \Theta$ such that $h_{\theta^*}\in \mcH_I$. We will then be interested in finding an equilibrium $(\theta^*, \sigma^*)$ of the zero-sum game defined by the empirical loss $L_n(\theta, \sigma)$.

We also assume the latter loss function is convex in $\theta$.
For instance the latter would be true if $\rho(z; h_\theta)$ is a linear function of $\theta$ (e.g. if $\rho(z; h)$ is linear in $h$ - as in the IV regression example - and $h_\theta(w) = \ldot{\theta}{\phi(w)}$, where $\phi(w)$ is some high-dimensional featurization of the input to the model $h$).
In the experimental part we will take the implied algorithms and apply them to deep neural network representations of $h$, where the featurization $\phi(w)$ is also learned by the modeler via the first few layers of the neural network. The latter setting does not obey the convexity assumption, however training such deep representations via first order gradient descent methods is known to perform well empirically. 

\subsection{Computation: Simultaneous No-Regret Dynamics}
To solve the zero-sum game, we will invoke simultaneous no-regret dynamics.
Specifically we can use online projected gradient descent for the modeler and the Hedge algorithm for the critic, generating the following dynamics:
\begin{equation}\label{eqn:dynamics}
\begin{aligned}
    \theta_{t+1} &~= \Pi_{\Theta}\left(\theta_{t} - \eta_m \nabla_{\theta} L_n(\theta_t, \sigma_t)\right)\\
    \sigma_{f, t+1} &~\propto~ \sigma_{f, t} \cdot \exp\left\{\eta_c \left(\E_n\left[\rho(z; h_{\theta_t}) f(x)\right]\right)^2 \right\}
\end{aligned}
\end{equation}
where $\Pi_{\Theta}(\theta) = \arg\min_{\theta^*\in \Theta} \|\theta - \theta^*\|$, denotes the projection of point $\theta$ onto the space $\Theta$ of parameter and $\eta_m,\eta_c$ are hyper-parameters to be set appropriately.
Then the pair of the average model $\theta^*=\frac{1}{T}\sum_{t=1}^T \theta_t$ and $\sigma^*=\frac{1}{T}\sum_{t=1}^T \sigma_t$, would correspond to an $\epsilon$-equilibrium of the game. The latter follows by standard arguments on solving zero-sum games using no-regret dynamics \cite{Freund1999} (see also the recent general formulations of this result \cite{Shalev2012, Rakhlin2013b, Syrgkanis}). Applying these results yields the following theorem:
\begin{theorem}\label{thm:dynamics}
Suppose that the loss function $L_n(\theta, \sigma)$ is convex in $\theta$ and for all $\theta\in \Theta$, $\|\theta\|\leq B$ and $\sup_{\sigma \in \Delta(\mcF)}\|\nabla_\theta L_n(\theta, \sigma)\|\leq L$. Moreover, suppose that $\sup_{z, x, \theta\in \Theta, f \in \mcF}|\rho(z; h_\theta) f(x)|\leq H$. Then by setting $\eta_m = \frac{B}{L\sqrt{2T}}$ and $\eta_c = \frac{\sqrt{\log(d)}}{H^2\sqrt{2T}}$, we are guaranteed that after $T$ iterations of the training dynamics, the average solutions $\theta^* = \frac{1}{T} \sum_{t=1}^T \theta_t$ and $\sigma^* = \frac{1}{T}\sum_{t=1}^T \sigma_t$ are an $\frac{H^2\sqrt{2\log(|\mcF|)} + BL\sqrt{2}}{\sqrt{T}}$-approximate equilibrium of the zero-sum game defined by $L_n$.
\end{theorem}

\paragraph{Stochastic gradient for the modeler.} In fact the latter would also hold with high-probability if we replaced the gradients in the above equations with unbiased estimates of these gradients. Interestingly, for the loss function $L_n(\theta, \mu)$ we can compute unbiased estimates of the gradient with respect to $\theta$ by using two samples from the distribution of data $z$, since the gradient takes the form:
\begin{align*}
    2 \sum_{f} \sigma_f \cdot \E_n[\rho(z; h_\theta) f(x)] \cdot \nabla_{\theta}  \E_n[\rho(z; h_\theta) f(x)]
\end{align*}
Thus we see that an unbiased estimate of the gradient with respect to $\theta$ takes the form:
\begin{align*}
    \hat{\nabla}_{\theta, t} = 2 \sum_{f} \sigma_f \cdot  \rho(z; h_\theta) f(x) f(\tilde{x})\nabla_{\theta}  \rho(\tilde{z}; h_\theta)
\end{align*}
where $z$ and $\tilde{z}$ are two independent random samples from the data. The latter unbiased estimate only requires two samples from the empirical distribution and can significantly speed up computation during training time. Similarly, for variance reduction, instead of two independent samples, we can also use two batches of independent samples of some mini-batch size $B$.

\paragraph{Bounded gradient bias for the critic.} Unfortunately, the update of the critic does not admit a stochastic version since the expectation over the data $z$ is under the square. Hence, an update for the critic requires calculating the moment violation over a large sample of data, so that concentration inequalities kick in and the bias in the gradient is of negligible size. Typically that would be of order $1/\sqrt{B}$ if we use a data set of $B$ samples. Alternatively, we can use the whole set of samples for the updates of the critic. Fortunately, the update of the critic does not involve calculating gradients of a deep neural net, rather it only requires evaluations of a neural net over a large sample set. If the latter is also costly to implement, then we can interleave multiple steps of the modeler in between one step of the critic.

\subsection{Sample Complexity and Generalization Error}

The dynamics of the previous section are guaranteed to find an equilibrium of the game defined by the empirical distribution loss $L_n$. However, we are interested in connecting an equilibrium of the empirical game with the population game and showing that any equilibrium of the empirical game will also satisfy the population moments of Equation \eqref{eqn:cond-moments} to within some vanishing error.
We require some conditions on the sample complexity of the hypothesis space $\mcH_{\Theta} =\{h_\theta: \theta \in \Theta\}$. 
Recall the Rademacher complexity of a function class $\mcG$ and sample $S=\{z_1,\ldots, z_n\}$:
\begin{equation}
    R(G, S) = \E_{\xi}\left[\sup_{g\in \mcG} \frac{1}{n}\sum_{i=1}^n \xi_i \cdot g(z_i)\right]
\end{equation}
where $\xi_i$ are Rademacher random variables which take values $\{-1, 1\}$ with equal probability.

\begin{theorem}\label{thm:uniform-conv}
Suppose that the Rademcaher complexity of the function class $\mcA = \{\rho(\cdot;h_{\theta})f(\cdot): \theta \in \Theta, f\in \mcF\}$ is upper bounded by $\mcR$ and that the conditions of Theorem \ref{thm:dynamics} hold. Then with probability $1-\delta$, the averages $\theta^*=\frac{1}{T}\sum_{t=1}^T \theta_t$ and $w^*=\frac{1}{T}\sum_{t=1}^T w_t$, correspond to an $\epsilon$-equilibrium of the population game defined by Equation \eqref{eqn:minimax} for $\epsilon=O\left(\frac{H^2\sqrt{2\log(|\mcF|)} + BL\sqrt{2}}{\sqrt{T}} + H\mcR + H^2\sqrt{\frac{\log(1/\delta)}{n}}\right)$. Therefore: $
    d(h_{\theta^*}, \mcH_I) \leq \kappa(\gamma(\sqrt{\epsilon}))$.
\end{theorem}

For the case of neural networks, classical results of \citet{Anthony2009} bound their Rademacher complexity as a function of the number of gates and the number of connections used. Moreover, as we show in the supplementary material for the case of Lipschitz moments, applying the latter theorem to the class of uniform Kernel test functions centered at a uniform grid of appropriate size, can lead to convergence rates of the order of $n^{-1/(2(d+1))}\times \max(\sqrt{d\log(d)}, \sqrt{r\log(r)}, \log(1/\delta))$, where $r$ is the number of parameters of the modeler, with probability $1-\delta$. We remark that an alternative route to connecting the empirical equilibrium with the population one, is to use the fact that our training process is stable in the sense that it employs gradient descent and the hedge algorithm for each player. Both of these algorithms are special cases of the Follow-the-Regularized-Leader algorithms, which are known to be algorithms that are stable with respect to each data point. We defer a more refined analysis of our theoretical results via stability to future work.

\section{The Algorithm: Adversarial GMM}

\begin{algorithm}[htpb]
    \begin{algorithmic}[1]
  \STATE{\textbf{Hyperparameters}: Step sizes $\eta_m, \eta_c$, mini-batch sizes $B_m, B_c$, number of models $M$ to use for averaging, a test function generator $G$.}
  \STATE{\textbf{Input}: data $S=\{(z_1, x_1), \ldots, (z_n, x_n)\}$}
  \STATE{Generate set of test functions $\mcF = G(S)$}
  \FOR{$t = 1 ,\ldots , T$}
    \STATE{Randomly sample with replacement from $S$ three batches of samples $S_1, S_2, S_3$, with sizes $B_m, B_m, B_c$ corrspondingly.}
    
    \STATE{Let $E_S$ denote an expectation with respect to the empirical distribution in a sample $S$. Construct estimates of the gradient of the modeler:
    \begin{align*}
        \hat{\nabla}_{\theta, t} =~& 2 \sum_{f \in \mcF} \sigma_{f,t} \cdot  \E_{S_1}\left[\rho(z; h_{\theta_t}) f(x)\right] \cdot \\
        &~~~~~~~~~~~~~~~~~~~~~~~~\E_{S_2}\left[f(x)\nabla_{\theta}  \rho(z; h_\theta)\right]
     \end{align*}
      and the utility of the critic:
      \begin{align*}
        \hat{U}_{f,t} =~& \left(\E_{S_3}\left[ \rho(z; h_{\theta_t}) f(x)\right]\right)^2, ~\forall f\in\mcF
    \end{align*}}
    \STATE{{\bf Modeler Step.} Take a gradient step for the modeler using projected gradient descent (or any first order algorithm like Adam):
    \begin{align*}
            \theta_{t+1} =~& \Pi_{\Theta}\left(\theta_{t} - \eta_m \hat{\nabla}_{\theta, t}\right)
    \end{align*}}
    \STATE{{\bf Critic Step.} Take a Hedge step for the critic:
    \begin{align*}
            \sigma_{f, t+1} \propto~& \sigma_{f, t} \cdot \exp\left\{\eta_c \hat{U}_{f,t} \right\}, ~\forall f\in \mcF
    \end{align*}}
    \STATE{{\bf Critic Jitter.} If each $f\in \mcF$ is parameterized via some parameter $w_f \in W$, then take a gradient step for $w_f$:
    \begin{align*}
    \nabla_{w_f, t} =~& 2 \E_{S_1}\left[\rho(z; h_{\theta_t}) f(x; w_f)\right] \cdot \\
        &~~~~~~~~~~~~~~~~~~~~~~~~\E_{S_2}\left[\rho(z; h_\theta) \nabla_{w_f} f(x; w_f)\right]\\ 
    w_{f,t+1} =~& \Pi_{W}\left(w_{f, t} + \eta_w \nabla_{w_f, t}\right)
    \end{align*}}
    \ENDFOR
\STATE{Return the average model $h^* = \frac{1}{M}\sum_{t\in I} h_t$, where $I$ is a set of $M$ randomly chosen steps during training.}
\caption{AdversarialGMM}
  \label{alg:deepgmm}
\end{algorithmic}
\end{algorithm}

\begin{algorithm}[htpb]
\begin{algorithmic}[1]
\STATE{\textbf{Hyperparameters}: Number of clusters $K$, minimum radius size $r$.}
  \STATE{\textbf{Input}: data $S=\{(z_1, x_1), \ldots, (z_n, x_n)\}$}
  \STATE{Cluster the data-points $(x_1,\ldots, x_n)$ into $K$ clusters using $K$-means.}
  \STATE{For each cluster $i\in [K]$ let $f_i:\mcX\rightarrow R$ be a Gaussian Kernel:
  \begin{equation}
      f_i(x) = \frac{1}{(2\pi\sigma^2)^{1/2d}} \exp\left\{-\frac{\|x - x_i\|_W^2}{2\sigma^2}\right\}
  \end{equation}
  with standard deviation $\sigma$ equals to twice the distance to the $r$-th closest data point to the centroid of cluster $i$, with respect to a $W$-matrix weight norm:
  \begin{equation}
  \|x\|_W^2 = x^T W x
  \end{equation}
  for some positive definite matrix $W=VV^T$.}
  \STATE{Return $\mcF = \{f_i: i\in [K]\}$}
\end{algorithmic}
\caption{KMeans based test function generation}\label{alg:test}
\end{algorithm}

Given the theory laid out in the previous sections we are now ready to state the main algorithm that we deployed in practice. In the previous section, we developed formal rates of convergence of the model estimated by our simultaneous training dynamics, as a function of the class of test functions $\mcF$ used. We also instantiated this class of test functions for the case of Lipschitz moments for the class of uniform kernels centered on a grid of points and derived worst-case concrete rates of convergence. However, in practice, taking a grid of points might be a computationally very heavy approach and might also not use the structure of the data. For this reason we propose that instead of discretizing the whole space of $\mcX$ we rather cluster the data points into small clusters based on some criterion and then place local kernels around these clusters.

The clustering approach that worked well in the experiments was to construct a K-means clustering of the data points $x_1, \ldots, x_n$ (i.e. the conditioning variables). Then consider the set of functions that correspond to Gaussian kernels around each centroid of the cluster and with standard deviation large enough to encompass the points in the cluster.\footnote{Alternatively, a uniform kernel that puts equal weight on the points of each cluster and zero weight outside can also be used. We found experimentally that the hard boundaries of the uniform kernel lead to poorer performance.} This gives rise to our main Algorithm \ref{alg:deepgmm}, with Subprocess \ref{alg:test}.

\paragraph{Alternative Designs of $\mcF$.} We propose several alternative approaches to constructing the set of test functions $\mcF$ that might perform better dependend on the structure of the moment problem and the parameterization of the hypothesis space:
\begin{itemize}
    \item {\bf Random Data-Points.} Instead of constructing a $K$-means clustering, one could simply take $K$ random data points as centroids and then construct gaussian or other kernels around these data points so as to cover a minimum sized neighborhood of $r$ other points
    \item {\bf Random Forest.} To take advantage of the structure of the problem we build a random forest by regressing the variables $z$ on the variables $x$. This will essentially uncover the dimensions of $x$ that really have any effect on $z$ and hence are useful in identifying the model $h$. Subsequently, we construct a test function for each leaf of the random forest which corresponds to a local kernel at the center of the leaf and covering the data-points of the leaf. One could either use a Gaussian kernel constructed based on the minimum enclosing ellipsoid of the points in the leaf or a uniform kernel on the leaf points.
    \item {\bf Sieve-based test functions.} An alternative to the local test functions, is to construct test functions that correspond to a sieve approximation of any function. One such sieve is the polynomial sieve where the test functions are simply $\mcF = \{x^d: d\in \{0, \ldots, K\}$ for some upper bound degree $K$ (potentially dividing by a normalizing constant so that each test function takes values in the same range). The latter corresponds to the standard approach of semi-parametric estimation in the econometrics literature (see e.g. the recent work on two-step sieve GMM estimation by \cite{Chen2015})
    \item {\bf Neural net based test functions.}  Yet another alternative is to be fully flexible in the creation of the test functions, by representing them as neural networks $f(\cdot; w_1), \ldots, f(\cdot; w_m)$. We want these different test functions to provide a good cover of the test function space. Hence, we can create a secondary game between the neural network critics, e.g. by adding a term in the loss of the critics that represents some metric of similarity with other critics. The latter would incentivize the critics to approximate different combinations of moments. Subsequently a meta-critic will optimize the weight placed on each critic using Hedge. The latter is closer in spirit to Wasserstein GANs \cite{arjovsky2017wasserstein}, which also approximate the class of all $1$-Lipschitz functions with a neural net representation.
\end{itemize}

\section{Experimental Evaluation}

We applied our AdversarialGMM algorithm to the problem of non-parametric instrumental variable regression from Section \ref{sec:iv-example}.

\paragraph{Data Generating Processes.} In each experiment we generated 1000 data points.
We analyzed the experimental performance for the following two data generating processes:

\paragraph{DGP  $1$:}
\begin{align*}
    w =~& (1-\gamma) x_1 + \gamma e + \zeta\\
    y =~&  h_0(w) + e + \delta \\
    e \sim~& \mcN(0, 2),~~ 
    x \sim \mcN(0, 2I_d)\\
    \zeta \sim~& \mcN(0, 0.1),~~
    \delta \sim \mcN(0, 0.1) 
\end{align*}

\paragraph{DGP  $2$:}
\begin{align*}
    w =~& (1-\gamma) \left(x_1 \cdot 1\{x_1>0\} + x_2 \cdot 1\{x_2<0\}\right)
    + \gamma e + \zeta\\
    y =~&  h_0(w) + e + \delta \\
    e \sim~& \mcN(0, 2),~~ 
    x \sim \mcN(0, 2 I_d)\\
    \zeta \sim~& \mcN(0, 0.1),~~
    \delta \sim \mcN(0, 0.1) 
\end{align*}

In other words, we generate $d$ instruments drawn from independent normal distributions. However, in DGP $1$ only the first instrument has any effect on the treatment, while in DGP $2$ only the first and the second instruments have any effect on the treatment. In DGP $1$ the effect of the first instrument is linear. In DGP $2$, the first instrument affects the treatment when it is positive and the second instrument when it is negative. The constant $\gamma$ captures the \emph{strength of the instrument}, since it controls the relative weight between the instrument and the confounding factor $e$. The goal of the multiple instruments is to see if the adversarial GMM algorithm will identify the relevant instruments through learning the appropriate weighting matrix $W=VV^T$ of the Kernel norm. In particular, whether the matrix $V$ will project to the dimensions of the relevant instruments and ignore the rest.

We considered different types of functions for the true model: 
\begin{itemize}
\item $2$-d polynomial (2dpoly): $h_0(w)=-1.5 w + .9  w^2$
\item $3$-d polynomial (3dpoly): $h_0(w)=-1.5 w + .9 w^2 + w^3$
\item Absolute value function (abs): $h_0(w)=|w|$
\item Identity function (linear): $h_0(w)=w$
\item Sigmoid function (sigmoid): $h_0(w)=\frac{2}{1+e^{-2x}}$
\item Sin function (sin): $h_0(w) = \sin(x)$
\item Step function (step): $h_0(w)= 1 + 1.5 \cdot 1\{x \geq 0\}$
\item Random piece-wise linear function (rand\_pw): $h_0(w) = \sum_{i=1}^5 (a_i w + b_i) 1\{x_i\in [\tau_{i-1}, \tau_i]\}$, where $\tau_0=-2$ and $\tau_i$ are chosen uniformly at random over a grid of step $0.1$ in $[-2, 2]$, $a_i$ are chosen uniformly at random in $[-4, 4]$, $b_1$ is chosen uniformly at random in $[-1, 1]$ and $b_i$ for $i>1$ are chosen appropriately so that the whole function is continuous.
\end{itemize}

\begin{figure}
    \centering
    \subfigure[$K$-means clustering]{
        \centering
        \includegraphics[scale=.17]{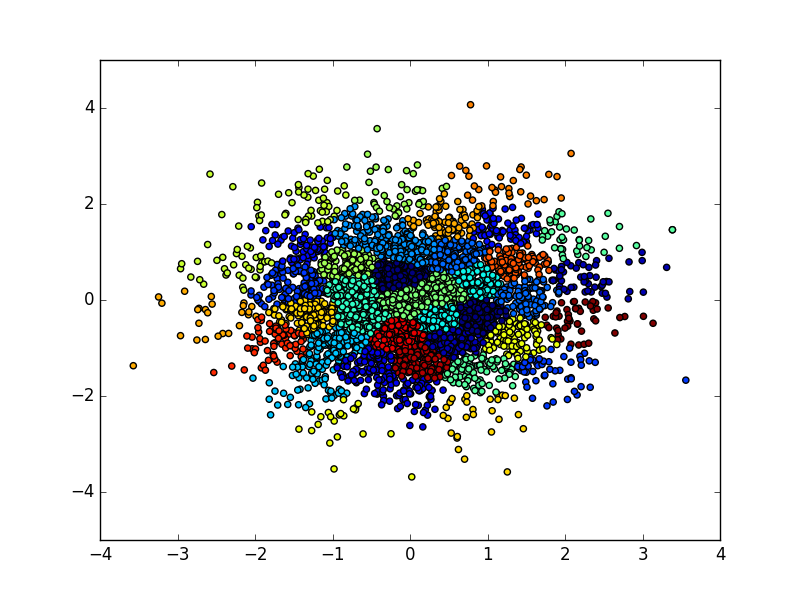}}
	\subfigure[Gaussian kernels centered at k-means clsuter centers]{
        \centering
        \includegraphics[scale=.17]{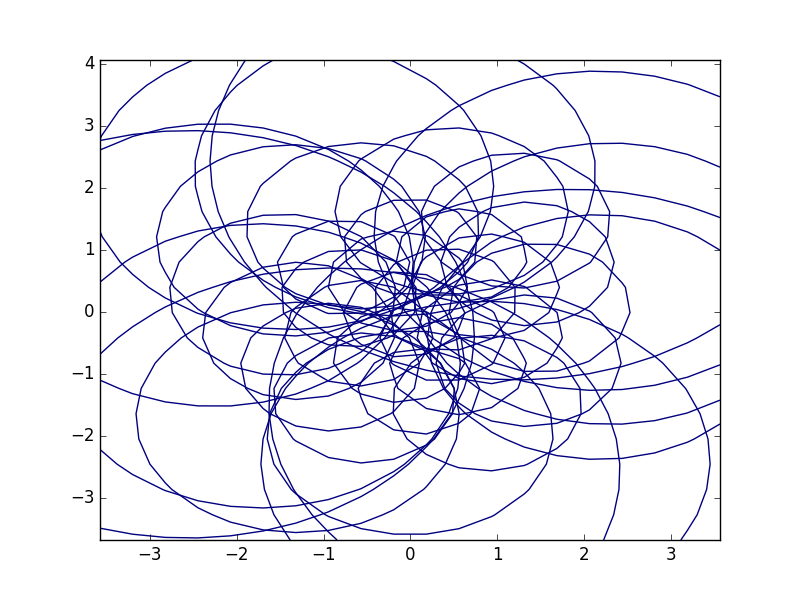}}
    \caption{$K$-means clustering of the two-dimensional space $\mcX$.}
    \label{fig:my_label}
\end{figure}

\paragraph{Training Hyper-Parameters.} We evaluated each of these functions with the same training process without further fine-tuning so that we can argue about their adaptivity to different types of true models.
We applied the adversarial GMM algorithm with $T=400$ steps of the training dynamics, where the modeler was trained with a batch-size of $B=100$ randomly sampled data points (with replacement) per gradient step, while the critic's loss was evaluated on the whole sample set of $n=1000$ points. We updated the critic after every iteration of the modeler and we also updated the weight matrix of the Kernel of the critic functions after every iteration. For the matrix $W=VV^T$, we used an $2\times d$ matrix $V$, i.e. we projected the $d$ instruments into a two-dimensional space that was learned via a gradient method (unless $d=1$ in which case we used $W=I$). Figure \ref{fig:my_label} gives an example of the gaussian Kernels generated by the algorithm. For stability and improved performance, we used the ADAM algorithm for all the gradient optimizations (modeler and critic jitter), rather than stochastic gradient descent. 

The modeler was represented as deep neural net with three hidden layers, each with width of $r=1000$ RELU (rectified linear units) gates. The critic was randomizing over the set of gaussian kernel functions centered on the centroids of a $K$-means clustering with $K=50$ performed on the $d$-dimensional $\mcX$ space. The learning rate of the modeler and critic were $\eta_m=\eta_w=0.007\approx 1/\sqrt{r T}$ and $\eta_c=0.11\approx 1/\sqrt{\log(K) T}$, correspondingly.

\paragraph{Benchmarks.} We compared our approach to a polynomial two-stage least squares (2SLSPoly) approach, where we represented $h_\theta(w)$ as a third degree polynomial and then estimated $w,w^2,w^3$ in separate first stage regressions on all interactions of $x_1$ and $x_2$ up to degree 3, regularizing to remove instability.  
We also compared it to running a direct polynomial regression (DirectPoly), direct neural net regression (DirectNN), standard linear two-stage least squares (2SLS), and DeepIV.

\begin{figure*}[htpb]
    \centering
    \subfigure[2dpoly]
    {\includegraphics[scale=.5]{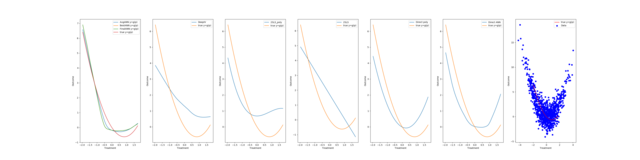}}
    \subfigure[3dpoly]
    {\includegraphics[scale=.5]{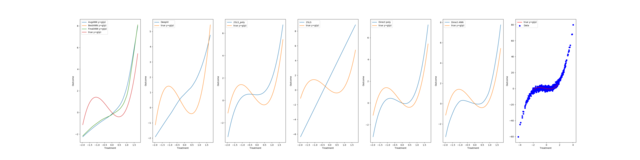}}
    \subfigure[abs]
    {\includegraphics[scale=.5]{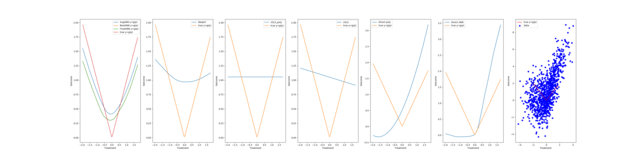}}
    \subfigure[linear]
    {\includegraphics[scale=.5]{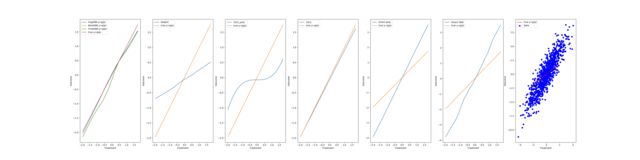}}
    \subfigure[sigmoid]
    {\includegraphics[scale=.5]{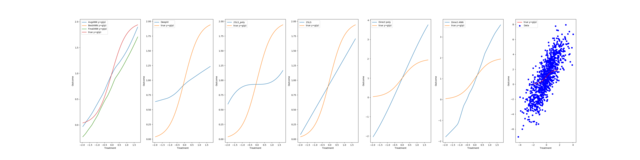}}
    \subfigure[sin]
    {\includegraphics[scale=.5]{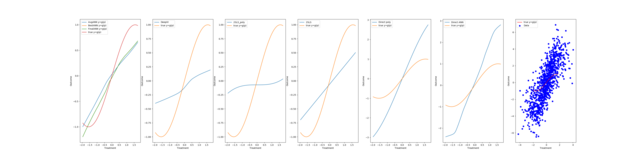}}
    \subfigure[step]
    {\includegraphics[scale=.5]{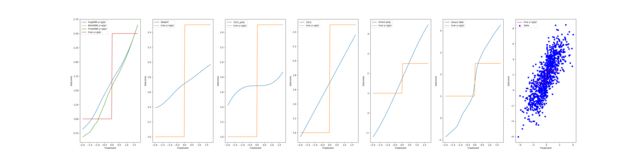}}
    \subfigure[rand\_pw]
    {\includegraphics[scale=.5]{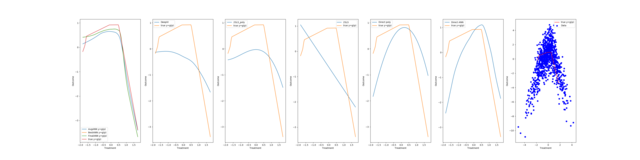}}
    \caption{Examples of fitted functions via each method (in order from left to right: AGMM (best, final, avg), DeepIV, 2SLSPoly, 2SLS, DirectPoly, DirectNN, data points). Data Generating Process: DGP $1$, Instrument strength: $\gamma=0.5$, Number of instruments: $d=1$, Number of samples: $n=1000$,  Training steps: $T=400$, Number of critics: $K=50$, Kernel radius: $r=50$ data points, Critic jitter: yes.}
    \label{fig:example-iv}
\end{figure*}

Figure \ref{fig:example-iv} shows qualitative examples of fitted functions using each of these methods and for each true functional model $h_0$.
We show three different outputs from our DNN: the average across all steps of the dynamics (avg), the final model (final), and the model that performed best in-sample moment violation (best).

\paragraph{Performance Metrics.} We repeated each experiment $M=100$ times on fresh samples of the data and we calculated the $R^2$ of each method in each of the runs on a set of test treatment points. More concretely, we generated a set of test points $\{w_1,\ldots, w_K\}$ and evaluated, the mean squared error of each estimator $\hat{h}$ \begin{equation}
    \MSE(\hat{h}) = \frac{1}{K}\sum_{i=1}^K \left(h_0(w_i) - \hat{h}(w_i)\right)^2
\end{equation} 
and the R-squared, which captures how much of the variance of the function was explained by our model: 
\begin{equation}
    R^2 = 1 - \frac{\MSE(\hat{h})}{\frac{1}{K}\sum_{i=1}^K \left(h_0(w_i) - \frac{1}{K} \sum_{t=1}^K h_0(w_{t})\right)^2}
\end{equation}

We created two types of test treatment points: i) in the first case we generated treatment points from the marginal distribution of the original data generating process, ii) in the second case we created a uniform grid of $100$ points between the $10$ and $90$ percentile of the treatment variable in the original set of sample points on which we trained our model.

\begin{figure*}[htpb]
    \centering
    \subfigure[DGP $1$]
    {\includegraphics[scale=.6]{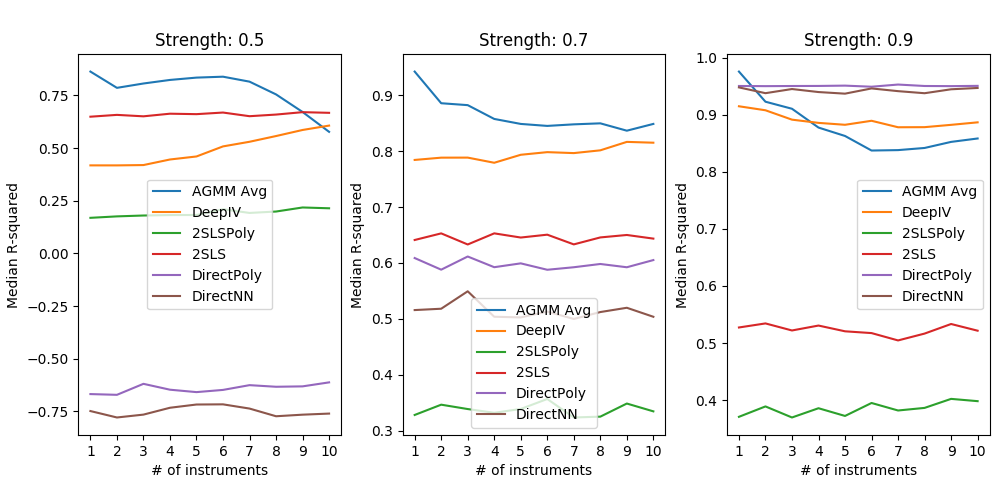}}
    \subfigure[DGP $2$]
    {\includegraphics[scale=.6]{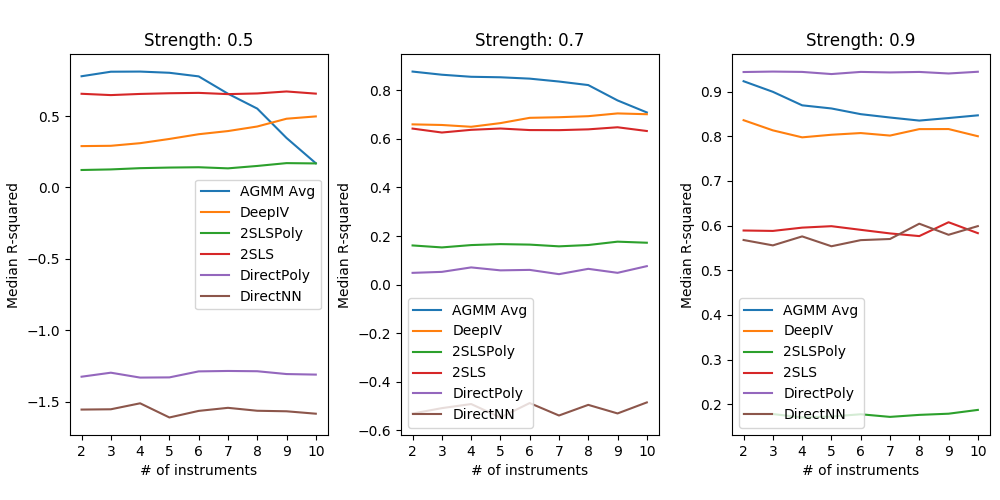}}
    \caption{Median $R^2$ of each method for a grid of $100$ treatment points used as test set and for all the models $h_0$ put together (except model rand\_pw). Performance is portrayed as a function of the number of instruments $d\in \{1,\ldots, 10\}$ and for different values of instrument strength $\gamma\in \{.5, .7, .9\}$. Number of samples: $n=1000$, Number of experiments: $M=700$, Training steps: $T=400$, Number of critics: $K=50$, Kernel radius: $r=50$ data points, Critic jitter: yes.}
   \label{fig:performance-dimension}
\end{figure*}

\begin{figure*}[htpb]
    \centering
    \subfigure[DGP $1$]
    {\includegraphics[scale=.6]{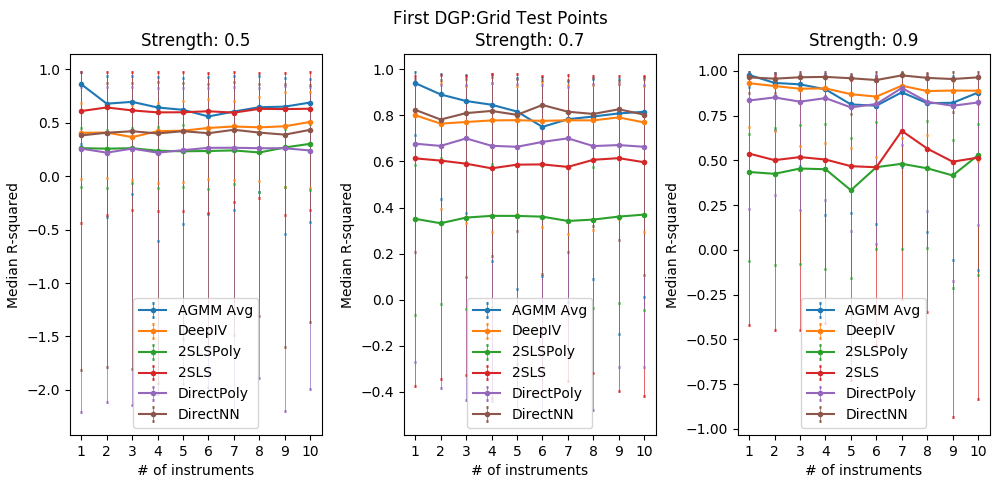}}
    \subfigure[DGP $2$]
    {\includegraphics[scale=.6]{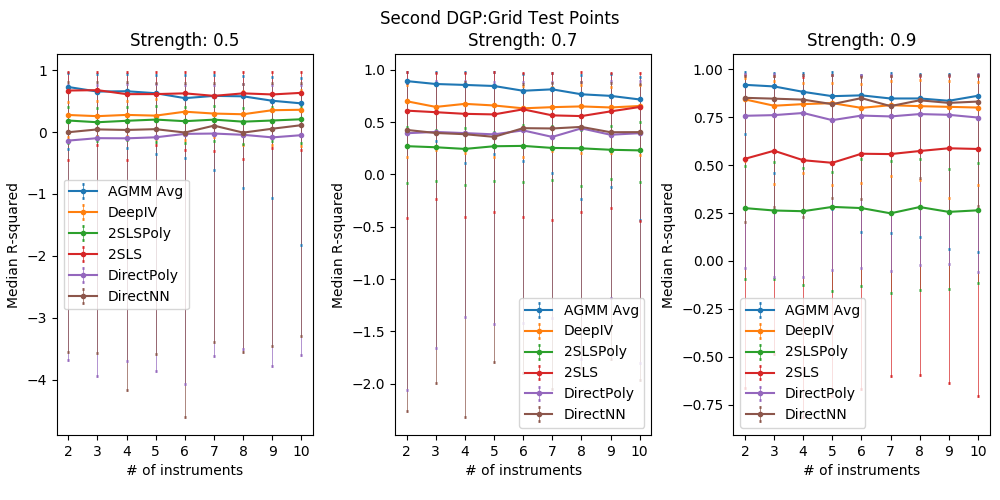}}
    \caption{Median $R^2$ and the $10-90$ percentiles of the $R^2$ of each method for a grid of $100$ treatment points used as test set and for true model rand\_pw. Performance is portrayed as a function of the number of instruments $d\in \{1,\ldots, 10\}$ and for different values of instrument strength $\gamma\in \{.5, .7, .9\}$. Number of samples: $n=1000$, Number of experiments: $M=100$, Training steps: $T=400$, Number of critics: $K=50$, Kernel radius: $r=50$ data points, Critic jitter: yes.}
   \label{fig:performance-dimension-rand-pw}
\end{figure*}

In Figures \ref{fig:performance} and \ref{fig:performance-2}, we give the median $R^2$ and the $5-95$ percentiles of the $R^2$ across the $M=100$ experiments for each of the two DGPs. Moreover, in Figure \ref{fig:performance-dimension}, shows how the $R^2$ performance of each method varies as the number of instruments $d$ increases. For this figure we merged together the experiments from all the different true models $h_0$ (except rand\_pw) and computed median $R^2$ across the $M=700$ total experiments. Finally in Figure \ref{fig:performance-dimension-rand-pw} we plot the median $R^2$ and the $10-90$ percentiles specifically for the random true model rand\_pw as the strength and dimension of the instruments varies.

In the Appendix we provide more detailed performance for each true model separately as well as a plethora of more detailed experimental results and figures.

\paragraph{Conclusion.} We find that the Adversarial GMM algorithm is consistently either the best performing method or within statistically insignificant different from the best performing method. It is not the best performing approach in models where a linear model approximates the truth well, in which case linear 2SLS outperforms it. Even then it several times outperforms 2SLS or is very close to its performance. The performance does degrade as the number of instruments $d$ increases, as expected. However, even up to $d=10$ it maintains great performance. We conclude that Adversarial GMM is an experimentally proven method for non-linear IV estimation, which performs well in a multitude of different underlying models without the need for any fine-tuning for the model at hand. 

\begin{figure*}[htpb]
\centering
\begin{footnotesize}
\begin{tabular}{c|c|c|c|c|c|c}
function & AGMM Avg & DeepIV & 2SLSpoly & 2SLS & DirectPoly & DirectNN\\
\hline\\
abs & {\bf 0.79 {\tiny (0.49, 0.93) } }&0.11 {\tiny (-0.09, 0.24) }&-0.16 {\tiny (-0.34, -0.02) }&-0.18 {\tiny (-0.38, -0.06) }&-3.25 {\tiny (-3.77, -2.65) }&-2.90 {\tiny (-3.78, -2.15) }\\
2dpoly & {\bf 0.97 {\tiny (0.91, 0.99) } }&0.49 {\tiny (0.32, 0.60) }&0.47 {\tiny (0.28, 0.59) }&0.56 {\tiny (0.42, 0.65) }&0.67 {\tiny (0.57, 0.72) }&0.66 {\tiny (0.55, 0.74) }\\
sigmoid & {\bf 0.90 {\tiny (0.60, 0.98) } }&0.53 {\tiny (0.30, 0.71) }&0.23 {\tiny (0.16, 0.33) }&0.89 {\tiny (0.79, 0.96) }&-1.16 {\tiny (-1.43, -0.82) }&-1.26 {\tiny (-1.91, -0.75) }\\
step & {\bf 0.67 {\tiny (0.41, 0.79) } }&0.40 {\tiny (0.22, 0.55) }&0.14 {\tiny (0.08, 0.21) }&0.66 {\tiny (0.58, 0.73) }&-1.05 {\tiny (-1.35, -0.74) }&-1.11 {\tiny (-1.63, -0.59) }\\
3dpoly & 0.21 {\tiny (-1.30, 0.78) }&-0.34 {\tiny (-1.31, -0.00) }&0.03 {\tiny (-0.59, 0.40) }&-9.57 {\tiny (-15.19, -5.14) }&{\bf 0.44 {\tiny (0.20, 0.65) } }&0.35 {\tiny (-0.16, 0.65) }\\
sin & {\bf 0.89 {\tiny (0.58, 0.98) } }&0.41 {\tiny (0.25, 0.58) }&0.12 {\tiny (0.04, 0.19) }&0.74 {\tiny (0.59, 0.85) }&-0.67 {\tiny (-0.96, -0.40) }&-0.94 {\tiny (-1.39, -0.55) }\\
linear & 0.97 {\tiny (0.90, 0.99) }&0.67 {\tiny (0.50, 0.79) }&0.43 {\tiny (0.35, 0.52) }&{\bf 1.00 {\tiny (0.98, 1.00) } }&0.00 {\tiny (-0.14, 0.14) }&0.00 {\tiny (-0.27, 0.25) }\\
rand\_pw & {\bf 0.86 {\tiny (0.04, 0.98) } }&0.41 {\tiny (-0.14, 0.71) }&0.26 {\tiny (-0.15, 0.48) }&0.61 {\tiny (-0.69, 0.99) }&0.26 {\tiny (-3.73, 0.88) }&0.38 {\tiny (-3.61, 0.88) }\\
\end{tabular}
\end{footnotesize}
\caption{Median $R^2$ and the $5-95$ percentiles of the $R^2$ of each method for a grid of $100$ treatment points used as test set. Boldface portrays the best performing method in terms of median $R^2$. Data Generating Process: DGP $1$, Instrument strength: $\gamma=0.5$, Number of instruments: $d=1$, Number of samples: $n=1000$, Number of experiments: $M=100$, Training steps: $T=400$, Number of critics: $K=50$, Kernel radius: $r=50$ data points, Critic jitter: yes.}
\label{fig:performance}
\end{figure*}

\begin{figure*}[htpb]
\centering
\begin{footnotesize}
\begin{tabular}{c|c|c|c|c|c|c}
function & AGMM Avg & DeepIV & 2SLSpoly & 2SLS & DirectPoly & DirectNN\\
\hline\\
abs & {\bf 0.67 {\tiny (0.17, 0.86) } }&0.08 {\tiny (-0.12, 0.22) }&-0.16 {\tiny (-0.41, -0.05) }&-0.20 {\tiny (-0.45, -0.06) }&-5.43 {\tiny (-6.51, -4.71) }&-5.33 {\tiny (-6.78, -4.27) }\\
2dpoly & {\bf 0.95 {\tiny (0.82, 0.98) } }&0.33 {\tiny (0.17, 0.44) }&0.33 {\tiny (0.09, 0.45) }&0.60 {\tiny (0.46, 0.68) }&0.45 {\tiny (0.33, 0.55) }&0.42 {\tiny (0.25, 0.56) }\\
sigmoid & 0.86 {\tiny (0.43, 0.96) }&0.39 {\tiny (0.21, 0.54) }&0.20 {\tiny (0.09, 0.30) }&{\bf 0.90 {\tiny (0.71, 0.96) } }&-2.02 {\tiny (-2.46, -1.53) }&-2.38 {\tiny (-3.27, -1.66) }\\
step & 0.59 {\tiny (0.15, 0.74) }&0.29 {\tiny (0.17, 0.44) }&0.11 {\tiny (0.03, 0.17) }&{\bf 0.66 {\tiny (0.47, 0.73) } }&-1.52 {\tiny (-1.93, -1.12) }&-1.74 {\tiny (-2.31, -1.19) }\\
3dpoly & -1.91 {\tiny (-4.67, -0.15) }&-0.68 {\tiny (-1.90, -0.04) }&-1.11 {\tiny (-2.02, -0.18) }&-13.88 {\tiny (-24.32, -8.76) }&{\bf -0.36 {\tiny (-1.02, 0.19) } }&-0.54 {\tiny (-1.47, 0.13) }\\
sin & {\bf 0.87 {\tiny (0.42, 0.96) } }&0.30 {\tiny (0.16, 0.47) }&0.11 {\tiny (-0.00, 0.18) }&0.76 {\tiny (0.51, 0.89) }&-1.41 {\tiny (-1.81, -1.00) }&-2.00 {\tiny (-2.60, -1.41) }\\
linear & 0.96 {\tiny (0.79, 1.00) }&0.50 {\tiny (0.32, 0.66) }&0.40 {\tiny (0.31, 0.48) }&{\bf 0.99 {\tiny (0.96, 1.00) } }&-0.57 {\tiny (-0.75, -0.37) }&-0.65 {\tiny (-0.98, -0.32) }\\
rand\_pw & {\bf 0.73 {\tiny (-0.53, 0.98) } }&0.28 {\tiny (-0.27, 0.58) }&0.19 {\tiny (-0.37, 0.44) }&0.67 {\tiny (-0.77, 0.99) }&-0.14 {\tiny (-6.99, 0.84) }&-0.00 {\tiny (-6.85, 0.83) }\\
\end{tabular}
\end{footnotesize}
\caption{Median $R^2$ and the $5-95$ percentiles of the $R^2$ of each method for a grid of $100$ treatment points used as test set. Boldface portrays the best performing method in terms of median $R^2$. Data Generating Process: DGP $2$, Instrument strength: $\gamma=0.5$, Number of instruments: $d=2$, Number of samples: $n=1000$, Number of experiments: $M=100$, Training steps: $T=400$, Number of critics: $K=50$, Kernel radius: $r=50$ data points, Critic jitter: yes.}
\label{fig:performance-2}
\end{figure*}

\bibliographystyle{icml2018}
\bibliography{refs}

\begin{thebibliography}{20}
\providecommand{\natexlab}[1]{#1}
\providecommand{\url}[1]{\texttt{#1}}
\expandafter\ifx\csname urlstyle\endcsname\relax
  \providecommand{\doi}[1]{doi: #1}\else
  \providecommand{\doi}{doi: \begingroup \urlstyle{rm}\Url}\fi

\bibitem[Anthony \& Bartlett(2009)Anthony and Bartlett]{Anthony2009}
Anthony, Martin and Bartlett, Peter~L.
\newblock \emph{Neural Network Learning: Theoretical Foundations}.
\newblock Cambridge University Press, New York, NY, USA, 1st edition, 2009.
\newblock ISBN 052111862X, 9780521118620.

\bibitem[Arjovsky et~al.(2017)Arjovsky, Chintala, and
  Bottou]{arjovsky2017wasserstein}
Arjovsky, Martin, Chintala, Soumith, and Bottou, L{\'e}on.
\newblock Wasserstein gan.
\newblock \emph{arXiv preprint arXiv:1701.07875}, 2017.

\bibitem[Athey et~al.(2016)Athey, Tibshirani, and Wager]{Athey2016}
Athey, Susan, Tibshirani, Julie, and Wager, Stefan.
\newblock Generalized random forests.
\newblock 2016.

\bibitem[Bartlett \& Mendelson(2003)Bartlett and Mendelson]{Bartlett2003}
Bartlett, Peter~L. and Mendelson, Shahar.
\newblock Rademacher and gaussian complexities: Risk bounds and structural
  results.
\newblock \emph{J. Mach. Learn. Res.}, 3:\penalty0 463--482, March 2003.
\newblock ISSN 1532-4435.

\bibitem[Chamberlain(1987)]{Chamberlain1987}
Chamberlain, Gary.
\newblock Asymptotic efficiency in estimation with conditional moment
  restrictions.
\newblock \emph{Journal of Econometrics}, 34\penalty0 (3):\penalty0 305 -- 334,
  1987.
\newblock ISSN 0304-4076.
\newblock \doi{https://doi.org/10.1016/0304-4076(87)90015-7}.

\bibitem[Chen \& Liao(2015)Chen and Liao]{Chen2015}
Chen, Xiaohong and Liao, Zhipeng.
\newblock Sieve semiparametric two-step gmm under weak dependence.
\newblock Cowles Foundation Discussion Papers 2012, Cowles Foundation for
  Research in Economics, Yale University, 2015.

\bibitem[Freund \& Schapire(1999)Freund and Schapire]{Freund1999}
Freund, Yoav and Schapire, Robert~E.
\newblock Adaptive game playing using multiplicative weights.
\newblock \emph{Games and Economic Behavior}, 29\penalty0 (1):\penalty0 79 --
  103, 1999.
\newblock ISSN 0899-8256.
\newblock \doi{https://doi.org/10.1006/game.1999.0738}.

\bibitem[Gallant \& Tauchen(1996)Gallant and Tauchen]{GallantTauchen96}
Gallant, A. and Tauchen, George.
\newblock Which moments to match?
\newblock \emph{Econometric Theory}, 12\penalty0 (04):\penalty0 657--681, 1996.

\bibitem[Goodfellow et~al.(2014)Goodfellow, Pouget-Abadie, Mirza, Xu,
  Warde-Farley, Ozair, Courville, and Bengio]{Goodfellow14}
Goodfellow, Ian, Pouget-Abadie, Jean, Mirza, Mehdi, Xu, Bing, Warde-Farley,
  David, Ozair, Sherjil, Courville, Aaron, and Bengio, Yoshua.
\newblock Generative adversarial nets.
\newblock In Ghahramani, Z., Welling, M., Cortes, C., Lawrence, N.~D., and
  Weinberger, K.~Q. (eds.), \emph{Advances in Neural Information Processing
  Systems 27}, pp.\  2672--2680. Curran Associates, Inc., 2014.

\bibitem[Goodfellow(2017)]{Goodfellow17}
Goodfellow, Ian~J.
\newblock {NIPS} 2016 tutorial: Generative adversarial networks.
\newblock \emph{CoRR}, abs/1701.00160, 2017.

\bibitem[Hansen(1982)]{Hansen1982}
Hansen, Lars~Peter.
\newblock Large sample properties of generalized method of moments estimators.
\newblock \emph{Econometrica}, 50\penalty0 (4):\penalty0 1029--1054, 1982.
\newblock ISSN 00129682, 14680262.

\bibitem[Hartford et~al.(2017)Hartford, Lewis, Leyton-Brown, and Taddy]{deepiv}
Hartford, Jason, Lewis, Greg, Leyton-Brown, Kevin, and Taddy, Matt.
\newblock Deep {IV}: A flexible approach for counterfactual prediction.
\newblock In Precup, Doina and Teh, Yee~Whye (eds.), \emph{Proceedings of the
  34th International Conference on Machine Learning}, volume~70 of
  \emph{Proceedings of Machine Learning Research}, pp.\  1414--1423,
  International Convention Centre, Sydney, Australia, 06--11 Aug 2017. PMLR.

\bibitem[{Kocaoglu} et~al.(2017){Kocaoglu}, {Snyder}, {Dimakis}, and
  {Vishwanath}]{Kocaoglu2017}
{Kocaoglu}, M., {Snyder}, C., {Dimakis}, A.~G., and {Vishwanath}, S.
\newblock {CausalGAN: Learning Causal Implicit Generative Models with
  Adversarial Training}.
\newblock \emph{ArXiv e-prints}, September 2017.

\bibitem[Koltchinskii \& Panchenko(2000)Koltchinskii and
  Panchenko]{Panchenko2000}
Koltchinskii, Vladimir and Panchenko, Dmitriy.
\newblock Rademacher processes and bounding the risk of function learning.
\newblock In Gin{\'e}, Evarist, Mason, David~M., and Wellner, Jon~A. (eds.),
  \emph{High Dimensional Probability II}, pp.\  443--457, Boston, MA, 2000.
  Birkh{\"a}user Boston.
\newblock ISBN 978-1-4612-1358-1.

\bibitem[Manski(1989)]{Manski1989}
Manski, Charles~F.
\newblock Anatomy of the selection problem.
\newblock \emph{The Journal of Human Resources}, 24\penalty0 (3):\penalty0
  343--360, 1989.
\newblock ISSN 0022166X.

\bibitem[Rakhlin \& Sridharan(2013)Rakhlin and Sridharan]{Rakhlin2013b}
Rakhlin, Alexander and Sridharan, Karthik.
\newblock Optimization, learning, and games with predictable sequences.
\newblock In \emph{Proceedings of the 26th International Conference on Neural
  Information Processing Systems - Volume 2}, NIPS'13, pp.\  3066--3074, USA,
  2013. Curran Associates Inc.

\bibitem[Shalev-Shwartz(2012)]{Shalev2012}
Shalev-Shwartz, Shai.
\newblock Online learning and online convex optimization.
\newblock \emph{Found. Trends Mach. Learn.}, 4\penalty0 (2):\penalty0 107--194,
  February 2012.
\newblock ISSN 1935-8237.
\newblock \doi{10.1561/2200000018}.

\bibitem[Shalev-Shwartz \& Ben-David(2014)Shalev-Shwartz and
  Ben-David]{Shalev2014}
Shalev-Shwartz, Shai and Ben-David, Shai.
\newblock \emph{Understanding Machine Learning: From Theory to Algorithms}.
\newblock Cambridge University Press, New York, NY, USA, 2014.
\newblock ISBN 1107057132, 9781107057135.

\bibitem[Stone(1982)]{Stone1982}
Stone, Charles~J.
\newblock Optimal global rates of convergence for nonparametric regression.
\newblock \emph{Ann. Statist.}, 10\penalty0 (4):\penalty0 1040--1053, 12 1982.
\newblock \doi{10.1214/aos/1176345969}.

\bibitem[Syrgkanis et~al.(2015)Syrgkanis, Agarwal, Luo, and
  Schapire]{Syrgkanis}
Syrgkanis, Vasilis, Agarwal, Alekh, Luo, Haipeng, and Schapire, Robert~E.
\newblock Fast convergence of regularized learning in games.
\newblock In \emph{Advances in Neural Information Processing Systems}, pp.\
  2989--2997, 2015.

\end{thebibliography}

\newpage
\onecolumn
\begin{appendix}

\newpage

\section{Omitted Proofs}

\subsection{Lipschitz Conditional Moments}
Let $\mcX=[0,1]^d$ and suppose that the function $\Psi(h, x)=\E[\rho(z; h) | x]$ is $\lambda$-Lipschitz with respect to $x$ and with respect to the $\|\cdot\|_{\infty}$ norm and that the density of the distribution of $x$ is lower bounded by $\mu>0$. Then consider the set of test functions corresponding to uniform Kernels around a set of grid points, i.e.: discretize the space $\mcX$ to a grid of multiples of some number $h$ (the bandwidth) and let $\mcX_{h}$ denote the discretized set of points. Then for each $x_0\in \mcX_{h}$, let $f(x; x_0) = 1\{\|x-x_0\|_{\infty}\leq h\}$. For any point $x^*$, let $x^*_{h}$ denote its closest grid point. Then observe that if $|\E[\rho(z; h) f(x; x_{h}^*)]|\leq \epsilon$, then:
\begin{align*}
    \epsilon \geq~& \left|\E[\rho(z;h) f(x; x^*_{h})]\right|\\ 
    \geq~& \left|\E\left[\E[\rho(z;h) |x] 1\{\|x-x^*_{h}\|_{\infty}\leq h\}]\right]\right|\\
    =~& \left|\E\left[\E[\rho(z;h) |x_h^*] 1\{\|x-x^*_{h}\|_{\infty}\leq h\}]\right] + \E\left[\left(\E[\rho(z;h) | x] - \E[\rho(z;h) |x_h^*]\right) 1\{\|x-x^*_{h}\|_{\infty}\leq h\}]\right]\right|\\
    \geq~& \left|\E[\rho(z;h) |x_h^*]\right| \Pr[\|x-x^*_h\|_{\infty}\leq h] - 
    \left|\E\left[\left(\E[\rho(z;h) | x] - \E[\rho(z;h) |x_h^*]\right) 1\{\|x-x^*_{h}\|_{\infty}\leq h\}]\right]\right|\\
    \geq~& \left|\E[\rho(z;h) |x_h^*]\right| \Pr[\|x-x^*_h\|_{\infty}\leq h] - 
    h\lambda \Pr[\|x-x^*_h\|_{\infty}\leq h]
\end{align*}
Hence, by re-arranging:
\begin{align*}
    \left|\E[\rho(z;h) |x_h^*]\right| \leq h\lambda + \frac{\epsilon}{\Pr[\|x-x^*_h\|_{\infty}\leq h]} \leq h\lambda + \frac{\epsilon}{\mu h^{d}}
\end{align*}
Thus the class of test functions $\mcF=\{f(x; x_0): x_0\in \mcX_{h}\}$ is a set of $\gamma$-test functions with $\gamma(\epsilon) = h\lambda + \frac{\epsilon}{\mu h^{d}}$. If we have a target $\epsilon$ in mind, we can set the optimal bandwidth $h=\left(\epsilon/\lambda\mu\right)^{1/(d+1)}$, to get $\gamma(\epsilon)=2\lambda^{d/(d+1)}\left(\epsilon/\mu\right)^{1/(d+1)}$. The latter slow convergence with respect to $d$ is a typical rate in non-parametric regression problems in $d$-dimensions \cite{Stone1982}.

Furthermore, we can show that this class also has good generalization properties. Specifically, observe that $\mcA = \{\rho(\cdot;h_{\theta})1\{\|\cdot - x_0\|_{\infty} \leq h\} : \theta \in \Theta, x_0 \in \mcX_h\}$. Further assume that $\rho(\cdot; h_\theta)\in [0,1]$ is a $\lambda$-Lipschitz function of $\theta$ and $\theta\in [0,1]^r$. Then by standard covering number arguments, the latter class $\mcA$ can be approximated to within $\epsilon$ by a finite class of functions of size $N=O((1/\epsilon)^r (1/h)^d))$. Therefore the Rademacher complexity is bounded by $O\left(\sqrt{\frac{r\log(r) + d\log(1/h)}{n}}\right)$ (see e.g. Lemma 27.5 of \cite{Shalev2014}). Moreover, $|\mcF| = (1/h)^d$ and we remind that $\gamma(\epsilon)=O(h + \frac{\epsilon}{h^d})$. Combining all the above we get that if we run the dynamics for $T=O(n)$ iterations, then we are guaranteed that the averages are an $\epsilon$-equilibrium of the population game for $\epsilon = O\left(\frac{\sqrt{d\log(1/h)}+\sqrt{rlog(r)} + \sqrt{\log(1/\delta)}}{\sqrt{n}}\right)$ with probability $1-\delta$ and that the conditional moment violations are upper bounded by $\gamma(\epsilon)$. Balancing $h$ appropriately, we get an error rate of the order of $n^{-1/(2(d+1))}$.

\subsection{Proof of Theorem \ref{thm:apx-id}}
For $h\in \mcH$ and $\sigma \in \Delta(\mcF)$ let $L(h, \sigma) = \E_{f\sim \sigma}\left[\left(\E\left[\rho(z; h) f(x)\right]\right)^2\right]$ denote the loss of the zero-sum game. Observe that if the modeler chooses an $h\in \mcH_I$, then he is guaranteed zero loss. Hence, the value of the game is zero. Subsequently, it is easy to see that at any $\epsilon$-equilibrium $(h^*, w^*)$: $\sup_{f\in \mcF} \left(\E\left[\rho(z; h) f(x)\right]\right)^2 \leq \epsilon \Rightarrow \sup_{f \in \mcF} \left|\E\left[\rho(z; h) f(x)\right]\right|\ \leq \sqrt{\epsilon}$. The latter also implies that this inequality holds for any convex combination of functions in $\mcF$, i.e. for any $\bar{f}\in \bar{\mcF}$: $\left|\E\left[\rho(z; h) \bar{f}(x)\right]\right|\leq \sqrt{\epsilon}$. Hence, by the triangle inequality and the property of $\gamma$-test functions, for any $x\in \mcX$: $\left|\E\left[\rho(z; h) | x\right]\right| \leq \gamma(\sqrt{\epsilon})$. The theorem then follows by Equation \eqref{eqn:distance-violation}.

\subsection{Proof of Theorem \ref{thm:dynamics}}
By Corollary 2.14 and 2.17 of \cite{Shalev2012} we have small regret. In particular:
\begin{align*}
    \frac{1}{T} \sum_{t=1}^T L_n(\theta_t, \sigma_t) \leq~& \inf_{\theta \in \Theta} \frac{1}{T} \sum_{t=1}^T L_n(\theta, \sigma_t) + \epsilon_1(T)\\
    \frac{1}{T} \sum_{t=1}^T L_n(\theta_t, \sigma_t) \geq~& \sup_{\sigma \in \Delta(\mcF)} \frac{1}{T} \sum_{t=1}^T L_n(\theta_t, \sigma) - \epsilon_2(T)
\end{align*}
for $\epsilon_1(T) = \frac{BL\sqrt{2}}{\sqrt{T}}$ and $\epsilon_2(T)=\frac{H^2\sqrt{2\log(|\mcF|)}}{\sqrt{T}}$. Subsequently, since the game is convex in $\theta$ and concave in $\sigma$, we can use the well-known fact that the average solutions $\theta^*$ and $\sigma^*$ is an $\epsilon_1(T)+\epsilon_2(T)$-approximate equilibrium (see e.g. \cite{Rakhlin2013b}).

\subsection{Proof of Theorem \ref{thm:uniform-conv}}

By the result of \cite{Panchenko2000,Bartlett2003} connecting Rademacher complexity and uniform convergence of empirical processes, we have with probability $1-\delta$:
\begin{equation*}
    \sup_{\theta \in \Theta, f\in \mcF} \left|\E_n\left[\rho(z; h_{\theta})f(x)\right] - \E\left[\rho(z; h_{\theta})f(x)\right]\right|\leq \Delta
    \triangleq  O\left(\mcR + H \sqrt{\frac{\log(1/\delta)}{n}}\right)
\end{equation*}
Thus since $(\theta^*, \sigma^*)$ is an $\epsilon$-equilibrium of the empirical game, where $\epsilon$ is given by Theorem \ref{thm:dynamics}, we get that it is also an $O(\epsilon+\Delta)$ equilibrium of the population game. More formally, let $L(\theta, \sigma) = \E_{f\sim \sigma}\left[\left(\E\left[\rho(z; h_{\theta})f(x)\right]\right)^2\right]$ be the loss of the population game. Then by the uniform  convergence property and the boundedness of the moment by $H$, we have with probability $1-\delta$: 
\begin{align*}
    \sup_{\theta, \sigma} |L_n(\theta, \sigma) - L(\theta, \sigma)| \leq 2H \Delta
\end{align*}
Subsequently, we can check that $\theta^*$ and $\sigma^*$ satisfy the approximate equilibrium conditions
\begin{align*}
    L(\theta^*, \sigma^*) \leq~& L_n(\theta^*, \sigma^*) + 2H\Delta\\
    \leq~& \inf_{\theta\in\Theta} L_n(\theta, \sigma^*) + \epsilon + 2H\Delta\\
    \leq~& \inf_{\theta\in \Theta} L(\theta, \sigma^*) + \epsilon + 4H\Delta
\end{align*}
Similarly for the maximizing player.

\section{Examples of Local Kernels}

We simulated a data generating process with two instruments. Instrument 1 affects the treatment only when it is negative and instrument 2 affects the treatment only when it is positive. We wanted to see the types of discretizations $\mcF_{\epsilon}$, i.e. local kernels
that the different algorithms we propose would have created. Below we depict the kernels based on, i) choosing random data points as centers and putting local gaussians around them with standard deviation equal to the distance of this point to the $k$-th nearest neighbor, where $k$ is a hyperparameter, ii) a random forest, where each leaf defines a local kernel, iii) a random forest where we put local gaussians at the centers of each leaf and with standard deviation equal to the maximum distance from the center of any data point in the leaf. The forest was trained by regressing the treatment on the instrument. We see that the forest correctly picks up that the treatment is more sensitive to the instruments in the lower quadrant and hence puts more local kernels around that area.

\begin{figure}[htpb]
    \centering
	\subfigure[Gaussians centered at random points]{
        \centering
        \includegraphics[scale=.17]{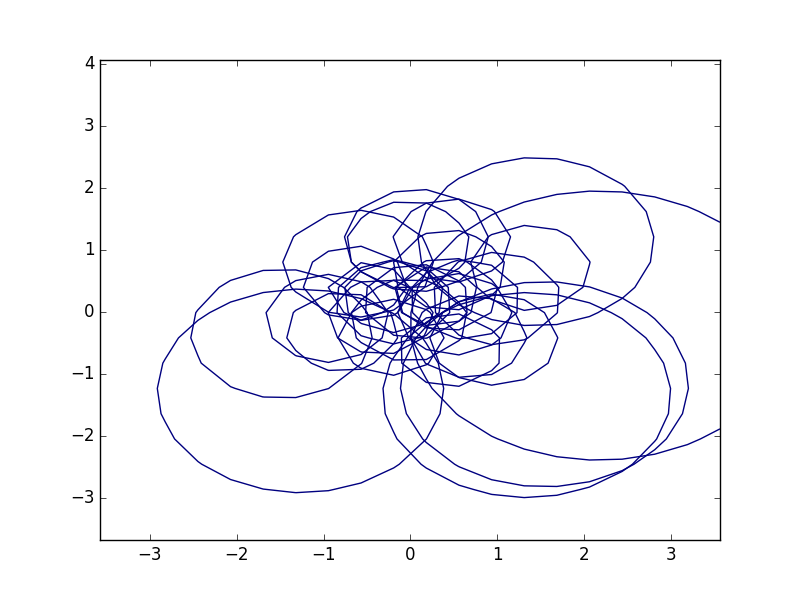}}
	\subfigure[Decision tree trained by regression of $w$ on $x$]{
        \centering
        \includegraphics[scale=.17]{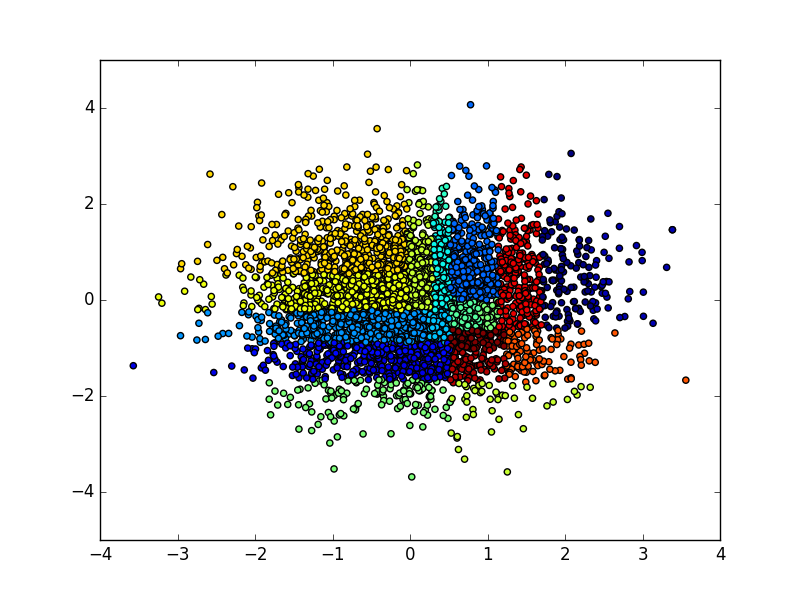}}
	\subfigure[Gaussian kernel centered on leafs of decision tree]{
        \centering
        \includegraphics[scale=.17]{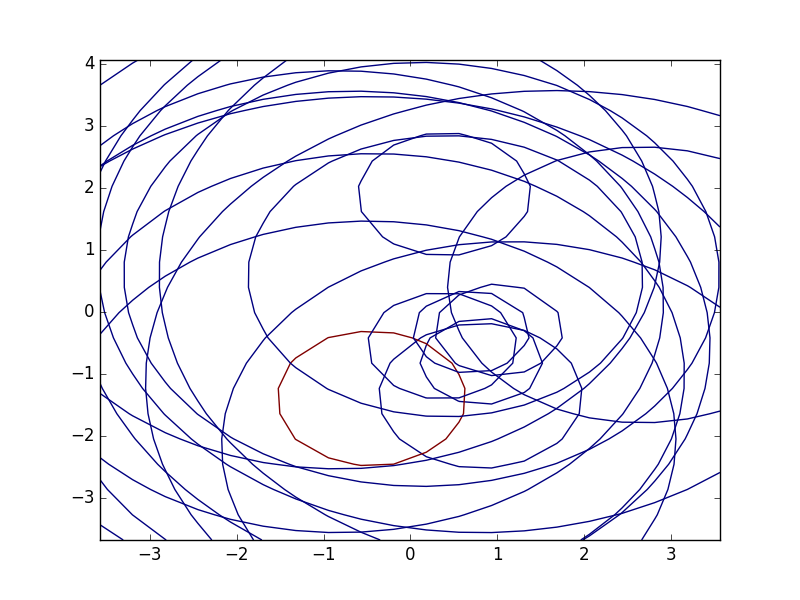}}
\end{figure}

\newpage

\section{Further Figures of Experimental Results}

\subsection{First DGP}

\subsubsection{Second Degree Polynomial}
\begin{figure*}[htpb]
    \centering
    \subfigure
    {\includegraphics[scale=1]{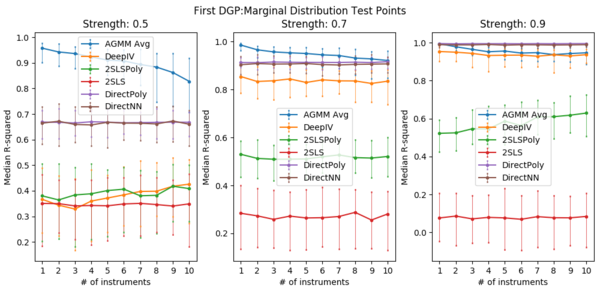}}
    \subfigure
    {\includegraphics[scale=1]{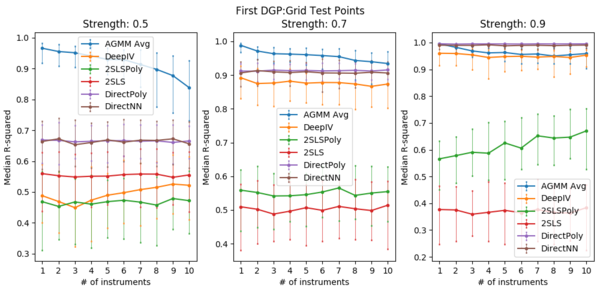}}
    \caption{Median and $10-90$ percentiles of $R^2$ across $100$ experiments as a function of number of instruments and instrument strength $\gamma$. $h_0(w)=-1.5 w + .9  w^2$. Number of samples: 1000, Training steps: 400, Number of critics: 50, Kernel radius: 50 data points, Critic jitter: yes}
    
\end{figure*}

\begin{figure*}[htpb]
    \centering
    \subfigure[Strength $0.5$]
    {\includegraphics[scale=1]{{example_figures/func_2dpoly_n_insts_1_n_steps_400_n_samples_1000_strength_0.5_jitter_1_n_crits_50_radius_50_dgp_two_0/benchmarks_15}.png}}
    \subfigure[Strength $0.7$]
    {\includegraphics[scale=1]{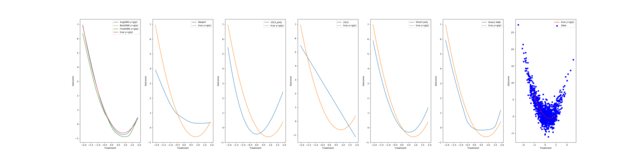}}
    \subfigure[Strength $0.9$]
    {\includegraphics[scale=1]{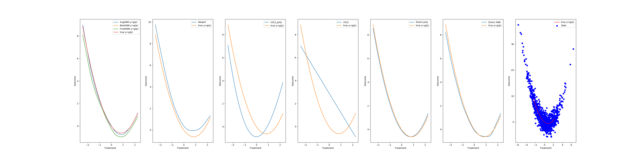}}    
    \caption{Examples of fitted functions via each method (in order from left to right: AGMM (best, final, avg), DeepIV, 2SLSPoly, 2SLS, DirectPoly, DirectNN, data points). $h_0(w)=-1.5 w + .9 w^2$, Number of instruments: 1, Number of samples: 1000,  Training steps: 400, Number of critics: 50, Kernel radius: 50 data points, Critic jitter: yes}
\end{figure*}

\begin{figure*}[htpb]
    \centering
    \subfigure[Strength $0.5$]
    {\includegraphics[scale=1]{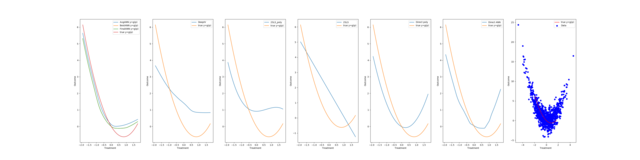}}
    \subfigure[Strength $0.7$]
    {\includegraphics[scale=1]{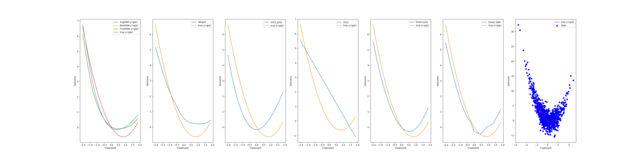}}
    \subfigure[Strength $0.9$]
    {\includegraphics[scale=1]{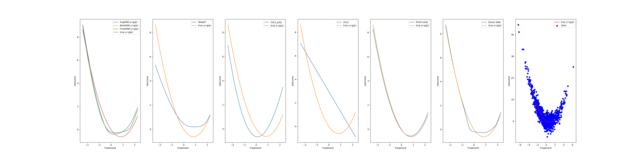}}    
    \caption{Examples of fitted functions via each method (in order from left to right: AGMM (best, final, avg), DeepIV, 2SLSPoly, 2SLS, DirectPoly, DirectNN, data points). $h_0(w)=-1.5 w + .9 w^2$, Number of instruments: 5, Number of samples: 1000,  Training steps: 400, Number of critics: 50, Kernel radius: 50 data points, Critic jitter: yes}
\end{figure*}

\newpage
\subsubsection{Third Degree Polynomial}
\begin{figure*}[htpb]
    \centering
    \subfigure
    {\includegraphics[scale=1]{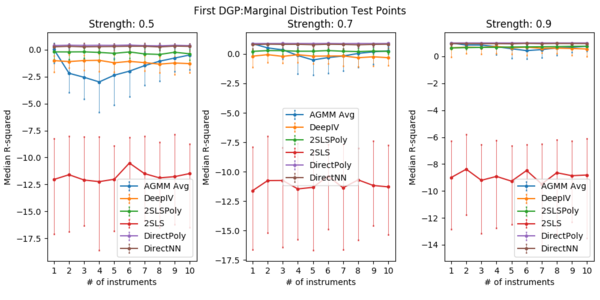}}
    \subfigure
    {\includegraphics[scale=1]{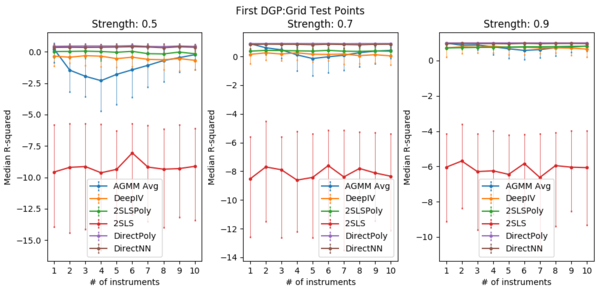}}
    \caption{Median and $10-90$ percentiles of $R^2$ across $100$ experiments as a function of number of instruments and instrument strength $\gamma$. $h_0(w)=-1.5 w + .9 w^2 + w^3$. Number of samples: 1000,  Training steps: 400, Number of critics: 50, Kernel radius: 50 data points, Critic jitter: yes}
\end{figure*}

\begin{figure*}[htpb]
    \centering
    \subfigure[Strength $0.5$]
    {\includegraphics[scale=1]{{example_figures/func_3dpoly_n_insts_1_n_steps_400_n_samples_1000_strength_0.5_jitter_1_n_crits_50_radius_50_dgp_two_0/benchmarks_15}.png}}
    \subfigure[Strength $0.7$]
    {\includegraphics[scale=1]{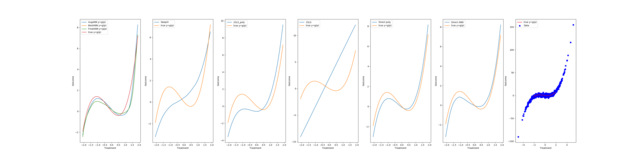}}
    \subfigure[Strength $0.9$]
    {\includegraphics[scale=1]{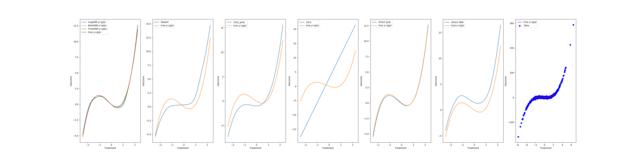}}    
    \caption{Examples of fitted functions via each method (in order from left to right: AGMM (best, final, avg), DeepIV, 2SLSPoly, 2SLS, DirectPoly, DirectNN, data points). $h_0(w)=-1.5 w + .9 w^2 + w^3$, Number of instruments: 1, Number of samples: 1000,  Training steps: 400, Number of critics: 50, Kernel radius: 50 data points, Critic jitter: yes}
\end{figure*}

\begin{figure*}[htpb]
    \centering
    \subfigure[Strength $0.5$]
    {\includegraphics[scale=1]{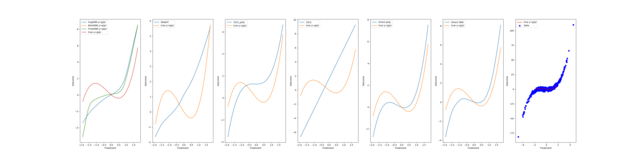}}
    \subfigure[Strength $0.7$]
    {\includegraphics[scale=1]{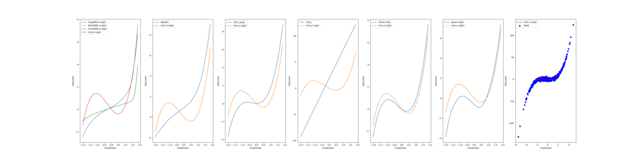}}
    \subfigure[Strength $0.9$]
    {\includegraphics[scale=1]{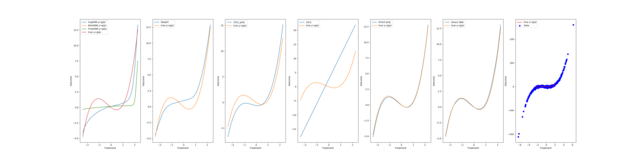}}    
    \caption{Examples of fitted functions via each method (in order from left to right: AGMM (best, final, avg), DeepIV, 2SLSPoly, 2SLS, DirectPoly, DirectNN, data points). $h_0(w)=-1.5 w + .9 w^2 + w^3$, Number of instruments: 5, Number of samples: 1000,  Training steps: 400, Number of critics: 50, Kernel radius: 50 data points, Critic jitter: yes}
\end{figure*}

\newpage
\subsubsection{Absolute Value}
\begin{figure*}[htpb]
    \centering
    \subfigure
    {\includegraphics[scale=1]{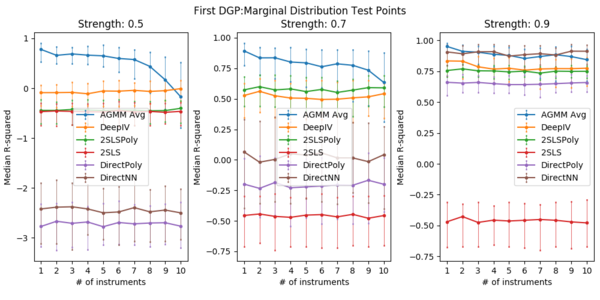}}
    \subfigure
    {\includegraphics[scale=1]{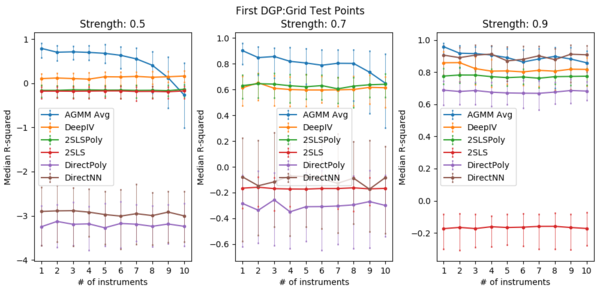}}
    \caption{Median $R^2$ as a function of number of instruments and instrument strength.$h_0(w)=|w|$. Number of samples: 1000,  Training steps: 400, Number of critics: 50, Kernel radius: 50 data points, Critic jitter: yes}
    
\end{figure*}

\begin{figure*}[htpb]
    \centering
    \subfigure[Strength $0.5$]
    {\includegraphics[scale=1]{{example_figures/func_abs_n_insts_1_n_steps_400_n_samples_1000_strength_0.5_jitter_1_n_crits_50_radius_50_dgp_two_0/benchmarks_15}.png}}
    \subfigure[Strength $0.7$]
    {\includegraphics[scale=1]{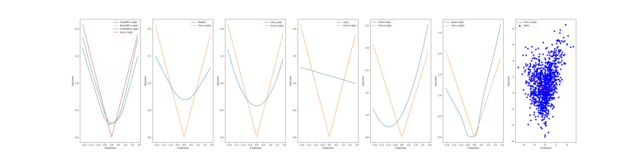}}
    \subfigure[Strength $0.9$]
    {\includegraphics[scale=1]{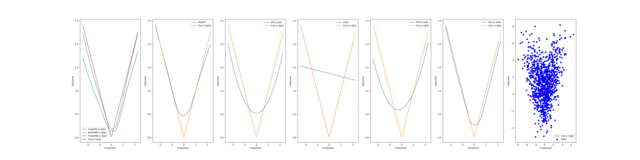}}    
    \caption{Examples of fitted functions via each method (in order from left to right: AGMM (best, final, avg), DeepIV, 2SLSPoly, 2SLS, DirectPoly, DirectNN, data points). $h_0(w)=|w|$, Number of instruments: 1, Number of samples: 1000,  Training steps: 400, Number of critics: 50, Kernel radius: 50 data points, Critic jitter: yes}
\end{figure*}

\begin{figure*}[htpb]
    \centering
    \subfigure[Strength $0.5$]
    {\includegraphics[scale=1]{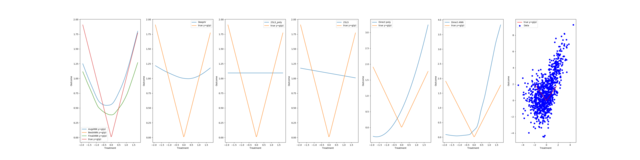}}
    \subfigure[Strength $0.7$]
    {\includegraphics[scale=1]{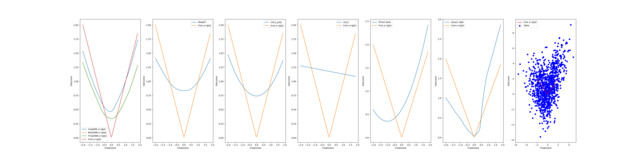}}
    \subfigure[Strength $0.9$]
    {\includegraphics[scale=1]{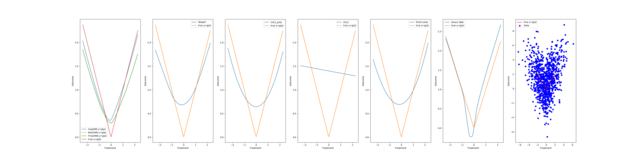}}    
    \caption{Examples of fitted functions via each method (in order from left to right: AGMM (best, final, avg), DeepIV, 2SLSPoly, 2SLS, DirectPoly, DirectNN, data points). $h_0(w)=|w|$, Number of instruments: 5, Number of samples: 1000,  Training steps: 400, Number of critics: 50, Kernel radius: 50 data points, Critic jitter: yes}
\end{figure*}

\newpage
\subsubsection{Identify Function}
\begin{figure*}[htpb]
    \centering
    \subfigure
    {\includegraphics[scale=1]{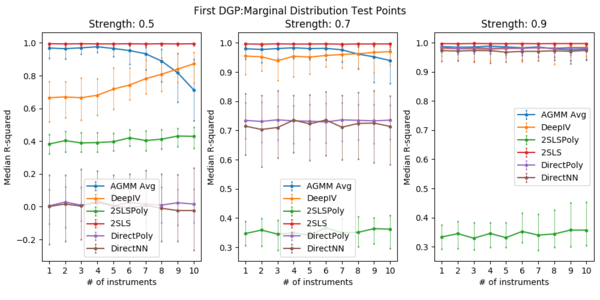}}
    \subfigure
    {\includegraphics[scale=1]{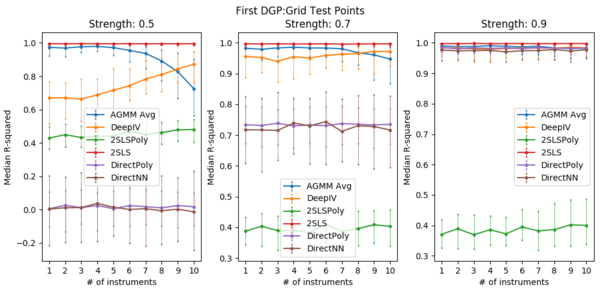}}
    \caption{Median and $10-90$ percentiles of $R^2$ across $100$ experiments as a function of number of instruments and instrument strength $\gamma$. $h_0(w) = x$. Number of samples: 1000,  Training steps: 400, Number of critics: 50, Kernel radius: 50 data points, Critic jitter: yes}
    
\end{figure*}

\begin{figure*}[htpb]
    \centering
    \subfigure[Strength $0.5$]
    {\includegraphics[scale=1]{{example_figures/func_linear_n_insts_1_n_steps_400_n_samples_1000_strength_0.5_jitter_1_n_crits_50_radius_50_dgp_two_0/benchmarks_15}.png}}
    \subfigure[Strength $0.7$]
    {\includegraphics[scale=1]{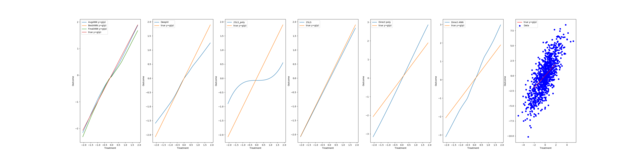}}
    \subfigure[Strength $0.9$]
    {\includegraphics[scale=1]{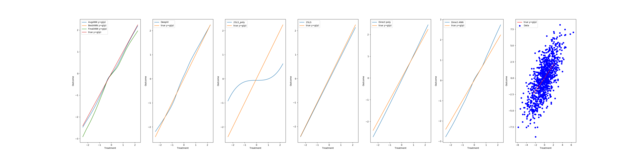}}    
    \caption{Examples of fitted functions via each method (in order from left to right: AGMM (best, final, avg), DeepIV, 2SLSPoly, 2SLS, DirectPoly, DirectNN, data points). $h_0(w)=w$, Number of instruments: 1, Number of samples: 1000,  Training steps: 400, Number of critics: 50, Kernel radius: 50 data points, Critic jitter: yes}
\end{figure*}

\begin{figure*}[htpb]
    \centering
    \subfigure[Strength $0.5$]
    {\includegraphics[scale=1]{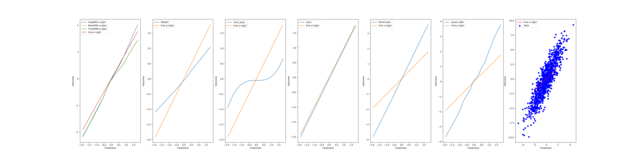}}
    \subfigure[Strength $0.7$]
    {\includegraphics[scale=1]{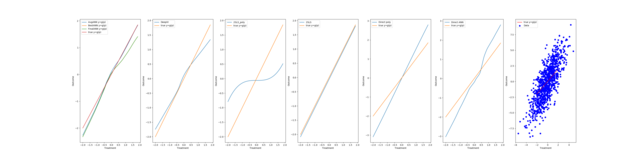}}
    \subfigure[Strength $0.9$]
    {\includegraphics[scale=1]{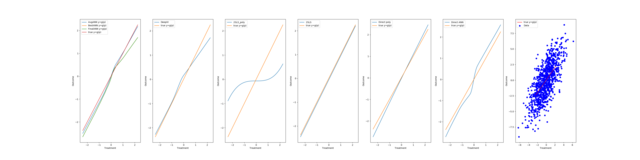}}    
    \caption{Examples of fitted functions via each method (in order from left to right: AGMM (best, final, avg), DeepIV, 2SLSPoly, 2SLS, DirectPoly, DirectNN, data points). $h_0(w)=w$, Number of instruments: 5, Number of samples: 1000,  Training steps: 400, Number of critics: 50, Kernel radius: 50 data points, Critic jitter: yes}
\end{figure*}
\newpage
\subsubsection{Sigmoid Function}
\begin{figure*}[htpb]
    \centering
    \subfigure
    {\includegraphics[scale=1]{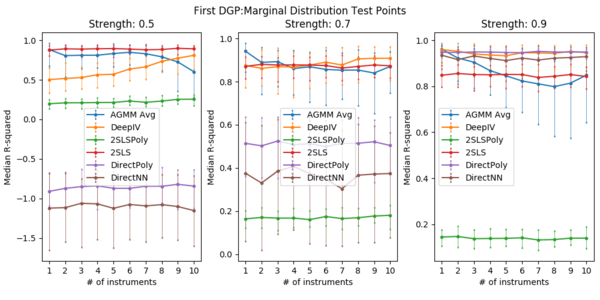}}
    \subfigure
    {\includegraphics[scale=1]{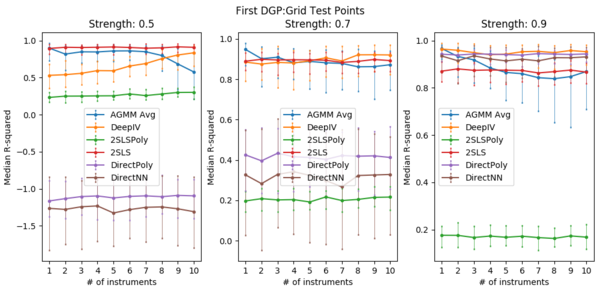}}
    \caption{Median and $10-90$ percentiles of $R^2$ across $100$ experiments as a function of number of instruments and instrument strength $\gamma$. $h_0(w)=\frac{2}{1+e^{-2x}}$. Number of samples: 1000,  Training steps: 400, Number of critics: 50, Kernel radius: 50 data points, Critic jitter: yes}
    
\end{figure*}

\begin{figure*}[htpb]
    \centering
    \subfigure[Strength $0.5$]
    {\includegraphics[scale=1]{{example_figures/func_sigmoid_n_insts_1_n_steps_400_n_samples_1000_strength_0.5_jitter_1_n_crits_50_radius_50_dgp_two_0/benchmarks_15}.png}}
    \subfigure[Strength $0.7$]
    {\includegraphics[scale=1]{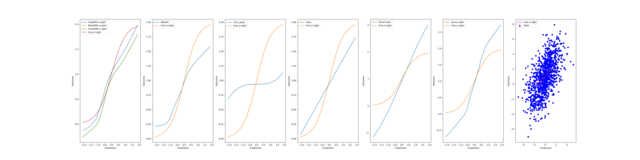}}
    \subfigure[Strength $0.9$]
    {\includegraphics[scale=1]{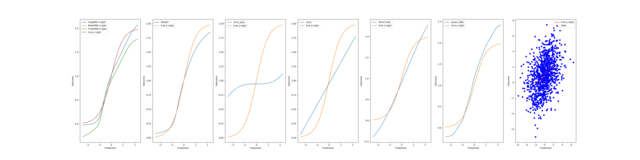}}    
    \caption{Examples of fitted functions via each method (in order from left to right: AGMM (best, final, avg), DeepIV, 2SLSPoly, 2SLS, DirectPoly, DirectNN, data points). $h_0(w)=\frac{2}{1+e^{-2x}}$, Number of instruments: 1, Number of samples: 1000,  Training steps: 400, Number of critics: 50, Kernel radius: 50 data points, Critic jitter: yes}
\end{figure*}

\begin{figure*}[htpb]
    \centering
    \subfigure[Strength $0.5$]
    {\includegraphics[scale=1]{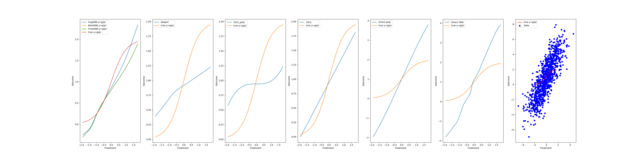}}
    \subfigure[Strength $0.7$]
    {\includegraphics[scale=1]{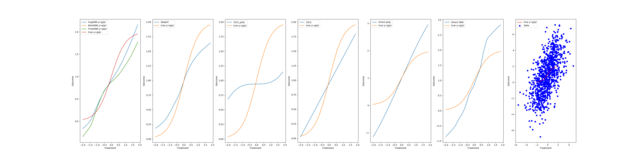}}
    \subfigure[Strength $0.9$]
    {\includegraphics[scale=1]{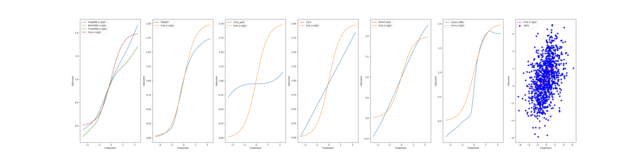}}    
    \caption{Examples of fitted functions via each method (in order from left to right: AGMM (best, final, avg), DeepIV, 2SLSPoly, 2SLS, DirectPoly, DirectNN, data points). $h_0(w)=\frac{2}{1+e^{-2x}}$, Number of instruments: 5, Number of samples: 1000,  Training steps: 400, Number of critics: 50, Kernel radius: 50 data points, Critic jitter: yes}
\end{figure*}
\newpage
\subsubsection{Sin Function}
\begin{figure*}[htpb]
    \centering
    \subfigure
    {\includegraphics[scale=1]{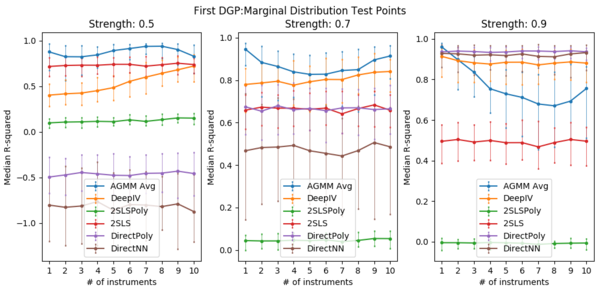}}
    \subfigure
    {\includegraphics[scale=1]{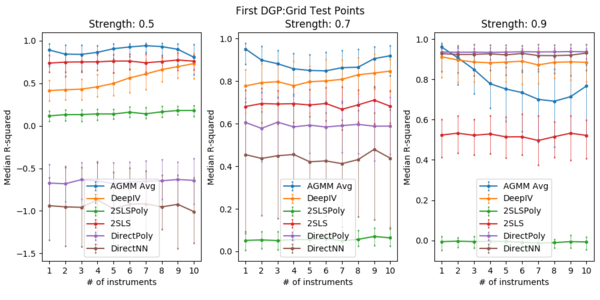}}
    \caption{Median and $10-90$ percentiles of $R^2$ across $100$ experiments as a function of number of instruments and instrument strength $\gamma$. $h_0(w) = \sin(x)$. Number of samples: 1000,  Training steps: 400, Number of critics: 50, Kernel radius: 50 data points, Critic jitter: yes}
    
\end{figure*}

\begin{figure*}[htpb]
    \centering
    \subfigure[Strength $0.5$]
    {\includegraphics[scale=1]{{example_figures/func_sin_n_insts_1_n_steps_400_n_samples_1000_strength_0.5_jitter_1_n_crits_50_radius_50_dgp_two_0/benchmarks_15}.png}}
    \subfigure[Strength $0.7$]
    {\includegraphics[scale=1]{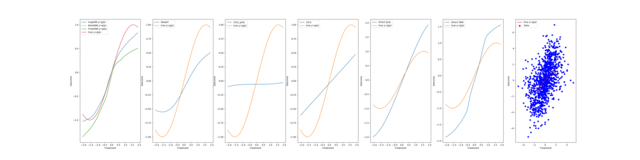}}
    \subfigure[Strength $0.9$]
    {\includegraphics[scale=1]{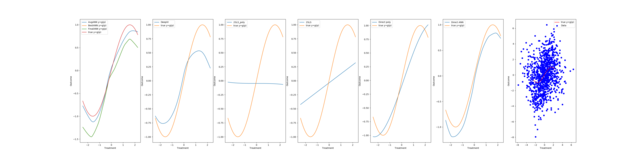}}    
    \caption{Examples of fitted functions via each method (in order from left to right: AGMM (best, final, avg), DeepIV, 2SLSPoly, 2SLS, DirectPoly, DirectNN, data points). $h_0(w)=sin(x)$, Number of instruments: 1, Number of samples: 1000,  Training steps: 400, Number of critics: 50, Kernel radius: 50 data points, Critic jitter: yes}
\end{figure*}

\begin{figure*}[htpb]
    \centering
    \subfigure[Strength $0.5$]
    {\includegraphics[scale=1]{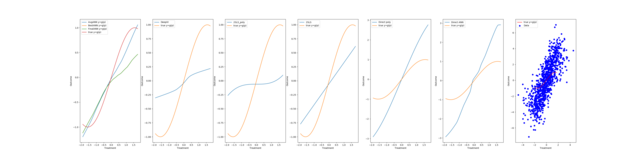}}
    \subfigure[Strength $0.7$]
    {\includegraphics[scale=1]{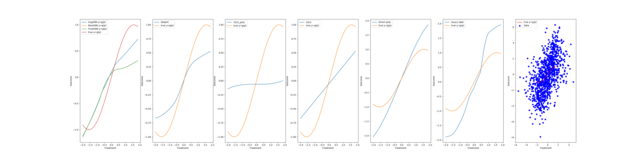}}
    \subfigure[Strength $0.9$]
    {\includegraphics[scale=1]{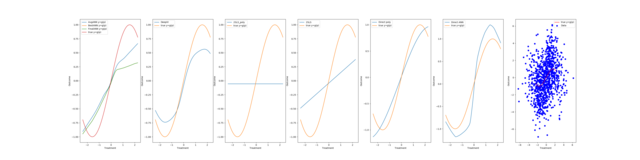}}    
    \caption{Examples of fitted functions via each method (in order from left to right: AGMM (best, final, avg), DeepIV, 2SLSPoly, 2SLS, DirectPoly, DirectNN, data points). $h_0(w)=sin(x)$, Number of instruments: 5, Number of samples: 1000,  Training steps: 400, Number of critics: 50, Kernel radius: 50 data points, Critic jitter: yes}
\end{figure*}

\newpage
\subsubsection{Step Function}
\begin{figure*}[htpb]
    \centering
    \subfigure
    {\includegraphics[scale=1]{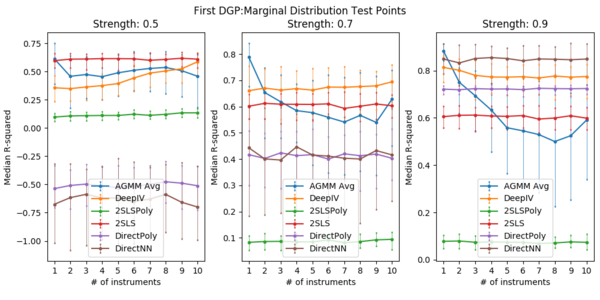}}
    \subfigure
    {\includegraphics[scale=1]{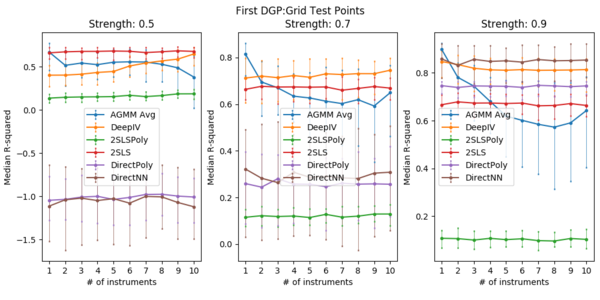}}
    \caption{Median and $10-90$ percentiles of $R^2$ across $100$ experiments as a function of number of instruments and instrument strength $\gamma$. $h_0(w)=1\{x<0\} + 2.5  1\{x \geq 0\}$. Number of samples: 1000,  Training steps: 400, Number of critics: 50, Kernel radius: 50 data points, Critic jitter: yes}
    
\end{figure*}

\begin{figure*}[htpb]
    \centering
    \subfigure[Strength $0.5$]
    {\includegraphics[scale=1]{{example_figures/func_step_n_insts_1_n_steps_400_n_samples_1000_strength_0.5_jitter_1_n_crits_50_radius_50_dgp_two_0/benchmarks_15}.png}}
    \subfigure[Strength $0.7$]
    {\includegraphics[scale=1]{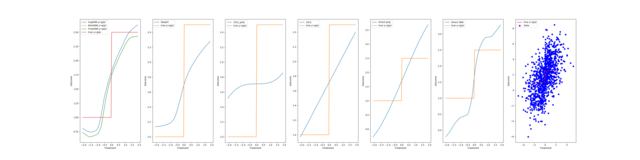}}
    \subfigure[Strength $0.9$]
    {\includegraphics[scale=1]{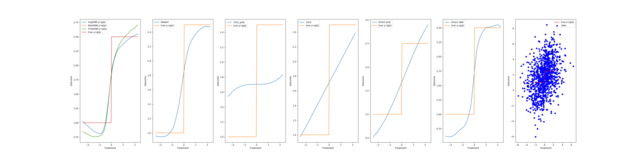}}    
    \caption{Examples of fitted functions via each method (in order from left to right: AGMM (best, final, avg), DeepIV, 2SLSPoly, 2SLS, DirectPoly, DirectNN, data points). $h_0(w)=1\{x<0\} + 2.5  1\{x \geq 0\}$, Number of instruments: 1, Number of samples: 1000,  Training steps: 400, Number of critics: 50, Kernel radius: 50 data points, Critic jitter: yes}
\end{figure*}

\begin{figure*}[htpb]
    \centering
    \subfigure[Strength $0.5$]
    {\includegraphics[scale=1]{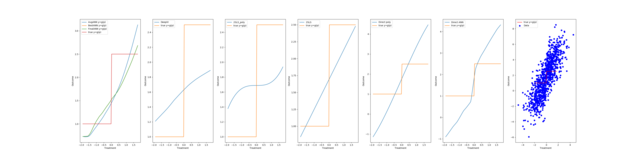}}
    \subfigure[Strength $0.7$]
    {\includegraphics[scale=1]{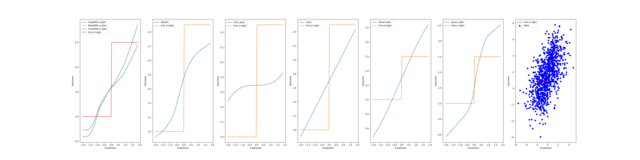}}
    \subfigure[Strength $0.9$]
    {\includegraphics[scale=1]{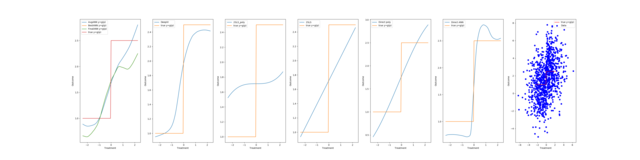}}    
    \caption{Examples of fitted functions via each method (in order from left to right: AGMM (best, final, avg), DeepIV, 2SLSPoly, 2SLS, DirectPoly, DirectNN, data points). $h_0(w)=1\{x<0\} + 2.5  1\{x \geq 0\}$, Number of instruments: 5, Number of samples: 1000,  Training steps: 400, Number of critics: 50, Kernel radius: 50 data points, Critic jitter: yes}
\end{figure*}

\newpage
\subsection{Second DGP}

\subsubsection{Second Degree Polynomial}
\begin{figure*}[htpb]
    \centering
    \subfigure
    {\includegraphics[scale=1]{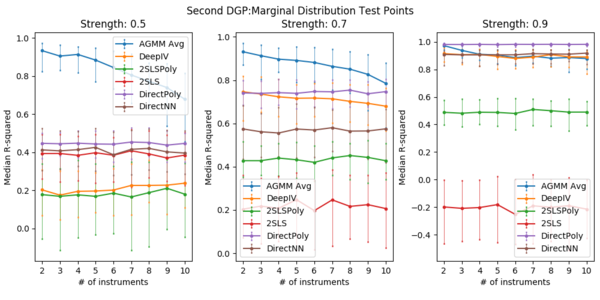}}
    \subfigure
    {\includegraphics[scale=1]{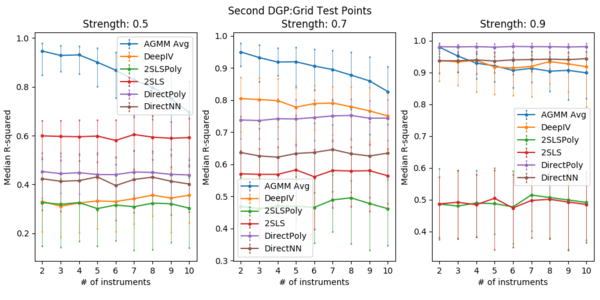}}
    \caption{Median and $10-90$ percentiles of $R^2$ across $100$ experiments as a function of number of instruments and instrument strength $\gamma$. $h_0(w)=-1.5 w + .9  w^2$. Number of samples: 1000,  Training steps: 400, Number of critics: 50, Kernel radius: 50 data points, Critic jitter: yes}
    
\end{figure*}

\begin{figure*}[htpb]
    \centering
    \subfigure[Strength $0.5$]
    {\includegraphics[scale=1]{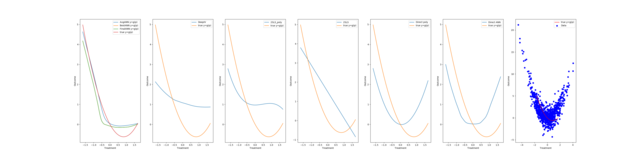}}
    \subfigure[Strength $0.7$]
    {\includegraphics[scale=1]{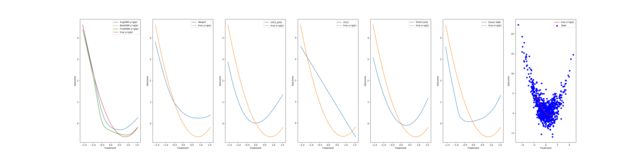}}
    \subfigure[Strength $0.9$]
    {\includegraphics[scale=1]{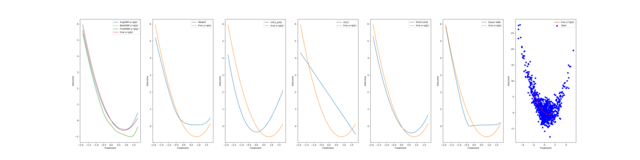}}    
    \caption{Examples of fitted functions via each method (in order from left to right: AGMM (best, final, avg), DeepIV, 2SLSPoly, 2SLS, DirectPoly, DirectNN, data points). $h_0(w)=-1.5 w + .9 w^2$, Number of instruments: 2, Number of samples: 1000,  Training steps: 400, Number of critics: 50, Kernel radius: 50 data points, Critic jitter: yes}
\end{figure*}

\begin{figure*}[htpb]
    \centering
    \subfigure[Strength $0.5$]
    {\includegraphics[scale=1]{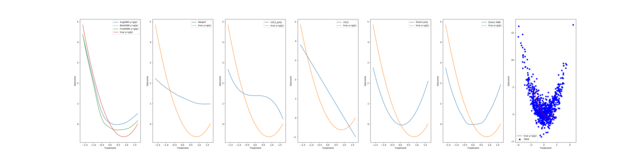}}
    \subfigure[Strength $0.7$]
    {\includegraphics[scale=1]{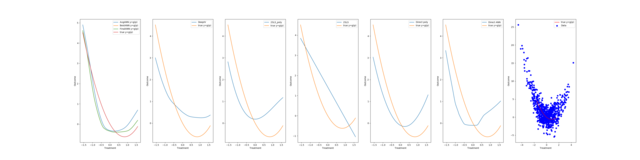}}
    \subfigure[Strength $0.9$]
    {\includegraphics[scale=1]{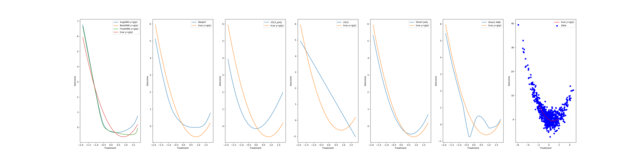}}    
    \caption{Examples of fitted functions via each method (in order from left to right: AGMM (best, final, avg), DeepIV, 2SLSPoly, 2SLS, DirectPoly, DirectNN, data points). $h_0(w)=-1.5 w + .9 w^2$, Number of instruments: 5, Number of samples: 1000,  Training steps: 400, Number of critics: 50, Kernel radius: 50 data points, Critic jitter: yes}
\end{figure*}

\newpage
\subsubsection{Third Degree Polynomial}
\begin{figure*}[htpb]
    \centering
    \subfigure
    {\includegraphics[scale=1]{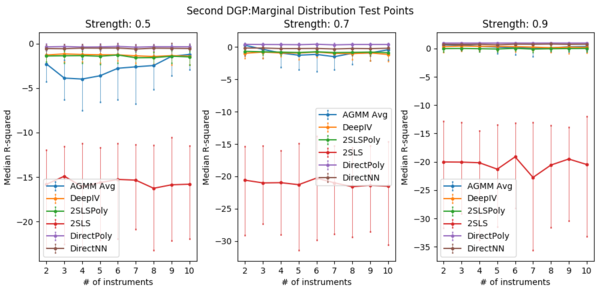}}
    \subfigure
    {\includegraphics[scale=1]{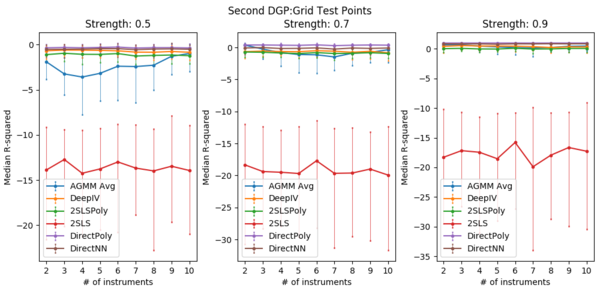}}
    \caption{Median and $10-90$ percentiles of $R^2$ across $100$ experiments as a function of number of instruments and instrument strength $\gamma$. $h_0(w)=-1.5 w + .9 w^2 + w^3$. Number of samples: 1000,  Training steps: 400, Number of critics: 50, Kernel radius: 50 data points, Critic jitter: yes}
    
\end{figure*}

\begin{figure*}[htpb]
    \centering
    \subfigure[Strength $0.5$]
    {\includegraphics[scale=1]{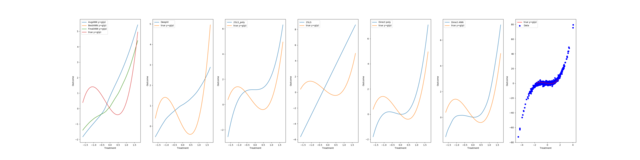}}
    \subfigure[Strength $0.7$]
    {\includegraphics[scale=1]{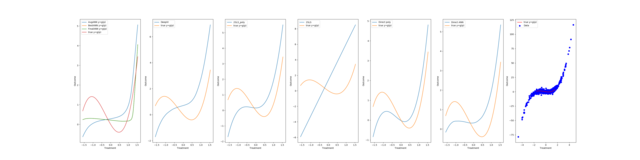}}
    \subfigure[Strength $0.9$]
    {\includegraphics[scale=1]{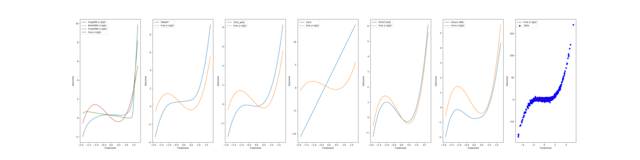}}    
    \caption{Examples of fitted functions via each method (in order from left to right: AGMM (best, final, avg), DeepIV, 2SLSPoly, 2SLS, DirectPoly, DirectNN, data points). $h_0(w)=-1.5 w + .9 w^2 + w^3$, Number of instruments: 2, Number of samples: 1000,  Training steps: 400, Number of critics: 50, Kernel radius: 50 data points, Critic jitter: yes}
\end{figure*}

\begin{figure*}[htpb]
    \centering
    \subfigure[Strength $0.5$]
    {\includegraphics[scale=1]{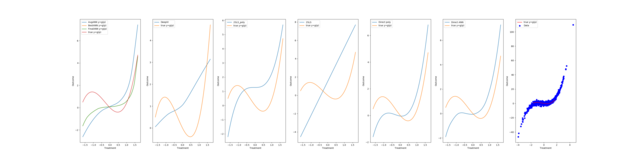}}
    \subfigure[Strength $0.7$]
    {\includegraphics[scale=1]{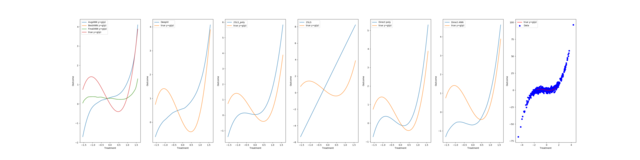}}
    \subfigure[Strength $0.9$]
    {\includegraphics[scale=1]{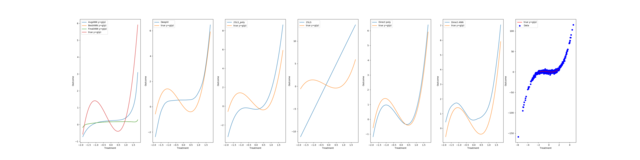}}    
    \caption{Examples of fitted functions via each method (in order from left to right: AGMM (best, final, avg), DeepIV, 2SLSPoly, 2SLS, DirectPoly, DirectNN, data points). $h_0(w)=-1.5 w + .9 w^2 + w^3$, Number of instruments: 5, Number of samples: 1000,  Training steps: 400, Number of critics: 50, Kernel radius: 50 data points, Critic jitter: yes}
\end{figure*}

\newpage
\subsubsection{absolute Value}
\begin{figure*}[htpb]
    \centering
    \subfigure
    {\includegraphics[scale=1]{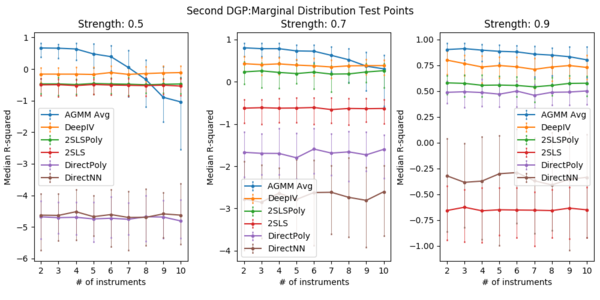}}
    \subfigure
    {\includegraphics[scale=1]{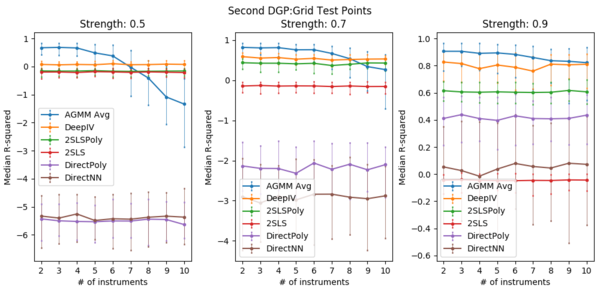}}
    \caption{Median and $10-90$ percentiles of $R^2$ across $100$ experiments as a function of number of instruments and instrument strength $\gamma$. $h_0(w)=|w|$. Number of samples: 1000,  Training steps: 400, Number of critics: 50, Kernel radius: 50 data points, Critic jitter: yes}
    
\end{figure*}

\begin{figure*}[htpb]
    \centering
    \subfigure[Strength $0.5$]
    {\includegraphics[scale=1]{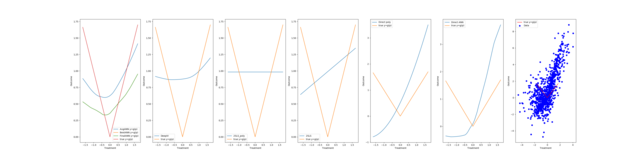}}
    \subfigure[Strength $0.7$]
    {\includegraphics[scale=1]{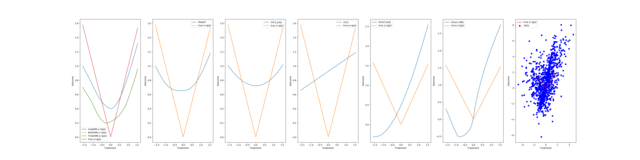}}
    \subfigure[Strength $0.9$]
    {\includegraphics[scale=1]{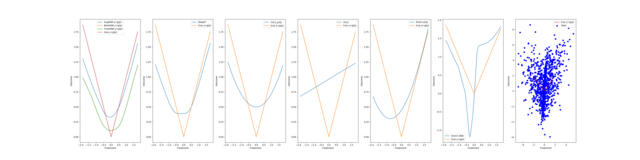}}    
    \caption{Examples of fitted functions via each method (in order from left to right: AGMM (best, final, avg), DeepIV, 2SLSPoly, 2SLS, DirectPoly, DirectNN, data points). $h_0(w)=|w|$, Number of instruments: 2, Number of samples: 1000,  Training steps: 400, Number of critics: 50, Kernel radius: 50 data points, Critic jitter: yes}
\end{figure*}

\begin{figure*}[htpb]
    \centering
    \subfigure[Strength $0.5$]
    {\includegraphics[scale=1]{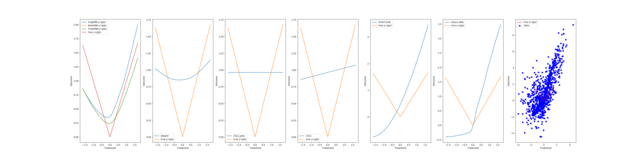}}
    \subfigure[Strength $0.7$]
    {\includegraphics[scale=1]{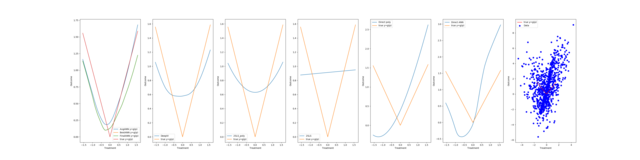}}
    \subfigure[Strength $0.9$]
    {\includegraphics[scale=1]{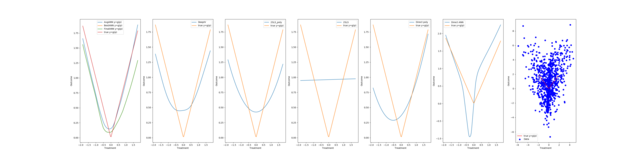}}    
    \caption{Examples of fitted functions via each method (in order from left to right: AGMM (best, final, avg), DeepIV, 2SLSPoly, 2SLS, DirectPoly, DirectNN, data points). $h_0(w)=|w|$, Number of instruments: 5, Number of samples: 1000,  Training steps: 400, Number of critics: 50, Kernel radius: 50 data points, Critic jitter: yes}
\end{figure*}

\newpage
\subsubsection{Identify Function}
\begin{figure*}[htpb]
    \centering
    \subfigure
    {\includegraphics[scale=1]{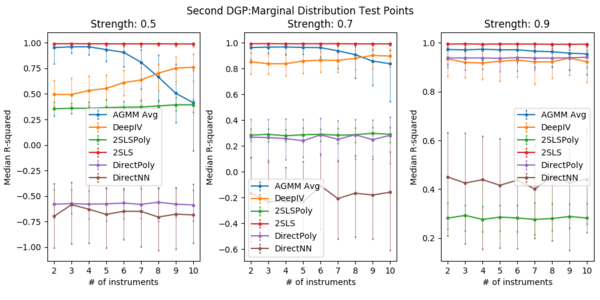}}
    \subfigure
    {\includegraphics[scale=1]{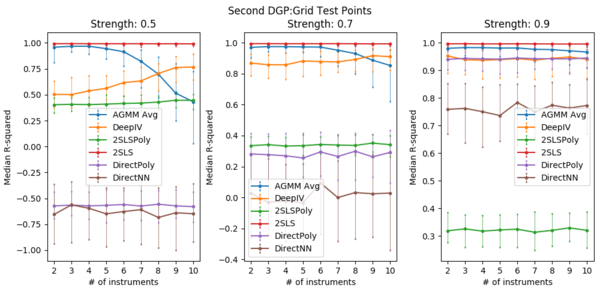}}
    \caption{Median and $10-90$ percentiles of $R^2$ across $100$ experiments as a function of number of instruments and instrument strength $\gamma$. $h_0(w) = x$. Number of samples: 1000,  Training steps: 400, Number of critics: 50, Kernel radius: 50 data points, Critic jitter: yes}
    
\end{figure*}

\begin{figure*}[htpb]
    \centering
    \subfigure[Strength $0.5$]
    {\includegraphics[scale=1]{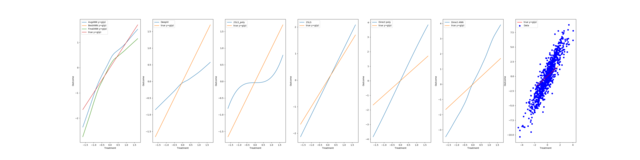}}
    \subfigure[Strength $0.7$]
    {\includegraphics[scale=1]{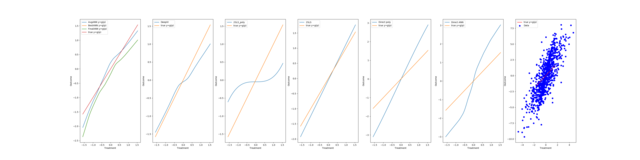}}
    \subfigure[Strength $0.9$]
    {\includegraphics[scale=1]{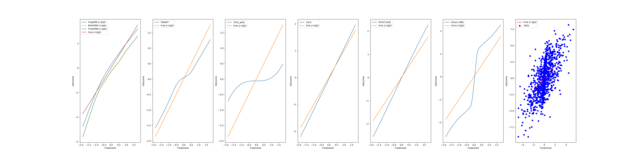}}    
    \caption{Examples of fitted functions via each method (in order from left to right: AGMM (best, final, avg), DeepIV, 2SLSPoly, 2SLS, DirectPoly, DirectNN, data points). $h_0(w)=w$, Number of instruments: 2, Number of samples: 1000,  Training steps: 400, Number of critics: 50, Kernel radius: 50 data points, Critic jitter: yes}
\end{figure*}

\begin{figure*}[htpb]
    \centering
    \subfigure[Strength $0.5$]
    {\includegraphics[scale=1]{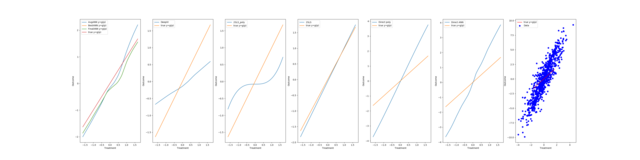}}
    \subfigure[Strength $0.7$]
    {\includegraphics[scale=1]{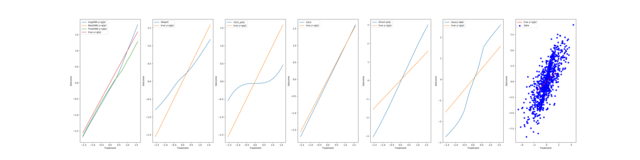}}
    \subfigure[Strength $0.9$]
    {\includegraphics[scale=1]{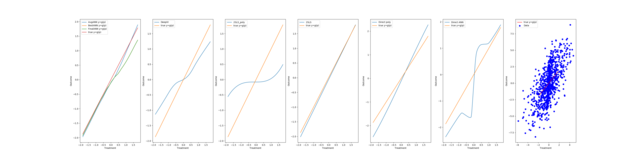}}    
    \caption{Examples of fitted functions via each method (in order from left to right: AGMM (best, final, avg), DeepIV, 2SLSPoly, 2SLS, DirectPoly, DirectNN, data points). $h_0(w)=w$, Number of instruments: 5, Number of samples: 1000,  Training steps: 400, Number of critics: 50, Kernel radius: 50 data points, Critic jitter: yes}
\end{figure*}
\newpage
\subsubsection{Sigmoid Function}
\begin{figure*}[htpb]
    \centering
    \subfigure
    {\includegraphics[scale=1]{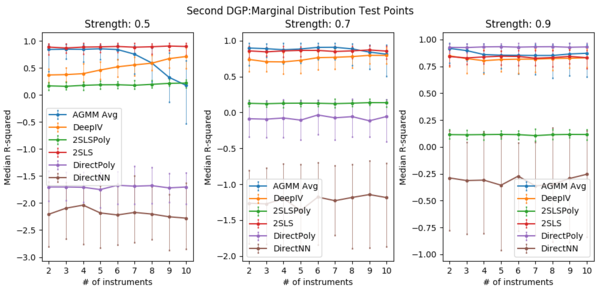}}
    \subfigure
    {\includegraphics[scale=1]{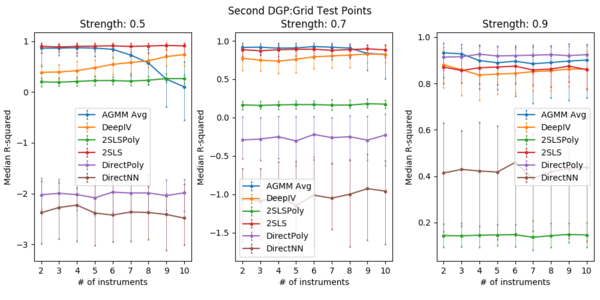}}
    \caption{Median and $10-90$ percentiles of $R^2$ across $100$ experiments as a function of number of instruments and instrument strength $\gamma$. $h_0(w)=\frac{2}{1+e^{-2x}}$. Number of samples: 1000,  Training steps: 400, Number of critics: 50, Kernel radius: 50 data points, Critic jitter: yes}
    
\end{figure*}

\begin{figure*}[htpb]
    \centering
    \subfigure[Strength $0.5$]
    {\includegraphics[scale=1]{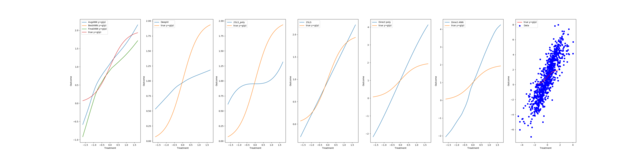}}
    \subfigure[Strength $0.7$]
    {\includegraphics[scale=1]{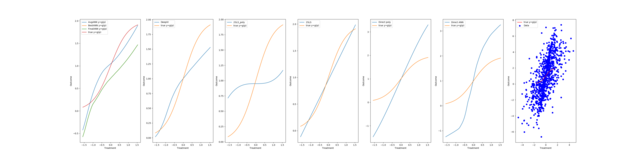}}
    \subfigure[Strength $0.9$]
    {\includegraphics[scale=1]{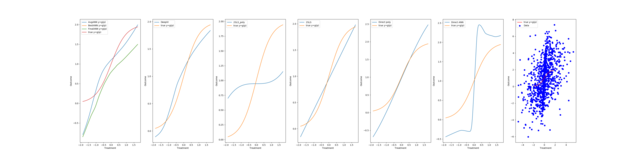}}    
    \caption{Examples of fitted functions via each method (in order from left to right: AGMM (best, final, avg), DeepIV, 2SLSPoly, 2SLS, DirectPoly, DirectNN, data points). $h_0(w)=\frac{2}{1+e^{-2x}}$, Number of instruments: 2, Number of samples: 1000,  Training steps: 400, Number of critics: 50, Kernel radius: 50 data points, Critic jitter: yes}
\end{figure*}

\begin{figure*}[htpb]
    \centering
    \subfigure[Strength $0.5$]
    {\includegraphics[scale=1]{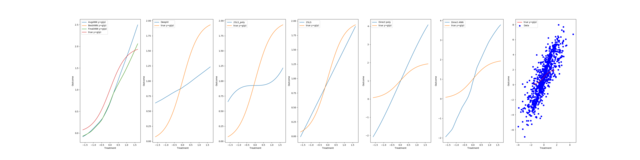}}
    \subfigure[Strength $0.7$]
    {\includegraphics[scale=1]{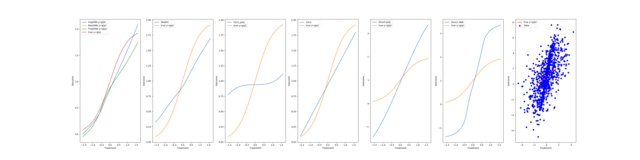}}
    \subfigure[Strength $0.9$]
    {\includegraphics[scale=1]{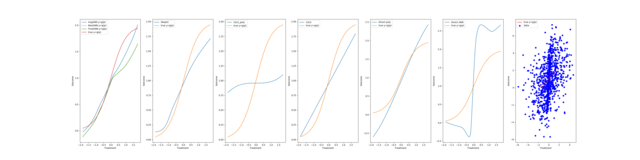}}    
    \caption{Examples of fitted functions via each method (in order from left to right: AGMM (best, final, avg), DeepIV, 2SLSPoly, 2SLS, DirectPoly, DirectNN, data points). $h_0(w)=\frac{2}{1+e^{-2x}}$, Number of instruments: 5, Number of samples: 1000,  Training steps: 400, Number of critics: 50, Kernel radius: 50 data points, Critic jitter: yes}
\end{figure*}

\newpage
\subsubsection{Sin Function}
\begin{figure*}[htpb]
    \centering
    \subfigure
    {\includegraphics[scale=1]{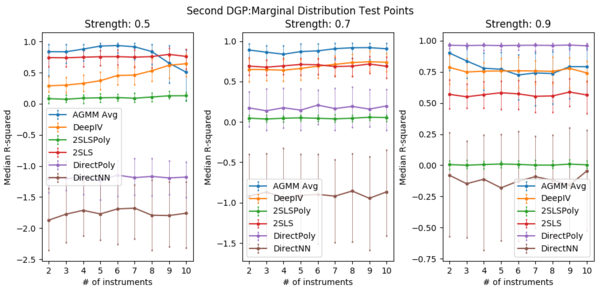}}
    \subfigure
    {\includegraphics[scale=1]{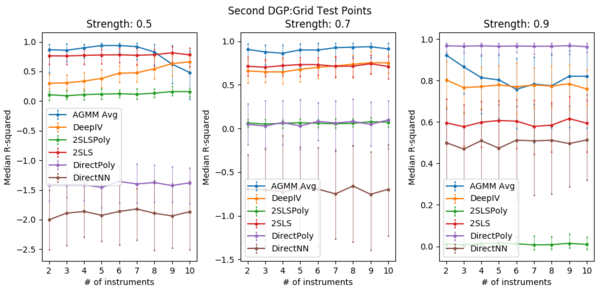}}
    \caption{Median and $10-90$ percentiles of $R^2$ across $100$ experiments as a function of number of instruments and instrument strength $\gamma$. $h_0(w) = \sin(x)$. Number of samples: 1000,  Training steps: 400, Number of critics: 50, Kernel radius: 50 data points, Critic jitter: yes}
    
\end{figure*}

\begin{figure*}[htpb]
    \centering
    \subfigure[Strength $0.5$]
    {\includegraphics[scale=1]{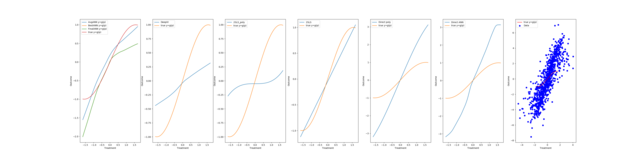}}
    \subfigure[Strength $0.7$]
    {\includegraphics[scale=1]{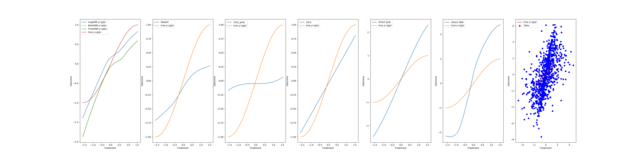}}
    \subfigure[Strength $0.9$]
    {\includegraphics[scale=1]{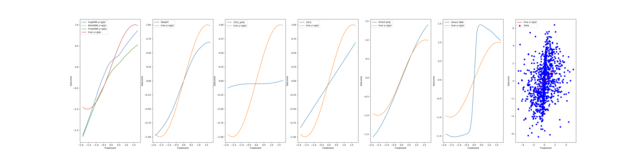}}    
    \caption{Examples of fitted functions via each method (in order from left to right: AGMM (best, final, avg), DeepIV, 2SLSPoly, 2SLS, DirectPoly, DirectNN, data points). $h_0(w)=sin(x)$, Number of instruments: 2, Number of samples: 1000,  Training steps: 400, Number of critics: 50, Kernel radius: 50 data points, Critic jitter: yes}
\end{figure*}

\begin{figure*}[htpb]
    \centering
    \subfigure[Strength $0.5$]
    {\includegraphics[scale=1]{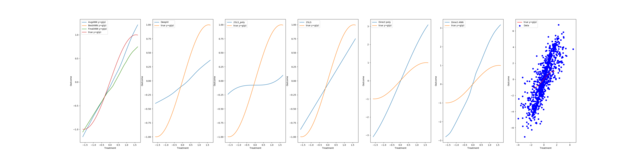}}
    \subfigure[Strength $0.7$]
    {\includegraphics[scale=1]{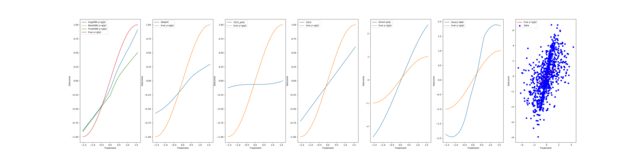}}
    \subfigure[Strength $0.9$]
    {\includegraphics[scale=1]{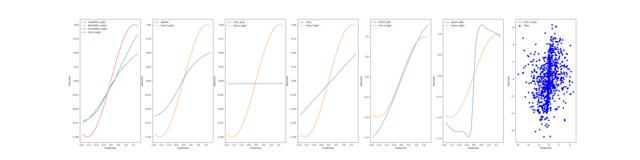}}    
    \caption{Examples of fitted functions via each method (in order from left to right: AGMM (best, final, avg), DeepIV, 2SLSPoly, 2SLS, DirectPoly, DirectNN, data points). $h_0(w)=sin(x)$, Number of instruments: 5, Number of samples: 1000,  Training steps: 400, Number of critics: 50, Kernel radius: 50 data points, Critic jitter: yes}
\end{figure*}

\newpage
\subsubsection{Step Function}
\begin{figure*}[htpb]
    \centering
    \subfigure
    {\includegraphics[scale=1]{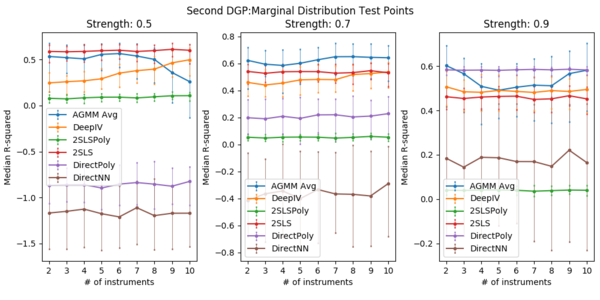}}
    \subfigure
    {\includegraphics[scale=1]{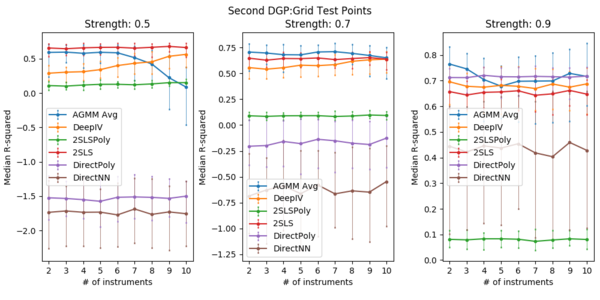}}
    \caption{Median and $10-90$ percentiles of $R^2$ across $100$ experiments as a function of number of instruments and instrument strength $\gamma$. $h_0(w)=1\{x<0\} + 2.5  1\{x \geq 0\}$. Number of samples: 1000,  Training steps: 400, Number of critics: 50, Kernel radius: 50 data points, Critic jitter: yes}
    
\end{figure*}

\begin{figure*}[htpb]
    \centering
    \subfigure[Strength $0.5$]
    {\includegraphics[scale=1]{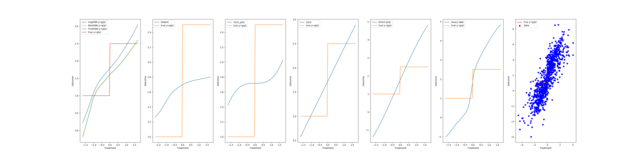}}
    \subfigure[Strength $0.7$]
    {\includegraphics[scale=1]{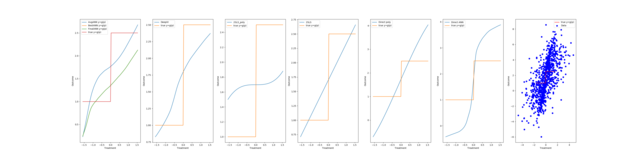}}
    \subfigure[Strength $0.9$]
    {\includegraphics[scale=1]{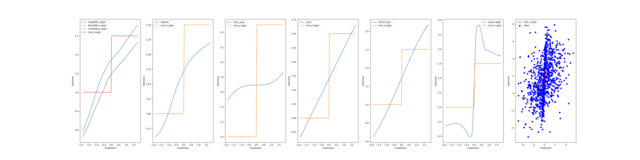}}    
    \caption{Examples of fitted functions via each method (in order from left to right: AGMM (best, final, avg), DeepIV, 2SLSPoly, 2SLS, DirectPoly, DirectNN, data points). $h_0(w)=1\{x<0\} + 2.5  1\{x \geq 0\}$, Number of instruments: 2, Number of samples: 1000,  Training steps: 400, Number of critics: 50, Kernel radius: 50 data points, Critic jitter: yes}
\end{figure*}

\begin{figure*}[htpb]
    \centering
    \subfigure[Strength $0.5$]
    {\includegraphics[scale=1]{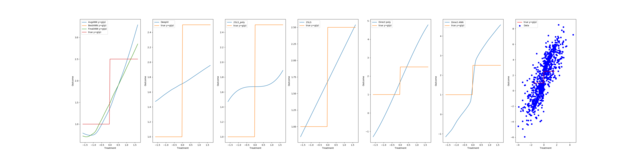}}
    \subfigure[Strength $0.7$]
    {\includegraphics[scale=1]{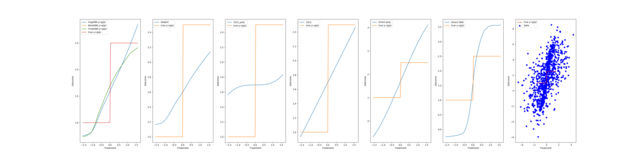}}
    \subfigure[Strength $0.9$]
    {\includegraphics[scale=1]{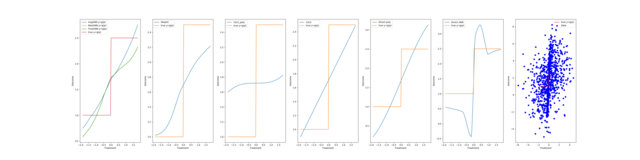}}    
    \caption{Examples of fitted functions via each method (in order from left to right: AGMM (best, final, avg), DeepIV, 2SLSPoly, 2SLS, DirectPoly, DirectNN, data points). $h_0(w)=1\{x<0\} + 2.5  1\{x \geq 0\}$, Number of instruments: 5, Number of samples: 1000,  Training steps: 400, Number of critics: 50, Kernel radius: 50 data points, Critic jitter: yes}
\end{figure*}

\newpage
\begin{center}
{\bf \Large Further Tables of Experimental Results}
\end{center}

\begin{center} \bf \Large First DGP \end{center}
\section{Marginal Distribution Test Points}
\subsection{Number of Instruments:1}
\subsubsection{Instrument Strength: 0.5}
\begin{figure}[H]
\begin{footnotesize}
\begin{tabular}{c|c|c|c|c|c|c}
function & AGMM Avg & DeepIV & 2SLSpoly & 2SLS & DirectPoly & DirectNN\\
\hline\\
abs & {\bf 0.78 {\tiny (0.46, 0.92) } }&-0.09 {\tiny (-0.44, 0.11) }&-0.44 {\tiny (-0.78, -0.20) }&-0.47 {\tiny (-0.81, -0.23) }&-2.77 {\tiny (-3.27, -2.20) }&-2.42 {\tiny (-3.36, -1.66) }\\
2dpoly & {\bf 0.96 {\tiny (0.89, 0.98) } }&0.37 {\tiny (0.18, 0.51) }&0.38 {\tiny (0.18, 0.52) }&0.35 {\tiny (0.15, 0.49) }&0.67 {\tiny (0.58, 0.72) }&0.66 {\tiny (0.56, 0.74) }\\
sigmoid & {\bf 0.88 {\tiny (0.56, 0.97) } }&0.50 {\tiny (0.28, 0.69) }&0.20 {\tiny (0.13, 0.28) }&0.88 {\tiny (0.77, 0.95) }&-0.91 {\tiny (-1.15, -0.59) }&-1.12 {\tiny (-1.73, -0.62) }\\
step & {\bf 0.61 {\tiny (0.34, 0.76) } }&0.36 {\tiny (0.19, 0.49) }&0.10 {\tiny (0.05, 0.15) }&0.60 {\tiny (0.52, 0.67) }&-0.54 {\tiny (-0.75, -0.30) }&-0.68 {\tiny (-1.11, -0.27) }\\
3dpoly & 0.02 {\tiny (-1.86, 0.71) }&-1.02 {\tiny (-2.27, -0.43) }&-0.21 {\tiny (-1.01, 0.22) }&-12.04 {\tiny (-18.23, -7.07) }&{\bf 0.38 {\tiny (0.15, 0.59) } }&0.26 {\tiny (-0.27, 0.58) }\\
sin & {\bf 0.88 {\tiny (0.56, 0.98) } }&0.40 {\tiny (0.24, 0.57) }&0.10 {\tiny (0.03, 0.16) }&0.72 {\tiny (0.58, 0.82) }&-0.50 {\tiny (-0.73, -0.23) }&-0.80 {\tiny (-1.28, -0.41) }\\
linear & 0.97 {\tiny (0.88, 0.99) }&0.67 {\tiny (0.49, 0.79) }&0.38 {\tiny (0.31, 0.47) }&{\bf 0.99 {\tiny (0.98, 1.00) } }&0.00 {\tiny (-0.14, 0.14) }&0.00 {\tiny (-0.29, 0.24) }\\
rand\_pw & {\bf 0.84 {\tiny (-0.01, 0.98) } }&0.34 {\tiny (-0.36, 0.70) }&0.20 {\tiny (-0.43, 0.43) }&0.54 {\tiny (-0.93, 0.98) }&0.26 {\tiny (-4.29, 0.88) }&0.43 {\tiny (-3.39, 0.90) }\\
\end{tabular}
\end{footnotesize}
\caption{Instrument strength: 0.5, Number of instruments: 1, Number of samples: 1000, Number of experiments: 100, Training steps: 400, Number of critics: 50, Kernel radius: 50 data points, Critic jitter: 1, Test points: Distribution}
\end{figure}
\subsubsection{Instrument Strength: 0.7}
\begin{figure}[H]
\begin{footnotesize}
\begin{tabular}{c|c|c|c|c|c|c}
function & AGMM Avg & DeepIV & 2SLSpoly & 2SLS & DirectPoly & DirectNN\\
\hline\\
abs & {\bf 0.89 {\tiny (0.72, 0.96) } }&0.53 {\tiny (0.25, 0.68) }&0.57 {\tiny (0.38, 0.69) }&-0.46 {\tiny (-0.77, -0.24) }&-0.20 {\tiny (-0.56, 0.08) }&0.07 {\tiny (-0.55, 0.36) }\\
2dpoly & {\bf 0.99 {\tiny (0.95, 1.00) } }&0.85 {\tiny (0.76, 0.91) }&0.53 {\tiny (0.40, 0.59) }&0.28 {\tiny (0.08, 0.42) }&0.91 {\tiny (0.89, 0.94) }&0.90 {\tiny (0.84, 0.95) }\\
sigmoid & {\bf 0.94 {\tiny (0.82, 0.98) } }&0.87 {\tiny (0.73, 0.96) }&0.16 {\tiny (0.11, 0.21) }&0.87 {\tiny (0.80, 0.93) }&0.51 {\tiny (0.33, 0.67) }&0.38 {\tiny (-0.00, 0.66) }\\
step & {\bf 0.79 {\tiny (0.65, 0.87) } }&0.66 {\tiny (0.53, 0.79) }&0.08 {\tiny (0.05, 0.11) }&0.60 {\tiny (0.54, 0.66) }&0.42 {\tiny (0.26, 0.53) }&0.44 {\tiny (0.12, 0.62) }\\
3dpoly & 0.87 {\tiny (0.75, 0.95) }&-0.19 {\tiny (-1.33, 0.43) }&0.22 {\tiny (-0.10, 0.48) }&-11.61 {\tiny (-18.61, -6.98) }&{\bf 0.88 {\tiny (0.81, 0.93) } }&0.79 {\tiny (0.33, 0.93) }\\
sin & {\bf 0.95 {\tiny (0.84, 0.98) } }&0.78 {\tiny (0.64, 0.91) }&0.04 {\tiny (-0.01, 0.08) }&0.66 {\tiny (0.55, 0.74) }&0.67 {\tiny (0.52, 0.78) }&0.47 {\tiny (0.09, 0.77) }\\
linear & 0.98 {\tiny (0.93, 0.99) }&0.95 {\tiny (0.87, 0.98) }&0.35 {\tiny (0.29, 0.40) }&{\bf 1.00 {\tiny (0.98, 1.00) } }&0.73 {\tiny (0.65, 0.81) }&0.71 {\tiny (0.57, 0.84) }\\
rand\_pw & {\bf 0.93 {\tiny (0.55, 0.99) } }&0.77 {\tiny (0.16, 0.96) }&0.29 {\tiny (-0.12, 0.64) }&0.54 {\tiny (-0.89, 0.98) }&0.66 {\tiny (-0.78, 0.95) }&0.83 {\tiny (-0.04, 0.97) }\\
\end{tabular}
\end{footnotesize}
\caption{Instrument strength: 0.7, Number of instruments: 1, Number of samples: 1000, Number of experiments: 100, Training steps: 400, Number of critics: 50, Kernel radius: 50 data points, Critic jitter: 1, Test points: Distribution}
\end{figure}
\subsubsection{Instrument Strength: 0.9}
\begin{figure}[H]
\begin{footnotesize}
\begin{tabular}{c|c|c|c|c|c|c}
function & AGMM Avg & DeepIV & 2SLSpoly & 2SLS & DirectPoly & DirectNN\\
\hline\\
abs & {\bf 0.95 {\tiny (0.87, 0.98) } }&0.83 {\tiny (0.68, 0.93) }&0.76 {\tiny (0.69, 0.81) }&-0.47 {\tiny (-0.70, -0.28) }&0.66 {\tiny (0.53, 0.73) }&0.91 {\tiny (0.78, 0.97) }\\
2dpoly & 0.99 {\tiny (0.99, 1.00) }&0.95 {\tiny (0.88, 0.98) }&0.52 {\tiny (0.40, 0.60) }&0.08 {\tiny (-0.11, 0.25) }&{\bf 1.00 {\tiny (0.99, 1.00) } }&0.99 {\tiny (0.97, 1.00) }\\
sigmoid & {\bf 0.96 {\tiny (0.90, 0.99) } }&0.96 {\tiny (0.86, 0.99) }&0.14 {\tiny (0.09, 0.19) }&0.85 {\tiny (0.78, 0.90) }&0.95 {\tiny (0.91, 0.96) }&0.94 {\tiny (0.83, 0.98) }\\
step & {\bf 0.88 {\tiny (0.83, 0.92) } }&0.81 {\tiny (0.71, 0.87) }&0.08 {\tiny (0.04, 0.11) }&0.61 {\tiny (0.54, 0.67) }&0.72 {\tiny (0.69, 0.74) }&0.85 {\tiny (0.76, 0.93) }\\
3dpoly & 0.99 {\tiny (0.97, 1.00) }&0.63 {\tiny (-0.24, 0.88) }&0.63 {\tiny (0.49, 0.72) }&-9.00 {\tiny (-13.68, -5.62) }&{\bf 1.00 {\tiny (0.99, 1.00) } }&0.95 {\tiny (0.62, 0.99) }\\
sin & {\bf 0.96 {\tiny (0.89, 0.98) } }&0.91 {\tiny (0.79, 0.96) }&-0.01 {\tiny (-0.05, 0.02) }&0.49 {\tiny (0.37, 0.61) }&0.94 {\tiny (0.89, 0.96) }&0.93 {\tiny (0.81, 0.97) }\\
linear & 0.99 {\tiny (0.96, 0.99) }&0.98 {\tiny (0.94, 0.99) }&0.33 {\tiny (0.28, 0.39) }&{\bf 1.00 {\tiny (0.99, 1.00) } }&0.98 {\tiny (0.96, 1.00) }&0.97 {\tiny (0.93, 0.99) }\\
rand\_pw & {\bf 0.97 {\tiny (0.85, 1.00) } }&0.92 {\tiny (0.59, 0.98) }&0.39 {\tiny (-0.25, 0.72) }&0.49 {\tiny (-1.11, 0.98) }&0.81 {\tiny (-0.01, 0.99) }&0.96 {\tiny (0.81, 0.99) }\\
\end{tabular}
\end{footnotesize}
\caption{Instrument strength: 0.9, Number of instruments: 1, Number of samples: 1000, Number of experiments: 100, Training steps: 400, Number of critics: 50, Kernel radius: 50 data points, Critic jitter: 1, Test points: Distribution}
\end{figure}
\subsection{Number of Instruments:2}
\subsubsection{Instrument Strength: 0.5}
\begin{figure}[H]
\begin{footnotesize}
\begin{tabular}{c|c|c|c|c|c|c}
function & AGMM Avg & DeepIV & 2SLSpoly & 2SLS & DirectPoly & DirectNN\\
\hline\\
abs & {\bf 0.66 {\tiny (0.31, 0.87) } }&-0.09 {\tiny (-0.43, 0.11) }&-0.45 {\tiny (-0.79, -0.17) }&-0.46 {\tiny (-0.81, -0.27) }&-2.67 {\tiny (-3.44, -2.23) }&-2.39 {\tiny (-3.25, -1.59) }\\
2dpoly & {\bf 0.94 {\tiny (0.87, 0.98) } }&0.34 {\tiny (0.20, 0.49) }&0.36 {\tiny (0.11, 0.53) }&0.35 {\tiny (0.18, 0.46) }&0.67 {\tiny (0.58, 0.73) }&0.67 {\tiny (0.56, 0.76) }\\
sigmoid & 0.81 {\tiny (0.48, 0.94) }&0.52 {\tiny (0.30, 0.74) }&0.21 {\tiny (0.12, 0.27) }&{\bf 0.89 {\tiny (0.78, 0.95) } }&-0.87 {\tiny (-1.17, -0.63) }&-1.11 {\tiny (-1.76, -0.60) }\\
step & 0.46 {\tiny (0.11, 0.66) }&0.35 {\tiny (0.21, 0.48) }&0.11 {\tiny (0.05, 0.15) }&{\bf 0.61 {\tiny (0.50, 0.68) } }&-0.51 {\tiny (-0.75, -0.33) }&-0.62 {\tiny (-1.21, -0.27) }\\
3dpoly & -2.19 {\tiny (-4.30, -0.41) }&-1.11 {\tiny (-2.28, -0.37) }&-0.21 {\tiny (-1.24, 0.23) }&-11.60 {\tiny (-18.00, -7.49) }&{\bf 0.41 {\tiny (0.03, 0.60) } }&0.31 {\tiny (-0.23, 0.63) }\\
sin & {\bf 0.82 {\tiny (0.45, 0.95) } }&0.42 {\tiny (0.23, 0.60) }&0.11 {\tiny (0.03, 0.16) }&0.73 {\tiny (0.57, 0.83) }&-0.47 {\tiny (-0.72, -0.27) }&-0.83 {\tiny (-1.34, -0.25) }\\
linear & 0.96 {\tiny (0.88, 0.99) }&0.67 {\tiny (0.52, 0.81) }&0.40 {\tiny (0.31, 0.47) }&{\bf 0.99 {\tiny (0.98, 1.00) } }&0.03 {\tiny (-0.16, 0.14) }&0.02 {\tiny (-0.26, 0.25) }\\
rand\_pw & {\bf 0.61 {\tiny (-0.74, 0.94) } }&0.31 {\tiny (-0.33, 0.68) }&0.19 {\tiny (-0.37, 0.44) }&0.58 {\tiny (-0.89, 0.98) }&0.24 {\tiny (-4.09, 0.89) }&0.43 {\tiny (-3.71, 0.89) }\\
\end{tabular}
\end{footnotesize}
\caption{Instrument strength: 0.5, Number of instruments: 2, Number of samples: 1000, Number of experiments: 100, Training steps: 400, Number of critics: 50, Kernel radius: 50 data points, Critic jitter: 1, Test points: Distribution}
\end{figure}
\subsubsection{Instrument Strength: 0.7}
\begin{figure}[H]
\begin{footnotesize}
\begin{tabular}{c|c|c|c|c|c|c}
function & AGMM Avg & DeepIV & 2SLSpoly & 2SLS & DirectPoly & DirectNN\\
\hline\\
abs & {\bf 0.84 {\tiny (0.64, 0.94) } }&0.56 {\tiny (0.32, 0.68) }&0.60 {\tiny (0.38, 0.71) }&-0.44 {\tiny (-0.72, -0.28) }&-0.23 {\tiny (-0.59, 0.13) }&-0.02 {\tiny (-0.46, 0.38) }\\
2dpoly & {\bf 0.97 {\tiny (0.92, 0.98) } }&0.83 {\tiny (0.69, 0.91) }&0.51 {\tiny (0.40, 0.61) }&0.27 {\tiny (0.10, 0.41) }&0.91 {\tiny (0.88, 0.94) }&0.91 {\tiny (0.85, 0.95) }\\
sigmoid & {\bf 0.89 {\tiny (0.71, 0.97) } }&0.86 {\tiny (0.70, 0.97) }&0.17 {\tiny (0.11, 0.23) }&0.88 {\tiny (0.80, 0.93) }&0.50 {\tiny (0.31, 0.67) }&0.33 {\tiny (-0.05, 0.63) }\\
step & 0.65 {\tiny (0.46, 0.77) }&{\bf 0.67 {\tiny (0.46, 0.75) } }&0.09 {\tiny (0.05, 0.12) }&0.61 {\tiny (0.54, 0.66) }&0.40 {\tiny (0.26, 0.52) }&0.40 {\tiny (0.16, 0.64) }\\
3dpoly & 0.51 {\tiny (-0.22, 0.82) }&-0.06 {\tiny (-0.98, 0.36) }&0.28 {\tiny (-0.14, 0.52) }&-10.75 {\tiny (-16.15, -6.68) }&{\bf 0.88 {\tiny (0.81, 0.93) } }&0.83 {\tiny (0.34, 0.94) }\\
sin & {\bf 0.88 {\tiny (0.67, 0.97) } }&0.79 {\tiny (0.62, 0.89) }&0.04 {\tiny (-0.00, 0.08) }&0.67 {\tiny (0.56, 0.76) }&0.65 {\tiny (0.49, 0.80) }&0.48 {\tiny (0.12, 0.72) }\\
linear & 0.98 {\tiny (0.93, 1.00) }&0.95 {\tiny (0.88, 0.98) }&0.36 {\tiny (0.29, 0.41) }&{\bf 1.00 {\tiny (0.98, 1.00) } }&0.73 {\tiny (0.64, 0.81) }&0.70 {\tiny (0.53, 0.83) }\\
rand\_pw & {\bf 0.87 {\tiny (0.24, 0.98) } }&0.72 {\tiny (0.23, 0.95) }&0.29 {\tiny (-0.12, 0.69) }&0.55 {\tiny (-1.02, 0.98) }&0.66 {\tiny (-0.86, 0.96) }&0.80 {\tiny (0.11, 0.96) }\\
\end{tabular}
\end{footnotesize}
\caption{Instrument strength: 0.7, Number of instruments: 2, Number of samples: 1000, Number of experiments: 100, Training steps: 400, Number of critics: 50, Kernel radius: 50 data points, Critic jitter: 1, Test points: Distribution}
\end{figure}
\subsubsection{Instrument Strength: 0.9}
\begin{figure}[H]
\begin{footnotesize}
\begin{tabular}{c|c|c|c|c|c|c}
function & AGMM Avg & DeepIV & 2SLSpoly & 2SLS & DirectPoly & DirectNN\\
\hline\\
abs & {\bf 0.91 {\tiny (0.79, 0.97) } }&0.83 {\tiny (0.62, 0.92) }&0.77 {\tiny (0.68, 0.82) }&-0.43 {\tiny (-0.74, -0.27) }&0.65 {\tiny (0.55, 0.74) }&0.89 {\tiny (0.79, 0.96) }\\
2dpoly & 0.98 {\tiny (0.95, 0.99) }&0.95 {\tiny (0.82, 0.98) }&0.52 {\tiny (0.39, 0.62) }&0.09 {\tiny (-0.11, 0.25) }&{\bf 0.99 {\tiny (0.99, 1.00) } }&0.99 {\tiny (0.97, 1.00) }\\
sigmoid & 0.92 {\tiny (0.80, 0.98) }&{\bf 0.95 {\tiny (0.88, 0.99) } }&0.15 {\tiny (0.09, 0.20) }&0.86 {\tiny (0.79, 0.90) }&0.95 {\tiny (0.91, 0.96) }&0.92 {\tiny (0.79, 0.98) }\\
step & 0.75 {\tiny (0.60, 0.82) }&0.80 {\tiny (0.69, 0.85) }&0.08 {\tiny (0.04, 0.11) }&0.61 {\tiny (0.54, 0.66) }&0.72 {\tiny (0.69, 0.74) }&{\bf 0.83 {\tiny (0.71, 0.91) } }\\
3dpoly & 0.85 {\tiny (0.51, 0.91) }&0.69 {\tiny (-0.06, 0.88) }&0.65 {\tiny (0.47, 0.76) }&-8.39 {\tiny (-12.35, -5.19) }&{\bf 1.00 {\tiny (0.99, 1.00) } }&0.96 {\tiny (0.72, 1.00) }\\
sin & 0.90 {\tiny (0.72, 0.96) }&0.89 {\tiny (0.74, 0.96) }&-0.01 {\tiny (-0.04, 0.02) }&0.50 {\tiny (0.36, 0.60) }&{\bf 0.94 {\tiny (0.89, 0.96) } }&0.93 {\tiny (0.82, 0.98) }\\
linear & 0.99 {\tiny (0.95, 1.00) }&0.98 {\tiny (0.91, 0.99) }&0.34 {\tiny (0.28, 0.39) }&{\bf 1.00 {\tiny (0.99, 1.00) } }&0.98 {\tiny (0.96, 1.00) }&0.97 {\tiny (0.93, 0.99) }\\
rand\_pw & 0.93 {\tiny (0.42, 0.99) }&0.89 {\tiny (0.52, 0.99) }&0.38 {\tiny (-0.19, 0.75) }&0.43 {\tiny (-1.04, 0.98) }&0.83 {\tiny (0.00, 0.99) }&{\bf 0.95 {\tiny (0.84, 0.99) } }\\
\end{tabular}
\end{footnotesize}
\caption{Instrument strength: 0.9, Number of instruments: 2, Number of samples: 1000, Number of experiments: 100, Training steps: 400, Number of critics: 50, Kernel radius: 50 data points, Critic jitter: 1, Test points: Distribution}
\end{figure}
\subsection{Number of Instruments:3}
\subsubsection{Instrument Strength: 0.5}
\begin{figure}[H]
\begin{footnotesize}
\begin{tabular}{c|c|c|c|c|c|c}
function & AGMM Avg & DeepIV & 2SLSpoly & 2SLS & DirectPoly & DirectNN\\
\hline\\
abs & {\bf 0.69 {\tiny (0.45, 0.86) } }&-0.08 {\tiny (-0.49, 0.15) }&-0.43 {\tiny (-0.82, -0.18) }&-0.47 {\tiny (-0.83, -0.25) }&-2.71 {\tiny (-3.36, -2.16) }&-2.38 {\tiny (-3.30, -1.79) }\\
2dpoly & {\bf 0.94 {\tiny (0.85, 0.97) } }&0.33 {\tiny (0.14, 0.47) }&0.38 {\tiny (0.08, 0.51) }&0.34 {\tiny (0.17, 0.46) }&0.66 {\tiny (0.59, 0.73) }&0.66 {\tiny (0.57, 0.74) }\\
sigmoid & 0.81 {\tiny (0.58, 0.93) }&0.53 {\tiny (0.35, 0.72) }&0.21 {\tiny (0.13, 0.27) }&{\bf 0.89 {\tiny (0.76, 0.94) } }&-0.85 {\tiny (-1.14, -0.55) }&-1.06 {\tiny (-1.88, -0.63) }\\
step & 0.47 {\tiny (0.17, 0.66) }&0.36 {\tiny (0.22, 0.50) }&0.11 {\tiny (0.05, 0.14) }&{\bf 0.61 {\tiny (0.50, 0.67) } }&-0.50 {\tiny (-0.73, -0.29) }&-0.59 {\tiny (-1.13, -0.27) }\\
3dpoly & -2.57 {\tiny (-4.99, -0.49) }&-1.01 {\tiny (-2.43, -0.36) }&-0.20 {\tiny (-1.06, 0.14) }&-12.09 {\tiny (-17.91, -7.42) }&{\bf 0.39 {\tiny (0.09, 0.59) } }&0.26 {\tiny (-0.47, 0.62) }\\
sin & {\bf 0.82 {\tiny (0.58, 0.96) } }&0.42 {\tiny (0.25, 0.56) }&0.11 {\tiny (0.03, 0.16) }&0.73 {\tiny (0.56, 0.82) }&-0.45 {\tiny (-0.72, -0.20) }&-0.81 {\tiny (-1.33, -0.36) }\\
linear & 0.97 {\tiny (0.92, 0.99) }&0.67 {\tiny (0.48, 0.80) }&0.39 {\tiny (0.32, 0.46) }&{\bf 0.99 {\tiny (0.98, 1.00) } }&0.01 {\tiny (-0.14, 0.14) }&0.00 {\tiny (-0.23, 0.28) }\\
rand\_pw & {\bf 0.62 {\tiny (-0.69, 0.93) } }&0.34 {\tiny (-0.26, 0.69) }&0.18 {\tiny (-0.24, 0.45) }&0.55 {\tiny (-0.87, 0.98) }&0.25 {\tiny (-3.86, 0.88) }&0.47 {\tiny (-3.21, 0.89) }\\
\end{tabular}
\end{footnotesize}
\caption{Instrument strength: 0.5, Number of instruments: 3, Number of samples: 1000, Number of experiments: 100, Training steps: 400, Number of critics: 50, Kernel radius: 50 data points, Critic jitter: 1, Test points: Distribution}
\end{figure}
\subsubsection{Instrument Strength: 0.7}
\begin{figure}[H]
\begin{footnotesize}
\begin{tabular}{c|c|c|c|c|c|c}
function & AGMM Avg & DeepIV & 2SLSpoly & 2SLS & DirectPoly & DirectNN\\
\hline\\
abs & {\bf 0.84 {\tiny (0.62, 0.92) } }&0.53 {\tiny (0.31, 0.68) }&0.57 {\tiny (0.39, 0.71) }&-0.46 {\tiny (-0.83, -0.25) }&-0.19 {\tiny (-0.55, 0.07) }&0.00 {\tiny (-0.39, 0.38) }\\
2dpoly & {\bf 0.96 {\tiny (0.90, 0.98) } }&0.84 {\tiny (0.73, 0.90) }&0.51 {\tiny (0.39, 0.58) }&0.26 {\tiny (0.11, 0.39) }&0.91 {\tiny (0.87, 0.94) }&0.91 {\tiny (0.86, 0.94) }\\
sigmoid & {\bf 0.89 {\tiny (0.69, 0.95) } }&0.87 {\tiny (0.71, 0.97) }&0.17 {\tiny (0.10, 0.22) }&0.88 {\tiny (0.78, 0.92) }&0.52 {\tiny (0.33, 0.67) }&0.39 {\tiny (-0.03, 0.69) }\\
step & 0.62 {\tiny (0.43, 0.76) }&{\bf 0.66 {\tiny (0.53, 0.76) } }&0.09 {\tiny (0.05, 0.12) }&0.61 {\tiny (0.53, 0.66) }&0.42 {\tiny (0.27, 0.53) }&0.40 {\tiny (0.13, 0.64) }\\
3dpoly & 0.35 {\tiny (-1.10, 0.78) }&-0.21 {\tiny (-0.97, 0.31) }&0.28 {\tiny (-0.11, 0.46) }&-10.75 {\tiny (-17.88, -7.58) }&{\bf 0.89 {\tiny (0.82, 0.93) } }&0.81 {\tiny (0.35, 0.93) }\\
sin & {\bf 0.86 {\tiny (0.68, 0.95) } }&0.80 {\tiny (0.66, 0.91) }&0.04 {\tiny (-0.01, 0.08) }&0.67 {\tiny (0.52, 0.75) }&0.68 {\tiny (0.51, 0.81) }&0.48 {\tiny (0.12, 0.78) }\\
linear & 0.98 {\tiny (0.93, 1.00) }&0.94 {\tiny (0.85, 0.98) }&0.35 {\tiny (0.26, 0.40) }&{\bf 1.00 {\tiny (0.98, 1.00) } }&0.74 {\tiny (0.65, 0.81) }&0.71 {\tiny (0.58, 0.85) }\\
rand\_pw & {\bf 0.84 {\tiny (0.02, 0.98) } }&0.73 {\tiny (0.14, 0.95) }&0.31 {\tiny (-0.12, 0.64) }&0.51 {\tiny (-0.82, 0.98) }&0.70 {\tiny (-0.67, 0.95) }&0.82 {\tiny (-0.20, 0.97) }\\
\end{tabular}
\end{footnotesize}
\caption{Instrument strength: 0.7, Number of instruments: 3, Number of samples: 1000, Number of experiments: 100, Training steps: 400, Number of critics: 50, Kernel radius: 50 data points, Critic jitter: 1, Test points: Distribution}
\end{figure}
\subsubsection{Instrument Strength: 0.9}
\begin{figure}[H]
\begin{footnotesize}
\begin{tabular}{c|c|c|c|c|c|c}
function & AGMM Avg & DeepIV & 2SLSpoly & 2SLS & DirectPoly & DirectNN\\
\hline\\
abs & 0.90 {\tiny (0.75, 0.96) }&0.79 {\tiny (0.62, 0.88) }&0.75 {\tiny (0.70, 0.81) }&-0.47 {\tiny (-0.75, -0.28) }&0.66 {\tiny (0.56, 0.72) }&{\bf 0.91 {\tiny (0.78, 0.98) } }\\
2dpoly & 0.96 {\tiny (0.92, 0.99) }&0.94 {\tiny (0.83, 0.97) }&0.55 {\tiny (0.43, 0.69) }&0.07 {\tiny (-0.08, 0.21) }&{\bf 0.99 {\tiny (0.99, 1.00) } }&0.99 {\tiny (0.97, 1.00) }\\
sigmoid & 0.90 {\tiny (0.75, 0.97) }&0.94 {\tiny (0.88, 0.99) }&0.14 {\tiny (0.08, 0.18) }&0.85 {\tiny (0.77, 0.90) }&{\bf 0.95 {\tiny (0.91, 0.96) } }&0.93 {\tiny (0.83, 0.98) }\\
step & 0.69 {\tiny (0.51, 0.81) }&0.78 {\tiny (0.65, 0.85) }&0.07 {\tiny (0.04, 0.10) }&0.61 {\tiny (0.53, 0.65) }&0.72 {\tiny (0.68, 0.74) }&{\bf 0.85 {\tiny (0.74, 0.92) } }\\
3dpoly & 0.84 {\tiny (0.37, 0.92) }&0.67 {\tiny (-0.08, 0.86) }&0.67 {\tiny (0.52, 0.76) }&-9.21 {\tiny (-14.50, -6.17) }&{\bf 1.00 {\tiny (0.99, 1.00) } }&0.94 {\tiny (0.41, 1.00) }\\
sin & 0.84 {\tiny (0.67, 0.93) }&0.88 {\tiny (0.75, 0.97) }&-0.01 {\tiny (-0.05, 0.02) }&0.49 {\tiny (0.34, 0.60) }&{\bf 0.94 {\tiny (0.87, 0.96) } }&0.92 {\tiny (0.80, 0.98) }\\
linear & 0.99 {\tiny (0.96, 1.00) }&0.98 {\tiny (0.93, 0.99) }&0.33 {\tiny (0.28, 0.41) }&{\bf 1.00 {\tiny (0.99, 1.00) } }&0.98 {\tiny (0.96, 1.00) }&0.97 {\tiny (0.93, 0.99) }\\
rand\_pw & 0.91 {\tiny (0.12, 0.98) }&0.88 {\tiny (0.45, 0.98) }&0.41 {\tiny (-0.20, 0.74) }&0.42 {\tiny (-1.03, 0.98) }&0.81 {\tiny (-0.03, 0.99) }&{\bf 0.96 {\tiny (0.69, 0.99) } }\\
\end{tabular}
\end{footnotesize}
\caption{Instrument strength: 0.9, Number of instruments: 3, Number of samples: 1000, Number of experiments: 100, Training steps: 400, Number of critics: 50, Kernel radius: 50 data points, Critic jitter: 1, Test points: Distribution}
\end{figure}
\subsection{Number of Instruments:5}
\subsubsection{Instrument Strength: 0.5}
\begin{figure}[H]
\begin{footnotesize}
\begin{tabular}{c|c|c|c|c|c|c}
function & AGMM Avg & DeepIV & 2SLSpoly & 2SLS & DirectPoly & DirectNN\\
\hline\\
abs & {\bf 0.65 {\tiny (0.32, 0.92) } }&-0.06 {\tiny (-0.39, 0.19) }&-0.45 {\tiny (-0.76, -0.08) }&-0.48 {\tiny (-0.82, -0.25) }&-2.78 {\tiny (-3.40, -2.18) }&-2.50 {\tiny (-3.29, -1.91) }\\
2dpoly & {\bf 0.91 {\tiny (0.84, 0.96) } }&0.37 {\tiny (0.20, 0.49) }&0.40 {\tiny (0.18, 0.50) }&0.34 {\tiny (0.14, 0.47) }&0.67 {\tiny (0.59, 0.72) }&0.67 {\tiny (0.55, 0.74) }\\
sigmoid & 0.83 {\tiny (0.58, 0.96) }&0.57 {\tiny (0.39, 0.75) }&0.21 {\tiny (0.14, 0.29) }&{\bf 0.89 {\tiny (0.76, 0.95) } }&-0.87 {\tiny (-1.18, -0.59) }&-1.12 {\tiny (-1.74, -0.59) }\\
step & 0.49 {\tiny (0.18, 0.68) }&0.39 {\tiny (0.27, 0.53) }&0.11 {\tiny (0.07, 0.16) }&{\bf 0.62 {\tiny (0.49, 0.67) } }&-0.53 {\tiny (-0.75, -0.29) }&-0.62 {\tiny (-1.13, -0.16) }\\
3dpoly & -2.37 {\tiny (-5.82, -0.46) }&-1.22 {\tiny (-2.17, -0.34) }&-0.34 {\tiny (-1.25, 0.19) }&-12.02 {\tiny (-18.81, -8.05) }&{\bf 0.40 {\tiny (0.10, 0.58) } }&0.29 {\tiny (-0.24, 0.59) }\\
sin & {\bf 0.89 {\tiny (0.71, 0.97) } }&0.48 {\tiny (0.32, 0.65) }&0.11 {\tiny (0.05, 0.18) }&0.74 {\tiny (0.58, 0.83) }&-0.48 {\tiny (-0.74, -0.21) }&-0.86 {\tiny (-1.48, -0.37) }\\
linear & 0.97 {\tiny (0.89, 0.99) }&0.72 {\tiny (0.51, 0.87) }&0.40 {\tiny (0.32, 0.46) }&{\bf 0.99 {\tiny (0.97, 1.00) } }&0.00 {\tiny (-0.16, 0.13) }&0.01 {\tiny (-0.25, 0.23) }\\
rand\_pw & 0.56 {\tiny (-0.91, 0.94) }&0.36 {\tiny (-0.42, 0.72) }&0.19 {\tiny (-0.43, 0.45) }&{\bf 0.58 {\tiny (-0.91, 0.99) } }&0.25 {\tiny (-3.70, 0.89) }&0.45 {\tiny (-2.79, 0.89) }\\
\end{tabular}
\end{footnotesize}
\caption{Instrument strength: 0.5, Number of instruments: 5, Number of samples: 1000, Number of experiments: 100, Training steps: 400, Number of critics: 50, Kernel radius: 50 data points, Critic jitter: 1, Test points: Distribution}
\end{figure}
\subsubsection{Instrument Strength: 0.7}
\begin{figure}[H]
\begin{footnotesize}
\begin{tabular}{c|c|c|c|c|c|c}
function & AGMM Avg & DeepIV & 2SLSpoly & 2SLS & DirectPoly & DirectNN\\
\hline\\
abs & {\bf 0.79 {\tiny (0.54, 0.94) } }&0.50 {\tiny (0.29, 0.68) }&0.56 {\tiny (0.36, 0.70) }&-0.45 {\tiny (-0.78, -0.25) }&-0.22 {\tiny (-0.55, 0.08) }&0.04 {\tiny (-0.39, 0.42) }\\
2dpoly & {\bf 0.95 {\tiny (0.90, 0.98) } }&0.83 {\tiny (0.74, 0.89) }&0.51 {\tiny (0.41, 0.60) }&0.26 {\tiny (0.11, 0.41) }&0.91 {\tiny (0.88, 0.94) }&0.91 {\tiny (0.85, 0.95) }\\
sigmoid & 0.87 {\tiny (0.62, 0.95) }&{\bf 0.88 {\tiny (0.73, 0.94) } }&0.16 {\tiny (0.11, 0.23) }&0.88 {\tiny (0.78, 0.92) }&0.51 {\tiny (0.34, 0.67) }&0.37 {\tiny (-0.10, 0.65) }\\
step & 0.58 {\tiny (0.26, 0.72) }&{\bf 0.66 {\tiny (0.55, 0.77) } }&0.08 {\tiny (0.05, 0.12) }&0.61 {\tiny (0.52, 0.66) }&0.42 {\tiny (0.30, 0.53) }&0.42 {\tiny (0.12, 0.58) }\\
3dpoly & -0.51 {\tiny (-2.26, 0.46) }&-0.19 {\tiny (-1.14, 0.35) }&0.23 {\tiny (-0.17, 0.51) }&-11.33 {\tiny (-17.56, -7.67) }&{\bf 0.88 {\tiny (0.81, 0.93) } }&0.78 {\tiny (0.44, 0.92) }\\
sin & {\bf 0.83 {\tiny (0.61, 0.95) } }&0.79 {\tiny (0.61, 0.89) }&0.04 {\tiny (-0.00, 0.09) }&0.66 {\tiny (0.52, 0.75) }&0.67 {\tiny (0.52, 0.80) }&0.47 {\tiny (0.15, 0.74) }\\
linear & 0.98 {\tiny (0.91, 1.00) }&0.95 {\tiny (0.88, 0.99) }&0.34 {\tiny (0.29, 0.41) }&{\bf 1.00 {\tiny (0.98, 1.00) } }&0.73 {\tiny (0.64, 0.82) }&0.72 {\tiny (0.52, 0.85) }\\
rand\_pw & 0.78 {\tiny (-0.25, 0.97) }&0.73 {\tiny (0.23, 0.95) }&0.32 {\tiny (-0.09, 0.62) }&0.51 {\tiny (-0.96, 0.98) }&0.65 {\tiny (-0.98, 0.96) }&{\bf 0.82 {\tiny (0.15, 0.97) } }\\
\end{tabular}
\end{footnotesize}
\caption{Instrument strength: 0.7, Number of instruments: 5, Number of samples: 1000, Number of experiments: 100, Training steps: 400, Number of critics: 50, Kernel radius: 50 data points, Critic jitter: 1, Test points: Distribution}
\end{figure}
\subsubsection{Instrument Strength: 0.9}
\begin{figure}[H]
\begin{footnotesize}
\begin{tabular}{c|c|c|c|c|c|c}
function & AGMM Avg & DeepIV & 2SLSpoly & 2SLS & DirectPoly & DirectNN\\
\hline\\
abs & {\bf 0.88 {\tiny (0.71, 0.97) } }&0.77 {\tiny (0.64, 0.89) }&0.74 {\tiny (0.67, 0.81) }&-0.46 {\tiny (-0.78, -0.25) }&0.64 {\tiny (0.55, 0.73) }&0.87 {\tiny (0.74, 0.97) }\\
2dpoly & 0.96 {\tiny (0.90, 0.98) }&0.93 {\tiny (0.83, 0.98) }&0.59 {\tiny (0.44, 0.69) }&0.08 {\tiny (-0.11, 0.28) }&{\bf 0.99 {\tiny (0.99, 1.00) } }&0.99 {\tiny (0.97, 1.00) }\\
sigmoid & 0.85 {\tiny (0.66, 0.94) }&0.93 {\tiny (0.83, 0.98) }&0.14 {\tiny (0.10, 0.19) }&0.85 {\tiny (0.76, 0.89) }&{\bf 0.95 {\tiny (0.91, 0.96) } }&0.91 {\tiny (0.80, 0.98) }\\
step & 0.56 {\tiny (0.26, 0.77) }&0.77 {\tiny (0.65, 0.83) }&0.07 {\tiny (0.04, 0.11) }&0.61 {\tiny (0.52, 0.66) }&0.72 {\tiny (0.68, 0.74) }&{\bf 0.85 {\tiny (0.71, 0.91) } }\\
3dpoly & 0.57 {\tiny (-0.28, 0.86) }&0.64 {\tiny (-0.26, 0.88) }&0.70 {\tiny (0.54, 0.81) }&-9.28 {\tiny (-13.28, -6.29) }&{\bf 1.00 {\tiny (0.99, 1.00) } }&0.94 {\tiny (0.37, 0.99) }\\
sin & 0.73 {\tiny (0.52, 0.87) }&0.89 {\tiny (0.75, 0.96) }&-0.01 {\tiny (-0.05, 0.03) }&0.49 {\tiny (0.32, 0.62) }&{\bf 0.94 {\tiny (0.87, 0.96) } }&0.92 {\tiny (0.80, 0.97) }\\
linear & 0.99 {\tiny (0.95, 1.00) }&0.98 {\tiny (0.93, 0.99) }&0.33 {\tiny (0.29, 0.41) }&{\bf 1.00 {\tiny (0.99, 1.00) } }&0.98 {\tiny (0.96, 1.00) }&0.97 {\tiny (0.91, 0.99) }\\
rand\_pw & 0.78 {\tiny (0.01, 0.98) }&0.87 {\tiny (0.47, 0.96) }&0.29 {\tiny (-0.28, 0.67) }&0.39 {\tiny (-1.02, 0.96) }&0.79 {\tiny (-0.06, 0.98) }&{\bf 0.96 {\tiny (0.62, 0.99) } }\\
\end{tabular}
\end{footnotesize}
\caption{Instrument strength: 0.9, Number of instruments: 5, Number of samples: 1000, Number of experiments: 32, Training steps: 400, Number of critics: 50, Kernel radius: 50 data points, Critic jitter: 1, Test points: Distribution}
\end{figure}
\subsection{Number of Instruments:10}
\subsubsection{Instrument Strength: 0.5}
\begin{figure}[H]
\begin{footnotesize}
\begin{tabular}{c|c|c|c|c|c|c}
function & AGMM Avg & DeepIV & 2SLSpoly & 2SLS & DirectPoly & DirectNN\\
\hline\\
abs & -0.17 {\tiny (-1.02, 0.56) }&{\bf -0.01 {\tiny (-0.29, 0.18) } }&-0.40 {\tiny (-0.76, -0.19) }&-0.46 {\tiny (-0.78, -0.23) }&-2.77 {\tiny (-3.29, -2.24) }&-2.50 {\tiny (-3.30, -1.94) }\\
2dpoly & {\bf 0.83 {\tiny (0.69, 0.94) } }&0.43 {\tiny (0.25, 0.55) }&0.41 {\tiny (0.22, 0.52) }&0.35 {\tiny (0.15, 0.50) }&0.67 {\tiny (0.59, 0.73) }&0.66 {\tiny (0.54, 0.74) }\\
sigmoid & 0.60 {\tiny (0.18, 0.87) }&0.81 {\tiny (0.58, 0.94) }&0.25 {\tiny (0.16, 0.32) }&{\bf 0.89 {\tiny (0.80, 0.94) } }&-0.84 {\tiny (-1.18, -0.60) }&-1.15 {\tiny (-1.71, -0.53) }\\
step & 0.46 {\tiny (0.15, 0.67) }&0.59 {\tiny (0.43, 0.70) }&0.13 {\tiny (0.08, 0.18) }&{\bf 0.61 {\tiny (0.53, 0.67) } }&-0.51 {\tiny (-0.79, -0.30) }&-0.70 {\tiny (-1.09, -0.28) }\\
3dpoly & -0.49 {\tiny (-2.07, 0.41) }&-1.30 {\tiny (-2.26, -0.41) }&-0.38 {\tiny (-1.43, 0.18) }&-11.49 {\tiny (-18.81, -7.93) }&{\bf 0.37 {\tiny (0.14, 0.58) } }&0.30 {\tiny (-0.20, 0.58) }\\
sin & {\bf 0.83 {\tiny (0.58, 0.97) } }&0.72 {\tiny (0.49, 0.88) }&0.15 {\tiny (0.07, 0.21) }&0.74 {\tiny (0.59, 0.83) }&-0.46 {\tiny (-0.76, -0.18) }&-0.87 {\tiny (-1.40, -0.38) }\\
linear & 0.71 {\tiny (0.45, 0.92) }&0.87 {\tiny (0.74, 0.96) }&0.43 {\tiny (0.34, 0.49) }&{\bf 0.99 {\tiny (0.98, 1.00) } }&0.02 {\tiny (-0.16, 0.13) }&-0.02 {\tiny (-0.35, 0.28) }\\
rand\_pw & {\bf 0.66 {\tiny (-1.10, 0.94) } }&0.48 {\tiny (-0.31, 0.80) }&0.22 {\tiny (-0.47, 0.45) }&0.55 {\tiny (-0.94, 0.98) }&0.25 {\tiny (-3.75, 0.89) }&0.49 {\tiny (-2.88, 0.89) }\\
\end{tabular}
\end{footnotesize}
\caption{Instrument strength: 0.5, Number of instruments: 10, Number of samples: 1000, Number of experiments: 100, Training steps: 400, Number of critics: 50, Kernel radius: 50 data points, Critic jitter: 1, Test points: Distribution}
\end{figure}
\subsubsection{Instrument Strength: 0.7}
\begin{figure}[H]
\begin{footnotesize}
\begin{tabular}{c|c|c|c|c|c|c}
function & AGMM Avg & DeepIV & 2SLSpoly & 2SLS & DirectPoly & DirectNN\\
\hline\\
abs & {\bf 0.63 {\tiny (0.21, 0.89) } }&0.54 {\tiny (0.23, 0.66) }&0.59 {\tiny (0.39, 0.71) }&-0.46 {\tiny (-0.74, -0.25) }&-0.20 {\tiny (-0.52, 0.04) }&0.04 {\tiny (-0.45, 0.30) }\\
2dpoly & {\bf 0.92 {\tiny (0.83, 0.97) } }&0.83 {\tiny (0.72, 0.89) }&0.52 {\tiny (0.42, 0.62) }&0.28 {\tiny (0.04, 0.42) }&0.92 {\tiny (0.88, 0.94) }&0.91 {\tiny (0.84, 0.95) }\\
sigmoid & 0.87 {\tiny (0.69, 0.96) }&{\bf 0.91 {\tiny (0.79, 0.98) } }&0.18 {\tiny (0.11, 0.24) }&0.87 {\tiny (0.80, 0.92) }&0.50 {\tiny (0.32, 0.67) }&0.37 {\tiny (0.03, 0.59) }\\
step & 0.63 {\tiny (0.39, 0.77) }&{\bf 0.69 {\tiny (0.59, 0.77) } }&0.09 {\tiny (0.05, 0.13) }&0.60 {\tiny (0.54, 0.66) }&0.40 {\tiny (0.26, 0.54) }&0.42 {\tiny (0.14, 0.66) }\\
3dpoly & 0.26 {\tiny (-0.89, 0.66) }&-0.32 {\tiny (-1.25, 0.29) }&0.22 {\tiny (-0.13, 0.51) }&-11.29 {\tiny (-17.23, -7.29) }&{\bf 0.88 {\tiny (0.80, 0.93) } }&0.82 {\tiny (0.25, 0.94) }\\
sin & {\bf 0.91 {\tiny (0.78, 0.98) } }&0.84 {\tiny (0.70, 0.92) }&0.05 {\tiny (-0.00, 0.10) }&0.66 {\tiny (0.55, 0.74) }&0.67 {\tiny (0.50, 0.79) }&0.49 {\tiny (0.06, 0.72) }\\
linear & 0.94 {\tiny (0.83, 0.99) }&0.97 {\tiny (0.90, 0.99) }&0.36 {\tiny (0.28, 0.43) }&{\bf 1.00 {\tiny (0.98, 1.00) } }&0.74 {\tiny (0.66, 0.81) }&0.71 {\tiny (0.54, 0.85) }\\
rand\_pw & 0.78 {\tiny (-0.32, 0.97) }&0.72 {\tiny (0.11, 0.93) }&0.33 {\tiny (-0.11, 0.66) }&0.50 {\tiny (-0.84, 0.98) }&0.66 {\tiny (-0.78, 0.96) }&{\bf 0.80 {\tiny (-0.16, 0.96) } }\\
\end{tabular}
\end{footnotesize}
\caption{Instrument strength: 0.7, Number of instruments: 10, Number of samples: 1000, Number of experiments: 100, Training steps: 400, Number of critics: 50, Kernel radius: 50 data points, Critic jitter: 1, Test points: Distribution}
\end{figure}
\subsubsection{Instrument Strength: 0.9}
\begin{figure}[H]
\begin{footnotesize}
\begin{tabular}{c|c|c|c|c|c|c}
function & AGMM Avg & DeepIV & 2SLSpoly & 2SLS & DirectPoly & DirectNN\\
\hline\\
abs & 0.84 {\tiny (0.66, 0.94) }&0.77 {\tiny (0.56, 0.88) }&0.75 {\tiny (0.68, 0.81) }&-0.48 {\tiny (-0.72, -0.28) }&0.66 {\tiny (0.57, 0.73) }&{\bf 0.91 {\tiny (0.78, 0.97) } }\\
2dpoly & 0.95 {\tiny (0.87, 0.99) }&0.94 {\tiny (0.83, 0.97) }&0.63 {\tiny (0.48, 0.75) }&0.08 {\tiny (-0.15, 0.26) }&{\bf 1.00 {\tiny (0.99, 1.00) } }&0.99 {\tiny (0.97, 1.00) }\\
sigmoid & 0.85 {\tiny (0.58, 0.96) }&0.95 {\tiny (0.87, 0.99) }&0.14 {\tiny (0.09, 0.19) }&0.84 {\tiny (0.78, 0.89) }&{\bf 0.95 {\tiny (0.90, 0.96) } }&0.93 {\tiny (0.82, 0.98) }\\
step & 0.59 {\tiny (0.24, 0.80) }&0.77 {\tiny (0.64, 0.84) }&0.07 {\tiny (0.03, 0.11) }&0.60 {\tiny (0.54, 0.65) }&0.72 {\tiny (0.69, 0.74) }&{\bf 0.85 {\tiny (0.75, 0.92) } }\\
3dpoly & 0.76 {\tiny (0.17, 0.90) }&0.55 {\tiny (-0.24, 0.83) }&0.75 {\tiny (0.60, 0.86) }&-8.82 {\tiny (-14.23, -5.54) }&{\bf 1.00 {\tiny (0.99, 1.00) } }&0.96 {\tiny (0.68, 0.99) }\\
sin & 0.76 {\tiny (0.33, 0.89) }&0.88 {\tiny (0.75, 0.95) }&-0.01 {\tiny (-0.05, 0.02) }&0.50 {\tiny (0.37, 0.58) }&{\bf 0.94 {\tiny (0.88, 0.96) } }&0.93 {\tiny (0.81, 0.98) }\\
linear & 0.98 {\tiny (0.91, 1.00) }&0.98 {\tiny (0.94, 1.00) }&0.36 {\tiny (0.29, 0.47) }&{\bf 1.00 {\tiny (0.99, 1.00) } }&0.98 {\tiny (0.96, 1.00) }&0.97 {\tiny (0.93, 0.99) }\\
rand\_pw & 0.85 {\tiny (-0.64, 0.98) }&0.84 {\tiny (0.20, 0.98) }&0.50 {\tiny (-0.31, 0.71) }&0.20 {\tiny (-0.93, 0.96) }&0.79 {\tiny (-0.09, 0.98) }&{\bf 0.96 {\tiny (0.64, 0.99) } }\\
\end{tabular}
\end{footnotesize}
\caption{Instrument strength: 0.9, Number of instruments: 10, Number of samples: 1000, Number of experiments: 24, Training steps: 400, Number of critics: 50, Kernel radius: 50 data points, Critic jitter: 1, Test points: Distribution}
\end{figure}
\section{Grid Test Points}
\subsection{Number of Instruments:1}
\subsubsection{Instrument Strength: 0.5}
\begin{figure}[H]
\begin{footnotesize}
\begin{tabular}{c|c|c|c|c|c|c}
function & AGMM Avg & DeepIV & 2SLSpoly & 2SLS & DirectPoly & DirectNN\\
\hline\\
abs & {\bf 0.79 {\tiny (0.49, 0.93) } }&0.11 {\tiny (-0.09, 0.24) }&-0.16 {\tiny (-0.34, -0.02) }&-0.18 {\tiny (-0.38, -0.06) }&-3.25 {\tiny (-3.77, -2.65) }&-2.90 {\tiny (-3.78, -2.15) }\\
2dpoly & {\bf 0.97 {\tiny (0.91, 0.99) } }&0.49 {\tiny (0.32, 0.60) }&0.47 {\tiny (0.28, 0.59) }&0.56 {\tiny (0.42, 0.65) }&0.67 {\tiny (0.57, 0.72) }&0.66 {\tiny (0.55, 0.74) }\\
sigmoid & {\bf 0.90 {\tiny (0.60, 0.98) } }&0.53 {\tiny (0.30, 0.71) }&0.23 {\tiny (0.16, 0.33) }&0.89 {\tiny (0.79, 0.96) }&-1.16 {\tiny (-1.43, -0.82) }&-1.26 {\tiny (-1.91, -0.75) }\\
step & {\bf 0.67 {\tiny (0.41, 0.79) } }&0.40 {\tiny (0.22, 0.55) }&0.14 {\tiny (0.08, 0.21) }&0.66 {\tiny (0.58, 0.73) }&-1.05 {\tiny (-1.35, -0.74) }&-1.11 {\tiny (-1.63, -0.59) }\\
3dpoly & 0.21 {\tiny (-1.30, 0.78) }&-0.34 {\tiny (-1.31, -0.00) }&0.03 {\tiny (-0.59, 0.40) }&-9.57 {\tiny (-15.19, -5.14) }&{\bf 0.44 {\tiny (0.20, 0.65) } }&0.35 {\tiny (-0.16, 0.65) }\\
sin & {\bf 0.89 {\tiny (0.58, 0.98) } }&0.41 {\tiny (0.25, 0.58) }&0.12 {\tiny (0.04, 0.19) }&0.74 {\tiny (0.59, 0.85) }&-0.67 {\tiny (-0.96, -0.40) }&-0.94 {\tiny (-1.39, -0.55) }\\
linear & 0.97 {\tiny (0.90, 0.99) }&0.67 {\tiny (0.50, 0.79) }&0.43 {\tiny (0.35, 0.52) }&{\bf 1.00 {\tiny (0.98, 1.00) } }&0.00 {\tiny (-0.14, 0.14) }&0.00 {\tiny (-0.27, 0.25) }\\
rand\_pw & {\bf 0.86 {\tiny (0.04, 0.98) } }&0.41 {\tiny (-0.14, 0.71) }&0.26 {\tiny (-0.15, 0.48) }&0.61 {\tiny (-0.69, 0.99) }&0.26 {\tiny (-3.73, 0.88) }&0.38 {\tiny (-3.61, 0.88) }\\
\end{tabular}
\end{footnotesize}
\caption{Instrument strength: 0.5, Number of instruments: 1, Number of samples: 1000, Number of experiments: 100, Training steps: 400, Number of critics: 50, Kernel radius: 50 data points, Critic jitter: 1, Test points: Grid}
\end{figure}
\subsubsection{Instrument Strength: 0.7}
\begin{figure}[H]
\begin{footnotesize}
\begin{tabular}{c|c|c|c|c|c|c}
function & AGMM Avg & DeepIV & 2SLSpoly & 2SLS & DirectPoly & DirectNN\\
\hline\\
abs & {\bf 0.90 {\tiny (0.75, 0.97) } }&0.61 {\tiny (0.43, 0.71) }&0.63 {\tiny (0.49, 0.72) }&-0.17 {\tiny (-0.36, -0.05) }&-0.28 {\tiny (-0.67, 0.02) }&-0.08 {\tiny (-0.73, 0.25) }\\
2dpoly & {\bf 0.99 {\tiny (0.96, 1.00) } }&0.89 {\tiny (0.81, 0.93) }&0.56 {\tiny (0.42, 0.64) }&0.51 {\tiny (0.33, 0.61) }&0.91 {\tiny (0.89, 0.94) }&0.91 {\tiny (0.85, 0.95) }\\
sigmoid & {\bf 0.95 {\tiny (0.85, 0.98) } }&0.89 {\tiny (0.75, 0.97) }&0.20 {\tiny (0.14, 0.26) }&0.89 {\tiny (0.82, 0.94) }&0.43 {\tiny (0.21, 0.60) }&0.33 {\tiny (-0.09, 0.64) }\\
step & {\bf 0.81 {\tiny (0.70, 0.88) } }&0.71 {\tiny (0.59, 0.82) }&0.12 {\tiny (0.07, 0.16) }&0.66 {\tiny (0.59, 0.72) }&0.26 {\tiny (0.05, 0.43) }&0.32 {\tiny (-0.05, 0.54) }\\
3dpoly & 0.89 {\tiny (0.78, 0.96) }&0.15 {\tiny (-0.73, 0.59) }&0.37 {\tiny (0.09, 0.59) }&-8.53 {\tiny (-14.56, -4.68) }&{\bf 0.90 {\tiny (0.83, 0.94) } }&0.83 {\tiny (0.46, 0.94) }\\
sin & {\bf 0.95 {\tiny (0.85, 0.98) } }&0.78 {\tiny (0.64, 0.92) }&0.05 {\tiny (-0.01, 0.09) }&0.68 {\tiny (0.57, 0.77) }&0.61 {\tiny (0.42, 0.73) }&0.45 {\tiny (0.05, 0.77) }\\
linear & 0.98 {\tiny (0.94, 0.99) }&0.96 {\tiny (0.87, 0.98) }&0.39 {\tiny (0.32, 0.45) }&{\bf 1.00 {\tiny (0.99, 1.00) } }&0.73 {\tiny (0.65, 0.81) }&0.72 {\tiny (0.59, 0.85) }\\
rand\_pw & {\bf 0.94 {\tiny (0.62, 0.99) } }&0.80 {\tiny (0.20, 0.96) }&0.35 {\tiny (-0.13, 0.68) }&0.61 {\tiny (-0.79, 0.98) }&0.68 {\tiny (-0.79, 0.95) }&0.82 {\tiny (-0.09, 0.96) }\\
\end{tabular}
\end{footnotesize}
\caption{Instrument strength: 0.7, Number of instruments: 1, Number of samples: 1000, Number of experiments: 100, Training steps: 400, Number of critics: 50, Kernel radius: 50 data points, Critic jitter: 1, Test points: Grid}
\end{figure}
\subsubsection{Instrument Strength: 0.9}
\begin{figure}[H]
\begin{footnotesize}
\begin{tabular}{c|c|c|c|c|c|c}
function & AGMM Avg & DeepIV & 2SLSpoly & 2SLS & DirectPoly & DirectNN\\
\hline\\
abs & {\bf 0.96 {\tiny (0.89, 0.98) } }&0.86 {\tiny (0.74, 0.95) }&0.78 {\tiny (0.71, 0.83) }&-0.17 {\tiny (-0.35, -0.07) }&0.69 {\tiny (0.57, 0.75) }&0.91 {\tiny (0.77, 0.97) }\\
2dpoly & 1.00 {\tiny (0.99, 1.00) }&0.96 {\tiny (0.90, 0.99) }&0.57 {\tiny (0.43, 0.65) }&0.38 {\tiny (0.21, 0.50) }&{\bf 1.00 {\tiny (0.99, 1.00) } }&0.99 {\tiny (0.97, 1.00) }\\
sigmoid & {\bf 0.97 {\tiny (0.91, 0.99) } }&0.96 {\tiny (0.87, 0.99) }&0.18 {\tiny (0.11, 0.22) }&0.87 {\tiny (0.81, 0.92) }&0.94 {\tiny (0.90, 0.96) }&0.93 {\tiny (0.85, 0.98) }\\
step & {\bf 0.90 {\tiny (0.84, 0.94) } }&0.85 {\tiny (0.75, 0.90) }&0.11 {\tiny (0.05, 0.15) }&0.67 {\tiny (0.60, 0.72) }&0.75 {\tiny (0.70, 0.77) }&0.86 {\tiny (0.73, 0.94) }\\
3dpoly & 0.99 {\tiny (0.98, 1.00) }&0.72 {\tiny (0.06, 0.92) }&0.71 {\tiny (0.60, 0.79) }&-6.05 {\tiny (-9.81, -3.65) }&{\bf 1.00 {\tiny (0.99, 1.00) } }&0.96 {\tiny (0.73, 0.99) }\\
sin & {\bf 0.96 {\tiny (0.90, 0.98) } }&0.91 {\tiny (0.78, 0.96) }&-0.01 {\tiny (-0.06, 0.03) }&0.52 {\tiny (0.39, 0.64) }&0.94 {\tiny (0.90, 0.96) }&0.93 {\tiny (0.80, 0.98) }\\
linear & 0.99 {\tiny (0.97, 1.00) }&0.98 {\tiny (0.95, 0.99) }&0.37 {\tiny (0.32, 0.43) }&{\bf 1.00 {\tiny (0.99, 1.00) } }&0.98 {\tiny (0.96, 1.00) }&0.98 {\tiny (0.94, 1.00) }\\
rand\_pw & {\bf 0.98 {\tiny (0.87, 1.00) } }&0.93 {\tiny (0.59, 0.98) }&0.44 {\tiny (-0.25, 0.74) }&0.54 {\tiny (-0.83, 0.98) }&0.83 {\tiny (0.03, 0.99) }&0.96 {\tiny (0.81, 1.00) }\\
\end{tabular}
\end{footnotesize}
\caption{Instrument strength: 0.9, Number of instruments: 1, Number of samples: 1000, Number of experiments: 100, Training steps: 400, Number of critics: 50, Kernel radius: 50 data points, Critic jitter: 1, Test points: Grid}
\end{figure}
\subsection{Number of Instruments:2}
\subsubsection{Instrument Strength: 0.5}
\begin{figure}[H]
\begin{footnotesize}
\begin{tabular}{c|c|c|c|c|c|c}
function & AGMM Avg & DeepIV & 2SLSpoly & 2SLS & DirectPoly & DirectNN\\
\hline\\
abs & {\bf 0.70 {\tiny (0.39, 0.88) } }&0.12 {\tiny (-0.11, 0.24) }&-0.16 {\tiny (-0.38, 0.00) }&-0.18 {\tiny (-0.39, -0.06) }&-3.13 {\tiny (-4.15, -2.66) }&-2.89 {\tiny (-3.71, -1.97) }\\
2dpoly & {\bf 0.96 {\tiny (0.89, 0.98) } }&0.47 {\tiny (0.36, 0.59) }&0.45 {\tiny (0.30, 0.59) }&0.55 {\tiny (0.42, 0.63) }&0.67 {\tiny (0.57, 0.74) }&0.67 {\tiny (0.55, 0.76) }\\
sigmoid & 0.82 {\tiny (0.57, 0.94) }&0.54 {\tiny (0.33, 0.75) }&0.25 {\tiny (0.14, 0.33) }&{\bf 0.91 {\tiny (0.80, 0.96) } }&-1.13 {\tiny (-1.47, -0.85) }&-1.28 {\tiny (-1.92, -0.78) }\\
step & 0.52 {\tiny (0.18, 0.70) }&0.40 {\tiny (0.26, 0.55) }&0.15 {\tiny (0.08, 0.21) }&{\bf 0.67 {\tiny (0.56, 0.73) } }&-1.04 {\tiny (-1.43, -0.76) }&-1.04 {\tiny (-1.74, -0.58) }\\
3dpoly & -1.46 {\tiny (-3.57, -0.09) }&-0.43 {\tiny (-1.13, 0.08) }&0.03 {\tiny (-0.72, 0.47) }&-9.20 {\tiny (-15.45, -5.27) }&{\bf 0.46 {\tiny (0.07, 0.67) } }&0.36 {\tiny (-0.16, 0.66) }\\
sin & {\bf 0.84 {\tiny (0.47, 0.96) } }&0.42 {\tiny (0.24, 0.61) }&0.13 {\tiny (0.04, 0.20) }&0.75 {\tiny (0.59, 0.85) }&-0.68 {\tiny (-0.97, -0.45) }&-0.95 {\tiny (-1.56, -0.37) }\\
linear & 0.97 {\tiny (0.89, 1.00) }&0.67 {\tiny (0.52, 0.81) }&0.45 {\tiny (0.35, 0.53) }&{\bf 0.99 {\tiny (0.98, 1.00) } }&0.03 {\tiny (-0.16, 0.14) }&0.01 {\tiny (-0.25, 0.23) }\\
rand\_pw & {\bf 0.68 {\tiny (-0.62, 0.95) } }&0.41 {\tiny (-0.09, 0.70) }&0.26 {\tiny (-0.17, 0.48) }&0.64 {\tiny (-0.71, 0.98) }&0.22 {\tiny (-3.51, 0.88) }&0.41 {\tiny (-3.48, 0.89) }\\
\end{tabular}
\end{footnotesize}
\caption{Instrument strength: 0.5, Number of instruments: 2, Number of samples: 1000, Number of experiments: 100, Training steps: 400, Number of critics: 50, Kernel radius: 50 data points, Critic jitter: 1, Test points: Grid}
\end{figure}
\subsubsection{Instrument Strength: 0.7}
\begin{figure}[H]
\begin{footnotesize}
\begin{tabular}{c|c|c|c|c|c|c}
function & AGMM Avg & DeepIV & 2SLSpoly & 2SLS & DirectPoly & DirectNN\\
\hline\\
abs & {\bf 0.85 {\tiny (0.64, 0.94) } }&0.65 {\tiny (0.47, 0.73) }&0.65 {\tiny (0.46, 0.75) }&-0.16 {\tiny (-0.32, -0.07) }&-0.34 {\tiny (-0.68, 0.06) }&-0.15 {\tiny (-0.61, 0.30) }\\
2dpoly & {\bf 0.97 {\tiny (0.94, 0.99) } }&0.87 {\tiny (0.77, 0.93) }&0.55 {\tiny (0.44, 0.65) }&0.50 {\tiny (0.37, 0.61) }&0.91 {\tiny (0.87, 0.94) }&0.91 {\tiny (0.85, 0.95) }\\
sigmoid & {\bf 0.90 {\tiny (0.75, 0.96) } }&0.88 {\tiny (0.74, 0.97) }&0.21 {\tiny (0.13, 0.27) }&0.90 {\tiny (0.82, 0.94) }&0.40 {\tiny (0.18, 0.61) }&0.28 {\tiny (-0.14, 0.59) }\\
step & 0.70 {\tiny (0.48, 0.80) }&{\bf 0.72 {\tiny (0.51, 0.80) } }&0.12 {\tiny (0.07, 0.16) }&0.68 {\tiny (0.60, 0.72) }&0.24 {\tiny (0.05, 0.43) }&0.28 {\tiny (-0.07, 0.58) }\\
3dpoly & 0.61 {\tiny (-0.04, 0.87) }&0.25 {\tiny (-0.47, 0.57) }&0.43 {\tiny (0.06, 0.63) }&-7.70 {\tiny (-12.40, -4.31) }&{\bf 0.90 {\tiny (0.83, 0.94) } }&0.86 {\tiny (0.48, 0.95) }\\
sin & {\bf 0.90 {\tiny (0.70, 0.97) } }&0.79 {\tiny (0.63, 0.90) }&0.05 {\tiny (-0.00, 0.10) }&0.69 {\tiny (0.58, 0.78) }&0.58 {\tiny (0.40, 0.75) }&0.44 {\tiny (0.05, 0.70) }\\
linear & 0.98 {\tiny (0.93, 1.00) }&0.95 {\tiny (0.88, 0.98) }&0.40 {\tiny (0.33, 0.46) }&{\bf 1.00 {\tiny (0.99, 1.00) } }&0.73 {\tiny (0.64, 0.81) }&0.72 {\tiny (0.54, 0.83) }\\
rand\_pw & {\bf 0.89 {\tiny (0.20, 0.98) } }&0.76 {\tiny (0.28, 0.96) }&0.33 {\tiny (-0.11, 0.69) }&0.60 {\tiny (-0.62, 0.98) }&0.67 {\tiny (-0.71, 0.95) }&0.78 {\tiny (0.15, 0.96) }\\
\end{tabular}
\end{footnotesize}
\caption{Instrument strength: 0.7, Number of instruments: 2, Number of samples: 1000, Number of experiments: 100, Training steps: 400, Number of critics: 50, Kernel radius: 50 data points, Critic jitter: 1, Test points: Grid}
\end{figure}
\subsubsection{Instrument Strength: 0.9}
\begin{figure}[H]
\begin{footnotesize}
\begin{tabular}{c|c|c|c|c|c|c}
function & AGMM Avg & DeepIV & 2SLSpoly & 2SLS & DirectPoly & DirectNN\\
\hline\\
abs & {\bf 0.92 {\tiny (0.80, 0.97) } }&0.86 {\tiny (0.71, 0.93) }&0.78 {\tiny (0.71, 0.84) }&-0.17 {\tiny (-0.35, -0.07) }&0.68 {\tiny (0.57, 0.76) }&0.89 {\tiny (0.76, 0.96) }\\
2dpoly & 0.98 {\tiny (0.96, 0.99) }&0.96 {\tiny (0.87, 0.99) }&0.58 {\tiny (0.45, 0.67) }&0.38 {\tiny (0.21, 0.51) }&{\bf 0.99 {\tiny (0.99, 1.00) } }&0.99 {\tiny (0.97, 1.00) }\\
sigmoid & 0.93 {\tiny (0.82, 0.98) }&{\bf 0.96 {\tiny (0.90, 0.99) } }&0.18 {\tiny (0.11, 0.24) }&0.88 {\tiny (0.81, 0.92) }&0.94 {\tiny (0.88, 0.96) }&0.91 {\tiny (0.79, 0.98) }\\
step & 0.78 {\tiny (0.63, 0.86) }&{\bf 0.83 {\tiny (0.73, 0.88) } }&0.11 {\tiny (0.06, 0.16) }&0.68 {\tiny (0.60, 0.72) }&0.74 {\tiny (0.69, 0.77) }&0.83 {\tiny (0.69, 0.93) }\\
3dpoly & 0.87 {\tiny (0.60, 0.94) }&0.78 {\tiny (0.16, 0.92) }&0.73 {\tiny (0.59, 0.82) }&-5.69 {\tiny (-9.21, -3.38) }&{\bf 1.00 {\tiny (0.99, 1.00) } }&0.97 {\tiny (0.79, 1.00) }\\
sin & 0.91 {\tiny (0.74, 0.96) }&0.90 {\tiny (0.76, 0.96) }&-0.00 {\tiny (-0.05, 0.02) }&0.53 {\tiny (0.38, 0.62) }&{\bf 0.93 {\tiny (0.90, 0.96) } }&0.92 {\tiny (0.80, 0.98) }\\
linear & 0.99 {\tiny (0.96, 1.00) }&0.98 {\tiny (0.92, 0.99) }&0.39 {\tiny (0.32, 0.44) }&{\bf 1.00 {\tiny (0.99, 1.00) } }&0.98 {\tiny (0.96, 1.00) }&0.97 {\tiny (0.93, 0.99) }\\
rand\_pw & 0.93 {\tiny (0.43, 0.99) }&0.92 {\tiny (0.53, 0.99) }&0.42 {\tiny (-0.20, 0.76) }&0.50 {\tiny (-0.88, 0.98) }&0.85 {\tiny (-0.05, 0.99) }&{\bf 0.96 {\tiny (0.86, 0.99) } }\\
\end{tabular}
\end{footnotesize}
\caption{Instrument strength: 0.9, Number of instruments: 2, Number of samples: 1000, Number of experiments: 100, Training steps: 400, Number of critics: 50, Kernel radius: 50 data points, Critic jitter: 1, Test points: Grid}
\end{figure}
\subsection{Number of Instruments:3}
\subsubsection{Instrument Strength: 0.5}
\begin{figure}[H]
\begin{footnotesize}
\begin{tabular}{c|c|c|c|c|c|c}
function & AGMM Avg & DeepIV & 2SLSpoly & 2SLS & DirectPoly & DirectNN\\
\hline\\
abs & {\bf 0.71 {\tiny (0.47, 0.87) } }&0.11 {\tiny (-0.10, 0.26) }&-0.15 {\tiny (-0.37, -0.00) }&-0.18 {\tiny (-0.38, -0.07) }&-3.19 {\tiny (-3.91, -2.65) }&-2.89 {\tiny (-3.94, -2.24) }\\
2dpoly & {\bf 0.95 {\tiny (0.89, 0.98) } }&0.45 {\tiny (0.31, 0.56) }&0.47 {\tiny (0.27, 0.57) }&0.55 {\tiny (0.43, 0.64) }&0.66 {\tiny (0.59, 0.73) }&0.65 {\tiny (0.56, 0.75) }\\
sigmoid & 0.85 {\tiny (0.61, 0.94) }&0.56 {\tiny (0.36, 0.73) }&0.25 {\tiny (0.16, 0.33) }&{\bf 0.91 {\tiny (0.78, 0.95) } }&-1.11 {\tiny (-1.42, -0.76) }&-1.24 {\tiny (-2.10, -0.79) }\\
step & 0.54 {\tiny (0.22, 0.71) }&0.41 {\tiny (0.26, 0.56) }&0.15 {\tiny (0.08, 0.20) }&{\bf 0.68 {\tiny (0.55, 0.73) } }&-1.01 {\tiny (-1.37, -0.70) }&-1.02 {\tiny (-1.69, -0.59) }\\
3dpoly & -1.95 {\tiny (-3.96, -0.15) }&-0.31 {\tiny (-1.31, 0.06) }&0.06 {\tiny (-0.57, 0.34) }&-9.14 {\tiny (-14.72, -5.14) }&{\bf 0.47 {\tiny (0.12, 0.63) } }&0.35 {\tiny (-0.23, 0.68) }\\
sin & {\bf 0.84 {\tiny (0.60, 0.96) } }&0.43 {\tiny (0.27, 0.56) }&0.13 {\tiny (0.05, 0.20) }&0.75 {\tiny (0.59, 0.84) }&-0.63 {\tiny (-0.93, -0.35) }&-0.96 {\tiny (-1.56, -0.43) }\\
linear & 0.98 {\tiny (0.93, 0.99) }&0.66 {\tiny (0.48, 0.80) }&0.43 {\tiny (0.36, 0.52) }&{\bf 1.00 {\tiny (0.98, 1.00) } }&0.01 {\tiny (-0.14, 0.14) }&0.01 {\tiny (-0.22, 0.27) }\\
rand\_pw & {\bf 0.70 {\tiny (-0.63, 0.94) } }&0.37 {\tiny (-0.09, 0.70) }&0.26 {\tiny (-0.10, 0.50) }&0.62 {\tiny (-0.48, 0.99) }&0.26 {\tiny (-3.44, 0.88) }&0.42 {\tiny (-3.24, 0.89) }\\
\end{tabular}
\end{footnotesize}
\caption{Instrument strength: 0.5, Number of instruments: 3, Number of samples: 1000, Number of experiments: 100, Training steps: 400, Number of critics: 50, Kernel radius: 50 data points, Critic jitter: 1, Test points: Grid}
\end{figure}
\subsubsection{Instrument Strength: 0.7}
\begin{figure}[H]
\begin{footnotesize}
\begin{tabular}{c|c|c|c|c|c|c}
function & AGMM Avg & DeepIV & 2SLSpoly & 2SLS & DirectPoly & DirectNN\\
\hline\\
abs & {\bf 0.86 {\tiny (0.65, 0.94) } }&0.61 {\tiny (0.46, 0.73) }&0.64 {\tiny (0.51, 0.74) }&-0.17 {\tiny (-0.38, -0.05) }&-0.26 {\tiny (-0.68, 0.01) }&-0.12 {\tiny (-0.56, 0.30) }\\
2dpoly & {\bf 0.96 {\tiny (0.92, 0.99) } }&0.88 {\tiny (0.78, 0.93) }&0.54 {\tiny (0.42, 0.62) }&0.49 {\tiny (0.38, 0.59) }&0.91 {\tiny (0.87, 0.93) }&0.91 {\tiny (0.86, 0.95) }\\
sigmoid & {\bf 0.91 {\tiny (0.77, 0.96) } }&0.88 {\tiny (0.74, 0.98) }&0.20 {\tiny (0.13, 0.27) }&0.90 {\tiny (0.81, 0.94) }&0.43 {\tiny (0.21, 0.60) }&0.33 {\tiny (-0.04, 0.65) }\\
step & 0.67 {\tiny (0.51, 0.79) }&{\bf 0.71 {\tiny (0.58, 0.81) } }&0.12 {\tiny (0.07, 0.16) }&0.67 {\tiny (0.59, 0.72) }&0.28 {\tiny (0.04, 0.42) }&0.26 {\tiny (-0.05, 0.56) }\\
3dpoly & 0.47 {\tiny (-0.67, 0.85) }&0.17 {\tiny (-0.58, 0.55) }&0.42 {\tiny (0.06, 0.61) }&-7.90 {\tiny (-14.04, -4.80) }&{\bf 0.90 {\tiny (0.84, 0.94) } }&0.85 {\tiny (0.54, 0.94) }\\
sin & {\bf 0.88 {\tiny (0.70, 0.96) } }&0.80 {\tiny (0.66, 0.91) }&0.05 {\tiny (-0.01, 0.10) }&0.69 {\tiny (0.55, 0.78) }&0.61 {\tiny (0.41, 0.76) }&0.45 {\tiny (0.09, 0.77) }\\
linear & 0.98 {\tiny (0.95, 1.00) }&0.94 {\tiny (0.85, 0.98) }&0.39 {\tiny (0.30, 0.46) }&{\bf 1.00 {\tiny (0.99, 1.00) } }&0.74 {\tiny (0.66, 0.82) }&0.72 {\tiny (0.60, 0.84) }\\
rand\_pw & {\bf 0.86 {\tiny (0.09, 0.99) } }&0.77 {\tiny (0.19, 0.95) }&0.36 {\tiny (-0.14, 0.67) }&0.59 {\tiny (-0.61, 0.99) }&0.70 {\tiny (-0.69, 0.95) }&0.81 {\tiny (-0.22, 0.97) }\\
\end{tabular}
\end{footnotesize}
\caption{Instrument strength: 0.7, Number of instruments: 3, Number of samples: 1000, Number of experiments: 100, Training steps: 400, Number of critics: 50, Kernel radius: 50 data points, Critic jitter: 1, Test points: Grid}
\end{figure}
\subsubsection{Instrument Strength: 0.9}
\begin{figure}[H]
\begin{footnotesize}
\begin{tabular}{c|c|c|c|c|c|c}
function & AGMM Avg & DeepIV & 2SLSpoly & 2SLS & DirectPoly & DirectNN\\
\hline\\
abs & {\bf 0.91 {\tiny (0.78, 0.97) } }&0.82 {\tiny (0.67, 0.90) }&0.78 {\tiny (0.72, 0.83) }&-0.17 {\tiny (-0.35, -0.08) }&0.69 {\tiny (0.58, 0.76) }&0.91 {\tiny (0.77, 0.98) }\\
2dpoly & 0.97 {\tiny (0.93, 0.99) }&0.95 {\tiny (0.85, 0.98) }&0.59 {\tiny (0.46, 0.72) }&0.36 {\tiny (0.24, 0.47) }&{\bf 0.99 {\tiny (0.99, 1.00) } }&0.99 {\tiny (0.97, 1.00) }\\
sigmoid & 0.92 {\tiny (0.80, 0.97) }&{\bf 0.95 {\tiny (0.89, 0.99) } }&0.17 {\tiny (0.11, 0.22) }&0.87 {\tiny (0.80, 0.92) }&0.94 {\tiny (0.89, 0.96) }&0.94 {\tiny (0.83, 0.98) }\\
step & 0.74 {\tiny (0.57, 0.84) }&0.82 {\tiny (0.71, 0.88) }&0.10 {\tiny (0.05, 0.14) }&0.67 {\tiny (0.60, 0.71) }&0.75 {\tiny (0.68, 0.77) }&{\bf 0.86 {\tiny (0.74, 0.93) } }\\
3dpoly & 0.88 {\tiny (0.47, 0.95) }&0.76 {\tiny (0.14, 0.91) }&0.75 {\tiny (0.63, 0.83) }&-6.29 {\tiny (-10.45, -3.75) }&{\bf 1.00 {\tiny (0.99, 1.00) } }&0.96 {\tiny (0.60, 1.00) }\\
sin & 0.85 {\tiny (0.68, 0.94) }&0.89 {\tiny (0.76, 0.97) }&-0.01 {\tiny (-0.05, 0.03) }&0.52 {\tiny (0.36, 0.63) }&{\bf 0.94 {\tiny (0.89, 0.96) } }&0.92 {\tiny (0.78, 0.98) }\\
linear & 0.99 {\tiny (0.97, 1.00) }&0.98 {\tiny (0.93, 1.00) }&0.37 {\tiny (0.31, 0.46) }&{\bf 1.00 {\tiny (0.99, 1.00) } }&0.98 {\tiny (0.96, 1.00) }&0.97 {\tiny (0.93, 0.99) }\\
rand\_pw & 0.92 {\tiny (0.17, 0.99) }&0.90 {\tiny (0.45, 0.98) }&0.45 {\tiny (-0.21, 0.75) }&0.52 {\tiny (-0.74, 0.98) }&0.83 {\tiny (-0.02, 0.99) }&{\bf 0.96 {\tiny (0.69, 0.99) } }\\
\end{tabular}
\end{footnotesize}
\caption{Instrument strength: 0.9, Number of instruments: 3, Number of samples: 1000, Number of experiments: 100, Training steps: 400, Number of critics: 50, Kernel radius: 50 data points, Critic jitter: 1, Test points: Grid}
\end{figure}
\subsection{Number of Instruments:5}
\subsubsection{Instrument Strength: 0.5}
\begin{figure}[H]
\begin{footnotesize}
\begin{tabular}{c|c|c|c|c|c|c}
function & AGMM Avg & DeepIV & 2SLSpoly & 2SLS & DirectPoly & DirectNN\\
\hline\\
abs & {\bf 0.68 {\tiny (0.43, 0.91) } }&0.15 {\tiny (-0.07, 0.30) }&-0.15 {\tiny (-0.34, 0.01) }&-0.17 {\tiny (-0.39, -0.06) }&-3.27 {\tiny (-3.97, -2.60) }&-2.97 {\tiny (-3.83, -2.37) }\\
2dpoly & {\bf 0.93 {\tiny (0.87, 0.97) } }&0.49 {\tiny (0.35, 0.59) }&0.47 {\tiny (0.30, 0.55) }&0.55 {\tiny (0.43, 0.65) }&0.67 {\tiny (0.58, 0.72) }&0.67 {\tiny (0.54, 0.74) }\\
sigmoid & 0.86 {\tiny (0.64, 0.96) }&0.59 {\tiny (0.43, 0.77) }&0.25 {\tiny (0.18, 0.34) }&{\bf 0.91 {\tiny (0.78, 0.95) } }&-1.12 {\tiny (-1.51, -0.79) }&-1.33 {\tiny (-1.99, -0.78) }\\
step & 0.55 {\tiny (0.28, 0.71) }&0.45 {\tiny (0.31, 0.58) }&0.15 {\tiny (0.09, 0.21) }&{\bf 0.68 {\tiny (0.56, 0.73) } }&-1.04 {\tiny (-1.38, -0.68) }&-1.03 {\tiny (-1.66, -0.45) }\\
3dpoly & -1.79 {\tiny (-4.77, -0.12) }&-0.54 {\tiny (-1.34, 0.11) }&-0.03 {\tiny (-0.80, 0.39) }&-9.36 {\tiny (-15.83, -5.84) }&{\bf 0.46 {\tiny (0.13, 0.65) } }&0.38 {\tiny (-0.08, 0.66) }\\
sin & {\bf 0.91 {\tiny (0.73, 0.98) } }&0.50 {\tiny (0.33, 0.67) }&0.14 {\tiny (0.07, 0.21) }&0.76 {\tiny (0.60, 0.85) }&-0.68 {\tiny (-0.99, -0.38) }&-0.98 {\tiny (-1.58, -0.48) }\\
linear & 0.97 {\tiny (0.92, 0.99) }&0.72 {\tiny (0.54, 0.87) }&0.44 {\tiny (0.36, 0.51) }&{\bf 0.99 {\tiny (0.98, 1.00) } }&0.00 {\tiny (-0.15, 0.13) }&0.01 {\tiny (-0.22, 0.23) }\\
rand\_pw & {\bf 0.62 {\tiny (-0.63, 0.95) } }&0.43 {\tiny (-0.19, 0.74) }&0.23 {\tiny (-0.22, 0.52) }&0.60 {\tiny (-0.63, 0.99) }&0.24 {\tiny (-3.87, 0.88) }&0.42 {\tiny (-3.23, 0.88) }\\
\end{tabular}
\end{footnotesize}
\caption{Instrument strength: 0.5, Number of instruments: 5, Number of samples: 1000, Number of experiments: 100, Training steps: 400, Number of critics: 50, Kernel radius: 50 data points, Critic jitter: 1, Test points: Grid}
\end{figure}
\subsubsection{Instrument Strength: 0.7}
\begin{figure}[H]
\begin{footnotesize}
\begin{tabular}{c|c|c|c|c|c|c}
function & AGMM Avg & DeepIV & 2SLSpoly & 2SLS & DirectPoly & DirectNN\\
\hline\\
abs & {\bf 0.81 {\tiny (0.60, 0.94) } }&0.60 {\tiny (0.43, 0.74) }&0.62 {\tiny (0.49, 0.74) }&-0.17 {\tiny (-0.35, -0.07) }&-0.31 {\tiny (-0.68, 0.03) }&-0.08 {\tiny (-0.54, 0.31) }\\
2dpoly & {\bf 0.96 {\tiny (0.92, 0.99) } }&0.88 {\tiny (0.81, 0.92) }&0.55 {\tiny (0.42, 0.64) }&0.51 {\tiny (0.37, 0.61) }&0.91 {\tiny (0.88, 0.94) }&0.91 {\tiny (0.85, 0.95) }\\
sigmoid & 0.89 {\tiny (0.66, 0.95) }&0.89 {\tiny (0.75, 0.95) }&0.19 {\tiny (0.14, 0.27) }&{\bf 0.90 {\tiny (0.81, 0.94) } }&0.41 {\tiny (0.24, 0.60) }&0.32 {\tiny (-0.16, 0.63) }\\
step & 0.63 {\tiny (0.33, 0.76) }&{\bf 0.72 {\tiny (0.60, 0.81) } }&0.11 {\tiny (0.07, 0.17) }&0.67 {\tiny (0.59, 0.72) }&0.26 {\tiny (0.08, 0.43) }&0.29 {\tiny (-0.07, 0.50) }\\
3dpoly & -0.14 {\tiny (-1.54, 0.62) }&0.16 {\tiny (-0.55, 0.58) }&0.38 {\tiny (0.03, 0.62) }&-8.43 {\tiny (-14.29, -5.05) }&{\bf 0.90 {\tiny (0.83, 0.94) } }&0.82 {\tiny (0.56, 0.94) }\\
sin & {\bf 0.85 {\tiny (0.65, 0.95) } }&0.80 {\tiny (0.62, 0.89) }&0.06 {\tiny (-0.00, 0.11) }&0.69 {\tiny (0.55, 0.78) }&0.59 {\tiny (0.42, 0.75) }&0.42 {\tiny (0.08, 0.72) }\\
linear & 0.98 {\tiny (0.94, 1.00) }&0.95 {\tiny (0.88, 0.99) }&0.38 {\tiny (0.32, 0.45) }&{\bf 1.00 {\tiny (0.99, 1.00) } }&0.73 {\tiny (0.64, 0.81) }&0.73 {\tiny (0.52, 0.86) }\\
rand\_pw & {\bf 0.82 {\tiny (-0.11, 0.97) } }&0.78 {\tiny (0.26, 0.95) }&0.36 {\tiny (-0.15, 0.65) }&0.59 {\tiny (-0.84, 0.98) }&0.66 {\tiny (-1.11, 0.96) }&0.80 {\tiny (0.13, 0.96) }\\
\end{tabular}
\end{footnotesize}
\caption{Instrument strength: 0.7, Number of instruments: 5, Number of samples: 1000, Number of experiments: 100, Training steps: 400, Number of critics: 50, Kernel radius: 50 data points, Critic jitter: 1, Test points: Grid}
\end{figure}
\subsubsection{Instrument Strength: 0.9}
\begin{figure}[H]
\begin{footnotesize}
\begin{tabular}{c|c|c|c|c|c|c}
function & AGMM Avg & DeepIV & 2SLSpoly & 2SLS & DirectPoly & DirectNN\\
\hline\\
abs & {\bf 0.89 {\tiny (0.76, 0.97) } }&0.81 {\tiny (0.69, 0.90) }&0.77 {\tiny (0.70, 0.83) }&-0.17 {\tiny (-0.35, -0.06) }&0.67 {\tiny (0.58, 0.76) }&0.87 {\tiny (0.73, 0.98) }\\
2dpoly & 0.96 {\tiny (0.92, 0.98) }&0.95 {\tiny (0.87, 0.98) }&0.63 {\tiny (0.47, 0.73) }&0.37 {\tiny (0.22, 0.49) }&{\bf 0.99 {\tiny (0.99, 1.00) } }&0.99 {\tiny (0.97, 1.00) }\\
sigmoid & 0.87 {\tiny (0.68, 0.95) }&{\bf 0.94 {\tiny (0.85, 0.98) } }&0.17 {\tiny (0.11, 0.23) }&0.87 {\tiny (0.79, 0.92) }&0.94 {\tiny (0.90, 0.96) }&0.91 {\tiny (0.81, 0.98) }\\
step & 0.62 {\tiny (0.33, 0.80) }&0.81 {\tiny (0.69, 0.86) }&0.10 {\tiny (0.07, 0.16) }&0.67 {\tiny (0.59, 0.72) }&0.74 {\tiny (0.69, 0.77) }&{\bf 0.85 {\tiny (0.70, 0.92) } }\\
3dpoly & 0.67 {\tiny (-0.02, 0.90) }&0.74 {\tiny (0.03, 0.93) }&0.77 {\tiny (0.64, 0.87) }&-6.45 {\tiny (-9.62, -4.13) }&{\bf 1.00 {\tiny (0.99, 1.00) } }&0.95 {\tiny (0.57, 1.00) }\\
sin & 0.75 {\tiny (0.55, 0.88) }&0.89 {\tiny (0.76, 0.96) }&-0.01 {\tiny (-0.06, 0.03) }&0.51 {\tiny (0.34, 0.64) }&{\bf 0.94 {\tiny (0.89, 0.96) } }&0.92 {\tiny (0.80, 0.98) }\\
linear & 0.99 {\tiny (0.96, 1.00) }&0.98 {\tiny (0.93, 0.99) }&0.37 {\tiny (0.32, 0.45) }&{\bf 1.00 {\tiny (0.99, 1.00) } }&0.98 {\tiny (0.97, 1.00) }&0.97 {\tiny (0.93, 1.00) }\\
rand\_pw & 0.81 {\tiny (0.09, 0.98) }&0.89 {\tiny (0.54, 0.97) }&0.33 {\tiny (-0.29, 0.69) }&0.51 {\tiny (-0.89, 0.97) }&0.82 {\tiny (0.00, 0.98) }&{\bf 0.96 {\tiny (0.63, 0.99) } }\\
\end{tabular}
\end{footnotesize}
\caption{Instrument strength: 0.9, Number of instruments: 5, Number of samples: 1000, Number of experiments: 32, Training steps: 400, Number of critics: 50, Kernel radius: 50 data points, Critic jitter: 1, Test points: Grid}
\end{figure}
\subsection{Number of Instruments:10}
\subsubsection{Instrument Strength: 0.5}
\begin{figure}[H]
\begin{footnotesize}
\begin{tabular}{c|c|c|c|c|c|c}
function & AGMM Avg & DeepIV & 2SLSpoly & 2SLS & DirectPoly & DirectNN\\
\hline\\
abs & -0.26 {\tiny (-1.20, 0.53) }&{\bf 0.16 {\tiny (-0.02, 0.32) } }&-0.13 {\tiny (-0.36, 0.01) }&-0.17 {\tiny (-0.38, -0.04) }&-3.24 {\tiny (-3.85, -2.60) }&-3.00 {\tiny (-3.97, -2.38) }\\
2dpoly & {\bf 0.84 {\tiny (0.70, 0.95) } }&0.52 {\tiny (0.36, 0.64) }&0.47 {\tiny (0.34, 0.58) }&0.56 {\tiny (0.40, 0.66) }&0.67 {\tiny (0.57, 0.74) }&0.66 {\tiny (0.54, 0.74) }\\
sigmoid & 0.58 {\tiny (0.09, 0.86) }&0.84 {\tiny (0.59, 0.95) }&0.30 {\tiny (0.20, 0.38) }&{\bf 0.91 {\tiny (0.82, 0.95) } }&-1.10 {\tiny (-1.48, -0.83) }&-1.31 {\tiny (-1.90, -0.68) }\\
step & 0.38 {\tiny (-0.08, 0.67) }&0.65 {\tiny (0.48, 0.75) }&0.19 {\tiny (0.11, 0.24) }&{\bf 0.68 {\tiny (0.59, 0.73) } }&-1.01 {\tiny (-1.38, -0.72) }&-1.12 {\tiny (-1.60, -0.60) }\\
3dpoly & -0.23 {\tiny (-1.66, 0.55) }&-0.67 {\tiny (-1.63, 0.07) }&-0.15 {\tiny (-0.90, 0.33) }&-9.12 {\tiny (-15.43, -5.94) }&{\bf 0.43 {\tiny (0.18, 0.65) } }&0.37 {\tiny (-0.09, 0.65) }\\
sin & {\bf 0.81 {\tiny (0.53, 0.97) } }&0.73 {\tiny (0.50, 0.89) }&0.18 {\tiny (0.09, 0.26) }&0.76 {\tiny (0.61, 0.85) }&-0.64 {\tiny (-0.97, -0.35) }&-1.01 {\tiny (-1.52, -0.44) }\\
linear & 0.73 {\tiny (0.50, 0.92) }&0.87 {\tiny (0.74, 0.96) }&0.48 {\tiny (0.38, 0.55) }&{\bf 0.99 {\tiny (0.98, 1.00) } }&0.02 {\tiny (-0.15, 0.12) }&-0.01 {\tiny (-0.32, 0.27) }\\
rand\_pw & {\bf 0.69 {\tiny (-1.27, 0.95) } }&0.51 {\tiny (-0.21, 0.85) }&0.30 {\tiny (-0.25, 0.51) }&0.63 {\tiny (-0.66, 0.98) }&0.24 {\tiny (-3.72, 0.89) }&0.43 {\tiny (-3.25, 0.89) }\\
\end{tabular}
\end{footnotesize}
\caption{Instrument strength: 0.5, Number of instruments: 10, Number of samples: 1000, Number of experiments: 100, Training steps: 400, Number of critics: 50, Kernel radius: 50 data points, Critic jitter: 1, Test points: Grid}
\end{figure}
\subsubsection{Instrument Strength: 0.7}
\begin{figure}[H]
\begin{footnotesize}
\begin{tabular}{c|c|c|c|c|c|c}
function & AGMM Avg & DeepIV & 2SLSpoly & 2SLS & DirectPoly & DirectNN\\
\hline\\
abs & {\bf 0.65 {\tiny (0.21, 0.91) } }&0.61 {\tiny (0.39, 0.71) }&0.64 {\tiny (0.50, 0.73) }&-0.17 {\tiny (-0.36, -0.05) }&-0.30 {\tiny (-0.65, 0.00) }&-0.08 {\tiny (-0.68, 0.21) }\\
2dpoly & {\bf 0.93 {\tiny (0.85, 0.98) } }&0.87 {\tiny (0.77, 0.92) }&0.55 {\tiny (0.44, 0.66) }&0.51 {\tiny (0.32, 0.61) }&0.92 {\tiny (0.88, 0.94) }&0.91 {\tiny (0.84, 0.95) }\\
sigmoid & 0.87 {\tiny (0.70, 0.95) }&{\bf 0.92 {\tiny (0.80, 0.98) } }&0.22 {\tiny (0.13, 0.29) }&0.89 {\tiny (0.82, 0.94) }&0.41 {\tiny (0.21, 0.61) }&0.33 {\tiny (-0.01, 0.54) }\\
step & 0.65 {\tiny (0.43, 0.79) }&{\bf 0.75 {\tiny (0.64, 0.82) } }&0.13 {\tiny (0.07, 0.18) }&0.67 {\tiny (0.59, 0.72) }&0.26 {\tiny (0.05, 0.44) }&0.31 {\tiny (0.01, 0.58) }\\
3dpoly & 0.42 {\tiny (-0.50, 0.78) }&0.04 {\tiny (-0.66, 0.53) }&0.37 {\tiny (0.06, 0.62) }&-8.35 {\tiny (-13.05, -4.90) }&{\bf 0.90 {\tiny (0.82, 0.94) } }&0.85 {\tiny (0.40, 0.95) }\\
sin & {\bf 0.92 {\tiny (0.79, 0.97) } }&0.85 {\tiny (0.70, 0.92) }&0.06 {\tiny (0.00, 0.12) }&0.68 {\tiny (0.58, 0.77) }&0.59 {\tiny (0.41, 0.74) }&0.44 {\tiny (0.03, 0.70) }\\
linear & 0.95 {\tiny (0.84, 0.99) }&0.97 {\tiny (0.91, 0.99) }&0.40 {\tiny (0.32, 0.48) }&{\bf 1.00 {\tiny (0.99, 1.00) } }&0.74 {\tiny (0.66, 0.81) }&0.72 {\tiny (0.54, 0.85) }\\
rand\_pw & {\bf 0.82 {\tiny (-0.26, 0.97) } }&0.77 {\tiny (0.14, 0.94) }&0.37 {\tiny (-0.13, 0.69) }&0.60 {\tiny (-0.67, 0.99) }&0.66 {\tiny (-0.78, 0.95) }&0.80 {\tiny (-0.06, 0.96) }\\
\end{tabular}
\end{footnotesize}
\caption{Instrument strength: 0.7, Number of instruments: 10, Number of samples: 1000, Number of experiments: 100, Training steps: 400, Number of critics: 50, Kernel radius: 50 data points, Critic jitter: 1, Test points: Grid}
\end{figure}
\subsubsection{Instrument Strength: 0.9}
\begin{figure}[H]
\begin{footnotesize}
\begin{tabular}{c|c|c|c|c|c|c}
function & AGMM Avg & DeepIV & 2SLSpoly & 2SLS & DirectPoly & DirectNN\\
\hline\\
abs & 0.86 {\tiny (0.66, 0.95) }&0.82 {\tiny (0.65, 0.90) }&0.77 {\tiny (0.70, 0.84) }&-0.17 {\tiny (-0.35, -0.06) }&0.68 {\tiny (0.58, 0.76) }&{\bf 0.91 {\tiny (0.77, 0.98) } }\\
2dpoly & 0.96 {\tiny (0.89, 0.99) }&0.95 {\tiny (0.87, 0.98) }&0.67 {\tiny (0.51, 0.78) }&0.38 {\tiny (0.17, 0.50) }&{\bf 1.00 {\tiny (0.99, 1.00) } }&0.99 {\tiny (0.97, 1.00) }\\
sigmoid & 0.87 {\tiny (0.65, 0.96) }&{\bf 0.95 {\tiny (0.89, 0.99) } }&0.17 {\tiny (0.11, 0.23) }&0.87 {\tiny (0.81, 0.91) }&0.94 {\tiny (0.90, 0.96) }&0.93 {\tiny (0.84, 0.98) }\\
step & 0.64 {\tiny (0.34, 0.81) }&0.81 {\tiny (0.69, 0.87) }&0.10 {\tiny (0.06, 0.15) }&0.66 {\tiny (0.60, 0.72) }&0.75 {\tiny (0.70, 0.77) }&{\bf 0.85 {\tiny (0.75, 0.94) } }\\
3dpoly & 0.83 {\tiny (0.37, 0.93) }&0.65 {\tiny (-0.21, 0.88) }&0.81 {\tiny (0.69, 0.90) }&-6.07 {\tiny (-10.31, -3.50) }&{\bf 1.00 {\tiny (0.99, 1.00) } }&0.97 {\tiny (0.78, 1.00) }\\
sin & 0.77 {\tiny (0.35, 0.90) }&0.88 {\tiny (0.77, 0.95) }&-0.01 {\tiny (-0.06, 0.02) }&0.52 {\tiny (0.39, 0.61) }&{\bf 0.94 {\tiny (0.90, 0.96) } }&0.93 {\tiny (0.79, 0.98) }\\
linear & 0.98 {\tiny (0.93, 1.00) }&0.98 {\tiny (0.94, 1.00) }&0.40 {\tiny (0.32, 0.50) }&{\bf 1.00 {\tiny (0.99, 1.00) } }&0.98 {\tiny (0.96, 1.00) }&0.98 {\tiny (0.94, 1.00) }\\
rand\_pw & 0.88 {\tiny (-0.61, 0.98) }&0.86 {\tiny (0.24, 0.98) }&0.52 {\tiny (-0.33, 0.73) }&0.44 {\tiny (-0.90, 0.97) }&0.81 {\tiny (-0.06, 0.98) }&{\bf 0.96 {\tiny (0.60, 0.99) } }\\
\end{tabular}
\end{footnotesize}
\caption{Instrument strength: 0.9, Number of instruments: 10, Number of samples: 1000, Number of experiments: 25, Training steps: 400, Number of critics: 50, Kernel radius: 50 data points, Critic jitter: 1, Test points: Grid}
\end{figure}

\newpage
\begin{center} \bf \Large Second DGP \end{center}
\section{Marginal Distribution Test Points}
\subsection{Number of Instruments:2}
\subsubsection{Instrument Strength: 0.5}
\begin{figure}[H]
\begin{footnotesize}
\begin{tabular}{c|c|c|c|c|c|c}
function & AGMM Avg & DeepIV & 2SLSpoly & 2SLS & DirectPoly & DirectNN\\
\hline\\
abs & {\bf 0.67 {\tiny (0.16, 0.85) } }&-0.16 {\tiny (-0.49, 0.08) }&-0.47 {\tiny (-0.91, -0.27) }&-0.50 {\tiny (-0.92, -0.28) }&-4.68 {\tiny (-5.62, -4.03) }&-4.63 {\tiny (-6.03, -3.59) }\\
2dpoly & {\bf 0.93 {\tiny (0.79, 0.98) } }&0.20 {\tiny (0.02, 0.33) }&0.18 {\tiny (-0.15, 0.36) }&0.39 {\tiny (0.24, 0.52) }&0.45 {\tiny (0.35, 0.55) }&0.41 {\tiny (0.27, 0.55) }\\
sigmoid & 0.84 {\tiny (0.35, 0.96) }&0.37 {\tiny (0.20, 0.52) }&0.17 {\tiny (0.08, 0.24) }&{\bf 0.88 {\tiny (0.69, 0.96) } }&-1.70 {\tiny (-2.13, -1.29) }&-2.21 {\tiny (-3.09, -1.44) }\\
step & 0.53 {\tiny (0.07, 0.70) }&0.25 {\tiny (0.14, 0.38) }&0.08 {\tiny (0.01, 0.12) }&{\bf 0.59 {\tiny (0.40, 0.67) } }&-0.87 {\tiny (-1.15, -0.55) }&-1.17 {\tiny (-1.65, -0.74) }\\
3dpoly & -2.29 {\tiny (-4.85, -0.39) }&-1.29 {\tiny (-2.61, -0.53) }&-1.40 {\tiny (-2.16, -0.59) }&-15.82 {\tiny (-26.59, -10.85) }&{\bf -0.38 {\tiny (-0.81, 0.07) } }&-0.57 {\tiny (-1.47, 0.08) }\\
sin & {\bf 0.84 {\tiny (0.40, 0.96) } }&0.29 {\tiny (0.15, 0.47) }&0.08 {\tiny (-0.00, 0.15) }&0.74 {\tiny (0.50, 0.87) }&-1.19 {\tiny (-1.58, -0.82) }&-1.87 {\tiny (-2.50, -1.31) }\\
linear & 0.95 {\tiny (0.79, 0.99) }&0.49 {\tiny (0.31, 0.65) }&0.35 {\tiny (0.27, 0.43) }&{\bf 0.99 {\tiny (0.96, 1.00) } }&-0.58 {\tiny (-0.76, -0.37) }&-0.70 {\tiny (-1.04, -0.35) }\\
rand\_pw & {\bf 0.67 {\tiny (-0.63, 0.97) } }&0.25 {\tiny (-0.42, 0.56) }&0.11 {\tiny (-0.62, 0.38) }&0.60 {\tiny (-0.83, 0.98) }&-0.09 {\tiny (-6.81, 0.83) }&0.05 {\tiny (-5.34, 0.84) }\\
\end{tabular}
\end{footnotesize}
\caption{Instrument strength: 0.5, Number of instruments: 2, Number of samples: 1000, Number of experiments: 100, Training steps: 400, Number of critics: 50, Kernel radius: 50 data points, Critic jitter: 1, Test points: Distribution}
\end{figure}
\subsubsection{Instrument Strength: 0.7}
\begin{figure}[H]
\begin{footnotesize}
%
\end{footnotesize}
\caption{Instrument strength: 0.7, Number of instruments: 2, Number of samples: 1000, Number of experiments: 100, Training steps: 400, Number of critics: 50, Kernel radius: 50 data points, Critic jitter: 1, Test points: Distribution}
\end{figure}
\subsubsection{Instrument Strength: 0.9}
\begin{figure}[H]
\begin{footnotesize}
%
\end{footnotesize}
\caption{Instrument strength: 0.9, Number of instruments: 2, Number of samples: 1000, Number of experiments: 100, Training steps: 400, Number of critics: 50, Kernel radius: 50 data points, Critic jitter: 1, Test points: Distribution}
\end{figure}
\subsection{Number of Instruments:3}
\subsubsection{Instrument Strength: 0.5}
\begin{figure}[H]
\begin{footnotesize}
%
\end{footnotesize}
\caption{Instrument strength: 0.5, Number of instruments: 3, Number of samples: 1000, Number of experiments: 100, Training steps: 400, Number of critics: 50, Kernel radius: 50 data points, Critic jitter: 1, Test points: Distribution}
\end{figure}
\subsubsection{Instrument Strength: 0.7}
\begin{figure}[H]
\begin{footnotesize}
%
\end{footnotesize}
\caption{Instrument strength: 0.7, Number of instruments: 3, Number of samples: 1000, Number of experiments: 100, Training steps: 400, Number of critics: 50, Kernel radius: 50 data points, Critic jitter: 1, Test points: Distribution}
\end{figure}
\subsubsection{Instrument Strength: 0.9}
\begin{figure}[H]
\begin{footnotesize}
%
\end{footnotesize}
\caption{Instrument strength: 0.9, Number of instruments: 3, Number of samples: 1000, Number of experiments: 100, Training steps: 400, Number of critics: 50, Kernel radius: 50 data points, Critic jitter: 1, Test points: Distribution}
\end{figure}
\subsection{Number of Instruments:5}
\subsubsection{Instrument Strength: 0.5}
\begin{figure}[H]
\begin{footnotesize}
%
\end{footnotesize}
\caption{Instrument strength: 0.5, Number of instruments: 5, Number of samples: 1000, Number of experiments: 100, Training steps: 400, Number of critics: 50, Kernel radius: 50 data points, Critic jitter: 1, Test points: Distribution}
\end{figure}
\subsubsection{Instrument Strength: 0.7}
\begin{figure}[H]
\begin{footnotesize}
%
\end{footnotesize}
\caption{Instrument strength: 0.7, Number of instruments: 5, Number of samples: 1000, Number of experiments: 100, Training steps: 400, Number of critics: 50, Kernel radius: 50 data points, Critic jitter: 1, Test points: Distribution}
\end{figure}
\subsubsection{Instrument Strength: 0.9}
\begin{figure}[H]
\begin{footnotesize}
%
\end{footnotesize}
\caption{Instrument strength: 0.9, Number of instruments: 5, Number of samples: 1000, Number of experiments: 100, Training steps: 400, Number of critics: 50, Kernel radius: 50 data points, Critic jitter: 1, Test points: Distribution}
\end{figure}
\subsection{Number of Instruments:10}
\subsubsection{Instrument Strength: 0.5}
\begin{figure}[H]
\begin{footnotesize}
%
\end{footnotesize}
\caption{Instrument strength: 0.5, Number of instruments: 10, Number of samples: 1000, Number of experiments: 100, Training steps: 400, Number of critics: 50, Kernel radius: 50 data points, Critic jitter: 1, Test points: Distribution}
\end{figure}
\subsubsection{Instrument Strength: 0.7}
\begin{figure}[H]
\begin{footnotesize}
%
\end{footnotesize}
\caption{Instrument strength: 0.7, Number of instruments: 10, Number of samples: 1000, Number of experiments: 100, Training steps: 400, Number of critics: 50, Kernel radius: 50 data points, Critic jitter: 1, Test points: Distribution}
\end{figure}
\subsubsection{Instrument Strength: 0.9}
\begin{figure}[H]
\begin{footnotesize}
%
\end{footnotesize}
\caption{Instrument strength: 0.9, Number of instruments: 10, Number of samples: 1000, Number of experiments: 100, Training steps: 400, Number of critics: 50, Kernel radius: 50 data points, Critic jitter: 1, Test points: Distribution}
\end{figure}
\section{Grid Test Points}
\subsection{Number of Instruments:2}
\subsubsection{Instrument Strength: 0.5}
\begin{figure}[H]
\begin{footnotesize}
%
\end{footnotesize}
\caption{Instrument strength: 0.5, Number of instruments: 2, Number of samples: 1000, Number of experiments: 100, Training steps: 400, Number of critics: 50, Kernel radius: 50 data points, Critic jitter: 1, Test points: Grid}
\end{figure}
\subsubsection{Instrument Strength: 0.7}
\begin{figure}[H]
\begin{footnotesize}
%
\end{footnotesize}
\caption{Instrument strength: 0.7, Number of instruments: 2, Number of samples: 1000, Number of experiments: 100, Training steps: 400, Number of critics: 50, Kernel radius: 50 data points, Critic jitter: 1, Test points: Grid}
\end{figure}
\subsubsection{Instrument Strength: 0.9}
\begin{figure}[H]
\begin{footnotesize}
%
\end{footnotesize}
\caption{Instrument strength: 0.9, Number of instruments: 2, Number of samples: 1000, Number of experiments: 100, Training steps: 400, Number of critics: 50, Kernel radius: 50 data points, Critic jitter: 1, Test points: Grid}
\end{figure}
\subsection{Number of Instruments:3}
\subsubsection{Instrument Strength: 0.5}
\begin{figure}[H]
\begin{footnotesize}
%
\end{footnotesize}
\caption{Instrument strength: 0.5, Number of instruments: 3, Number of samples: 1000, Number of experiments: 100, Training steps: 400, Number of critics: 50, Kernel radius: 50 data points, Critic jitter: 1, Test points: Grid}
\end{figure}
\subsubsection{Instrument Strength: 0.7}
\begin{figure}[H]
\begin{footnotesize}
%
\end{footnotesize}
\caption{Instrument strength: 0.7, Number of instruments: 3, Number of samples: 1000, Number of experiments: 100, Training steps: 400, Number of critics: 50, Kernel radius: 50 data points, Critic jitter: 1, Test points: Grid}
\end{figure}
\subsubsection{Instrument Strength: 0.9}
\begin{figure}[H]
\begin{footnotesize}
%
\end{footnotesize}
\caption{Instrument strength: 0.9, Number of instruments: 3, Number of samples: 1000, Number of experiments: 100, Training steps: 400, Number of critics: 50, Kernel radius: 50 data points, Critic jitter: 1, Test points: Grid}
\end{figure}
\subsection{Number of Instruments:5}
\subsubsection{Instrument Strength: 0.5}
\begin{figure}[H]
\begin{footnotesize}
%
\end{footnotesize}
\caption{Instrument strength: 0.5, Number of instruments: 5, Number of samples: 1000, Number of experiments: 100, Training steps: 400, Number of critics: 50, Kernel radius: 50 data points, Critic jitter: 1, Test points: Grid}
\end{figure}
\subsubsection{Instrument Strength: 0.7}
\begin{figure}[H]
\begin{footnotesize}
%
\end{footnotesize}
\caption{Instrument strength: 0.7, Number of instruments: 5, Number of samples: 1000, Number of experiments: 100, Training steps: 400, Number of critics: 50, Kernel radius: 50 data points, Critic jitter: 1, Test points: Grid}
\end{figure}
\subsubsection{Instrument Strength: 0.9}
\begin{figure}[H]
\begin{footnotesize}
%
\end{footnotesize}
\caption{Instrument strength: 0.9, Number of instruments: 5, Number of samples: 1000, Number of experiments: 100, Training steps: 400, Number of critics: 50, Kernel radius: 50 data points, Critic jitter: 1, Test points: Grid}
\end{figure}
\subsection{Number of Instruments:10}
\subsubsection{Instrument Strength: 0.5}
\begin{figure}[H]
\begin{footnotesize}
%
\end{footnotesize}
\caption{Instrument strength: 0.5, Number of instruments: 10, Number of samples: 1000, Number of experiments: 100, Training steps: 400, Number of critics: 50, Kernel radius: 50 data points, Critic jitter: 1, Test points: Grid}
\end{figure}
\subsubsection{Instrument Strength: 0.7}
\begin{figure}[H]
\begin{footnotesize}
%
\end{footnotesize}
\caption{Instrument strength: 0.7, Number of instruments: 10, Number of samples: 1000, Number of experiments: 100, Training steps: 400, Number of critics: 50, Kernel radius: 50 data points, Critic jitter: 1, Test points: Grid}
\end{figure}
\subsubsection{Instrument Strength: 0.9}
\begin{figure}[H]
\begin{footnotesize}
%
\end{footnotesize}
\caption{Instrument strength: 0.9, Number of instruments: 10, Number of samples: 1000, Number of experiments: 100, Training steps: 400, Number of critics: 50, Kernel radius: 50 data points, Critic jitter: 1, Test points: Grid}
\end{figure}

\end{appendix}
\end{document}